\documentclass[longauth, colorlinks=true, citecolor=brown, linkcolor=blue]{aa}  

\usepackage{graphicx}
\usepackage{txfonts}
\usepackage{multicol}
\usepackage{comment}
\usepackage{orcidlink}
\usepackage{colortbl}
\usepackage{enumitem}
\usepackage{siunitx} 
\usepackage[table,xcdraw]{xcolor}
\usepackage{amsfonts, amsmath, amssymb}
\usepackage{subcaption}

\usepackage{mathrsfs}  
\usepackage{upgreek}   
\usepackage{booktabs}
\usepackage{adjustbox}
\usepackage{siunitx}
\usepackage{rotating}
\usepackage{dblfloatfix}

\usepackage{lscape}             
\usepackage{placeins}           

\usepackage{svg}
\renewcommand{\arraystretch}{1.3} 

\usepackage{soul}

\newif\ifreferee
\refereetrue  

\let\oldtextbf\textbf
\renewcommand{\textbf}[1]{\ifreferee\oldtextbf{#1}\else#1\fi}

\usepackage{hyperref}

\begin{document} 

\authorrunning{Hazarika D. et al.}
\titlerunning{Multiband P-L relations and reddenning maps of Magellanic Clouds using S-PLUS classical Cepheids}
\title{Variables in S-PLUS:}
\subtitle{I. Multiband period-luminosity relations and reddening maps of the Magellanic Clouds using classical Cepheids}
\author{ 
  Debasish Hazarika\inst{\orcidlink{0000-0003-4379-6777}1,7}
  \corrauth{debasish.hazarika.22@alumnos.uda.cl}        
  \and Carlos E. Ferreira Lopes\inst{\orcidlink{0000-0002-8525-7977}1} \email{carlos.ferreira@uda.cl} \fnmsep\thanks{carlos.ferreira@uda.cl} \and
  Márcio Catelan\inst{\orcidlink{0000-0001-6003-8877}2,3} \and \\
  Felipe Almeida-Fernandes\inst{\orcidlink{0000-0002-8048-8717}5} \and
  Nicolás Rodriguez-Segovia\inst{\orcidlink{0000-0002-0125-1472}6} \and
  Vinicius M. Placco\inst{\orcidlink{0000-0003-4479-1265}7} \and
  Guilherme Limberg\inst{\orcidlink{0000-0002-9269-8287}8}\and \\
  Luis A. Gutiérrez-Soto\inst{\orcidlink{0000-0002-9891-8017}9} \and
  Vinicius Cordeiro\inst{\orcidlink{0000-0001-9079-9511}11}\and
  Marcelo Borges Fernandes\inst{\orcidlink{0000-0001-5740-2914}10} \and
  Pedro K. Humire\inst{\orcidlink{0000-0003-3537-4849}4} \and 
  William Schoenell\inst{\orcidlink{0000-0002-4064-7234}11}\and 
  Antonio Kanaan\inst{\orcidlink{0009-0007-8005-4541}12}\and
  Tiago Ribeiro\inst{\orcidlink{0000-0002-0138-1365}13} \and
  Claudia Mendes de Oliveira\inst{\orcidlink{0000-0002-5267-9065}4} 
}

\institute{
  Instituto de Astronom\'{i}a y Ciencias Planetarias, Universidad de Atacama, Copayapu 485, Copiap\'{o}, Chile 
  \and
  Instituto de Astrof\'{i}sica, Pontificia Universidad Cat\'{o}lica de Chile, Av. Vicu\~{n}a Mackenna 4860, 7820436 Macul, Santiago, Chile
  \and
  Centro de Astro-Ingenier\'{i}a, Pontificia Universidad Cat\'{o}lica de Chile, Av. Vicu\~{n}a Mackenna 4860, 7820436 Macul, Santiago, Chile 
  \and
  Universidade de S\~{a}o Paulo, Instituto de Astronomia, Geof\'{i}sica e Ciências Atmosf\'{e}ricas, Departamento de Astronomia, Rua do Mat\~{a}o, 1226, 05509-090, S\~{a}o Paulo, Brazil
  \and
  Instituto Nacional de Pesquisas Espaciais, Av. dos Astronautas 1758, Jardim da Granja,12227-010 S\~ao Jos\'e dos Campos, SP, Brazil 
  \and
  School of Science, University of New South Wales, Australian Defense Force Academy, Northcott Dr, Canberra, ACT 2600, Australia
  \and
  NSF NOIRLab, Tucson, AZ 85719, USA 
  \and
  Kavli Institute for Cosmological Physics, University of Chicago, Chicago, IL 60637, USA
  \and
  Instituto de Astrofísica de La Plata (CCT La Plata - CONICET - UNLP), B1900FWA, La Plata, Argentina
  \and
  Observatório Nacional, Rua General José Cristino 77, CEP: 20921-400, São Cristóvão, Rio de Janeiro, Brazil
  \and
  The Observatories of the Carnegie Institution for Science, 813 Santa Barbara St, Pasadena, CA 91101, USA
  \and
  Departamento de F\'{i}sica, Universidade Federal de Santa Catarina, Florian\'{o}polis, SC, 88040-900, Brazil
  \and
  Rubin Observatory Project Office, 950 N. Cherry Ave., Tucson, AZ 85719, USA
}

\date{Received:  xxx  ; Accepted:  xxx }


\abstract
{Classical Cepheids (CCs) are key standard candles for establishing the extragalactic distance scale through their period--luminosity (P--L) relations. While these relations are well calibrated in broadband photometric systems, their behavior in narrowband filters remains largely unexplored due to the lack of suitable observations. Combining broad- and narrowband photometry can probe wavelength-dependent extinction and provide improved constraints on the dust distribution.}
{Our goal is to derive multiband P--L relations for CCs in the Magellanic Clouds (MCs) and construct Cepheid-based reddening maps using the Southern Photometric Local Universe Survey (S-PLUS) bands, which includes five broad bands ($u,g,r,i,z$) and seven narrow bands centered on spectral features: the Balmer jump/[OII], Ca H+K, H$\delta$, CH g-band, Mg b triplet, H$\alpha$, and the Ca triplet.}
{We adjusted the single-epoch magnitudes of a total of more than 8000 CCs in the MCs by employing a mathematical framework to disentangle the effects of relative reddening and relative distance with respect to their host galaxies across all 12 bands. The P--L coefficients were obtained via iterative least-squares fitting and maximum likelihood optimization, with uncertainties estimated through a Bayesian Markov chain Monte Carlo (MCMC) analysis using separate fits for fundamental and overtone pulsators.}
{We present the first Cepheid P--L relations in the S-PLUS photometric system, providing independent and consistent estimates of distance and reddening. The dispersion in the P--L relations decreases systematically toward longer wavelengths, while the P-L slopes vary from about $-2.0$ to $-2.8$ in the bluest bands to about $-2.9$ to $-3.2$ in the reddest bands. The resulting reddening maps reveal spatially nonuniform extinction across both galaxies and recover known structures, such as 30 Doradus and the HI supergiant shell SGS 12 (LMC 3). The inferred geometries and orientations of the MCs are consistent with previous studies.}
{Our results support that reliable distances and reddening can be recovered even from single-epoch photometry when combined with multiband data and corrected using light-curve information from other bands, enabling future high-precision 3D mapping of the MCs.}

\keywords{stars: variables: Cepheids, Magellanic Clouds, stars: distances, methods: statistical, cosmology: distance scale}  

\maketitle
\nolinenumbers


\section{Introduction}
Cepheid variables serve as primary distance indicators in the first rung of the cosmic distance ladder through their well-established period--luminosity (P--L) \citep[Leavitt Law:][]{1908leavitt, 1912leavitt} and period-Wesenheit (PW) relations \citep{1982mado, Caputo2000}. Cepheids comprise two main subclasses, classical Cepheids (CCs) and type II Cepheids, which exhibit distinct light-curve morphologies that provide insights into their masses, ages, and evolutionary histories. CCs, also known as Population I or Delta Cepheid variables, undergo regular pulsations primarily in the fundamental (FU) and first-overtone (1O) modes. They are young massive stars with typical masses up to $\sim12$--$13,M_\odot$ and luminosities reaching $10^4$--$10^5,L_\odot$ \citep[e.g.,][and references therein]{2015catelanbook}. Their high luminosities make them powerful standard candles for measuring distances well beyond the Milky Way \citep{2024bono}, and for calibrating secondary indicators such as Type Ia supernovae. Consequently, accurate distance measurements to nearby calibrator galaxies such as the Large and Small Magellanic Clouds (LMC and SMC, together known as the MCs) are critical for empirical determinations of the Hubble constant ($H_0$) and for reducing associated systematic uncertainties \citep[e.g.,][]{2016riess,2018riess,2021hubble,2024hubble}. Precise distances to individual Cepheids across the MCs further enable reconstruction of their 3D structure, providing insight into disk geometry, depth, warp, and tidal features driven by interactions between the MCs and the Milky Way (MW) \citep[e.g.,][]{2016Jacy, 2017_OGLE_CC_MC_updated, 2022ripepi}.

The LMC and SMC are interacting satellite dwarf galaxies of the MW, located at distances of $\sim$50 \citep[][]{2019piet} and $\sim$60 kpc \citep[][]{2015bonno}, respectively. The LMC exhibits a barred disk structure with clear evidence of warping and tidal distortions \citep[][]{1972deva, 2001marel, 2018choi}, whereas the SMC shows a highly irregular morphology and significant line-of-sight depth, likely induced by repeated gravitational encounters with the LMC \citep[][]{2012subra, 2013nide, 2024alme}. Beyond distance-scale applications, the proximity of the MCs to the MW allows individual stars to be resolved across their full spatial extent, and the relatively modest foreground extinction, together with limited MW contamination, makes them ideal laboratories for studying stellar populations, star formation histories \citep[][]{2018rube, 2019joshi}, dust distribution, and the dynamical evolution of the Local Group.

Accurate characterization of the MCs requires careful treatment of interstellar reddening, which varies across different regions and is closely linked to star-forming environments. Reddening maps derived from tracers such as red clump stars \citep{Haschke2011,2018choi, 2021skowron} and Cepheid P--L relations \citep{niko2004, 2018deb_lmc, 2019deb_smc, 2019joshi} reveal clumpy and asymmetric dust distributions across both galaxies, with higher extinction in active regions such as 30 Doradus in the LMC and comparatively lower yet heterogeneous reddening in the SMC. Because extinction alters the observed P--L relation of CCs, spatially resolved corrections are essential for reliable distance and structural analyzes. Period-luminosity relations 
that employ the so-called Wesenheit indices, which are reddening-free by construction, are still affected by the choice of extinction law \citep{1982mado, FreedmanMadore2011, 2012madore}. On the other hand, multiwavelength analyzes of CCs were initially formulated by \citet{1985freed,1991freed} and later refined by \citet{2014rich}. This analysis method provides a robust framework to disentangle intrinsic luminosity from dust effects by fitting a reddening law to apparent distance moduli as a function of wavelength. This approach enables simultaneous estimation of distances and reddening for individual Cepheids, and increasing the number of photometric bands further strengthens the separation of the correlated effects of distance, reddening, age, and chemical composition \citep{niko2004}. Later, \citet{2016Inno} and \citet{2018deb_lmc, 2019deb_smc} demonstrated that such multiband analyzes produce tightened P--L relations, improved reddening maps, and robust reconstructions of the 3D geometry and viewing angles of both the LMC and SMC. Nevertheless, these methods generally consider a universal extinction law, and deviations from it may introduce residual systematic uncertainties.

Despite the growing number of multiwavelength surveys that extend into the near- and mid-IR \citep[e.g.,][]{2004Spitzer,2010VVV,2010WISE}, predominantly conducted in broadband filters, studies using narrow- and intermediate-band systems remain sparse for variability analysis. Although near-IR P--L relations benefit from reduced extinction and smaller intrinsic scatter \citep{2011storm,2012riess}, the smaller amplitudes and more symmetric light curves at longer wavelengths can make classification challenging \citep{2019dekany}. Alternatively, optical light curves remain central to variable star classification because of their larger amplitudes and distinctive morphologies \citep[e.g.,][]{2015wfcam,2019_gaiadr2_plr,2019gaia_ceph_rrly,2020ztfvariable,2023gaia_ceph, 2023riss, 2024tess, 2025ApJS..276...57G, 2025corot}.

Early studies using Walraven, Strömgren-like H$\beta$, and ultranarrow filters \citep{Pel1976, Eggen1985, Crawford1966} highlighted the potential of narrow bands to probe temperature variations and refine P--L relations. More recently, \citet{2025bras} introduced multiband light-curve templates from the optical to the mid-IR. However, Cepheid behavior in specialized narrowband systems such as the one employed by the Southern Photometric Local Universe Survey \citep[S-PLUS;][]{MendesOliveira-2019} remains largely unexplored. In the era of the Vera C. Rubin Observatory's Legacy Survey of Space and Time (LSST) \citep[][]{2023di_lsst, 2024lsst}, narrow- and intermediate-band studies are complementary for fully characterizing variable stars and their physical parameters. 

Recognizing this gap, we initiate the first study in the ``Variables in S-PLUS'' series, focusing on Cepheid variables in the MCs. The S-PLUS photometric system comprises five broadband filters ($u$, $g$, $r$, $i$, $z$), similar to those used in the Sloan Digital Sky Survey \citep[SDSS;][]{York-2000}, and seven narrowband filters ($J0378$, $J0395$, $J0410$, $J0430$, $J0515$, $J0660$, $J0861$) following the Javalambre photometric system \citep{MarinFranch:2012}, closely resembling the filter set used in the Javalambre Photometric Local Universe Survey (J-PLUS) \citep{2019cenarro}. These narrow bands are centered on key stellar spectral features: the Balmer jump/[OII] for $J0378$, Ca H+K for $J0395$, H$\delta$ for $J0410$, CH G-band for $J0430$, Mg b triplet for $J0515$, H$\alpha$ for $J0660$, and the Ca triplet for $J0861$. The $J0395$ filter, which measures the Ca H+K line, is highly sensitive to metallicity \citep[][]{2015squez}, and presents an opportunity to potentially disentangle the intricate effects of metallicity on the P--L relation. Additionally, the $J0515$ filter, used to detect Mg b, is particularly responsive to variations in surface gravity \citep{Geisler-1984, Majewski-2000}, and the $J0660$ filter, which traces H$\alpha$, is sensitive to the effective temperature and chromospheric activity of stars \citep[][and references therein]{2019giri}.

A practical consideration for variability studies with S-PLUS is that the survey provides single-epoch photometry, so each Cepheid is sampled at an essentially random pulsation phase. Recovering reliable mean magnitudes from such data is therefore nontrivial. Classical approaches involve (i) constructing template light curves from well-sampled time series in a given band and deriving the mean magnitudes of other bands by scaling the amplitudes and introducing a phase lag \citep[e.g.,][]{1988freed, 1996stet}; (ii) utilizing Fourier decomposition of well-sampled light curves in one band and parameterizing linear relationships between the Fourier coefficients across different bands \citep[e.g.,][]{1996jones, 2003ngeow, 2005tanvir, 2015inno, 2024sici}; or (iii) applying principal component analysis to the light curves \citep[e.g.,][]{2025bras}. An alternative approach, provided by \citet{niko2004},  models the observed magnitude with a phase-dependent correction term and enables statistical recovery of mean magnitudes from single-epoch data, making it well suited to the S-PLUS data in the absence of multi-epoch coverage for MCs CCs.

In this work, we construct multiband P--L relations for FU and 1O Cepheids in the MCs using single-epoch photometry for all 12 S-PLUS bands. We disentangle the effects of differential reddening and relative distance within each host galaxy by simultaneous fitting of the multiband distance moduli as a function of inverse effective wavelength, assuming the extinction law of \citet[][hereafter WC23]{2023wang}. While previous studies employing such a multiband approach were limited to at most seven broad bands, the present study spans 12 S-PLUS bands, many of them narrow bands, providing enhanced wavelength leverage, increasing the dimensionality of the analysis, and enabling improved separation of reddening and distance effects in the P--L relations. This framework further allows the reddening to be determined on a star-by-star basis, reducing reliance on external extinction maps and allowing the construction of spatially resolved reddening maps of the SMC and LMC. It also enables a detailed assessment of Cepheid behavior in narrowband filters, leading to improved constraints on extinction, stellar parameters, and distance estimates, particularly in regions affected by strong foreground and spatially variable dust.

This paper is structured as follows. In Sect. \ref{Sec:data}, we describe the photometric data used in this study and the criteria adopted for sample selection. Sect. \ref{Sec:Methodology} details the methodology employed to derive the P--L relation using multiband photometry from the S-PLUS survey, including our fitting procedures and error analysis. Sect. \ref{Sec:Results} provides a comprehensive discussion of the main findings, highlighting their implications in the context of distance scale, reddening maps, and stellar population studies. Finally, Sect. \ref{Sec:Conclusions} summarizes the main conclusions of our work.

\section{Data and sample selection}\label{Sec:data}

We utilized single-epoch photometry from S-PLUS Data Release 4 (DR4) \citep{Herpich:2024}, covering 3022.7 deg$^2$ across 1629 southern fields and spanning 3000--10000\,\AA\ in 12 optical filters. Observations were obtained with the 0.86\,m T80-South robotic telescope at Cerro Tololo Inter-American Observatory, Chile, which features a $1.4\times1.4$ deg$^2$ field of view and a plate scale of 0.55 arcsec\,pixel$^{-1}$ \citep{MendesOliveira-2019, Almeida-Fernandes-2022}. The DR4 provides aperture photometry for $\sim$4.6 million sources, mostly for MW, and point spread function (PSF)  photometry for $\sim$8.3 million point sources (S/N $>$ 3), including 150 fields in the MCs region. Each field is observed with three exposures per filter (36 exposures in total), which are co-added into a single science image per band \citep[see Table~5 of][]{MendesOliveira-2019}. This approach also facilitates obtaining closely spaced or nearly identical phase points for Cepheids for all three observations. To reduce the impact of bad pixels, slight dithering is implemented. While magnitudes are not corrected for interstellar extinction, airmass corrections are incorporated within the DR4 calibration pipeline \citep[see][for details regarding the calibration processes]{Herpich:2024}.

\begin{figure}[htbp]
\centering
\includegraphics[width=0.98\linewidth]{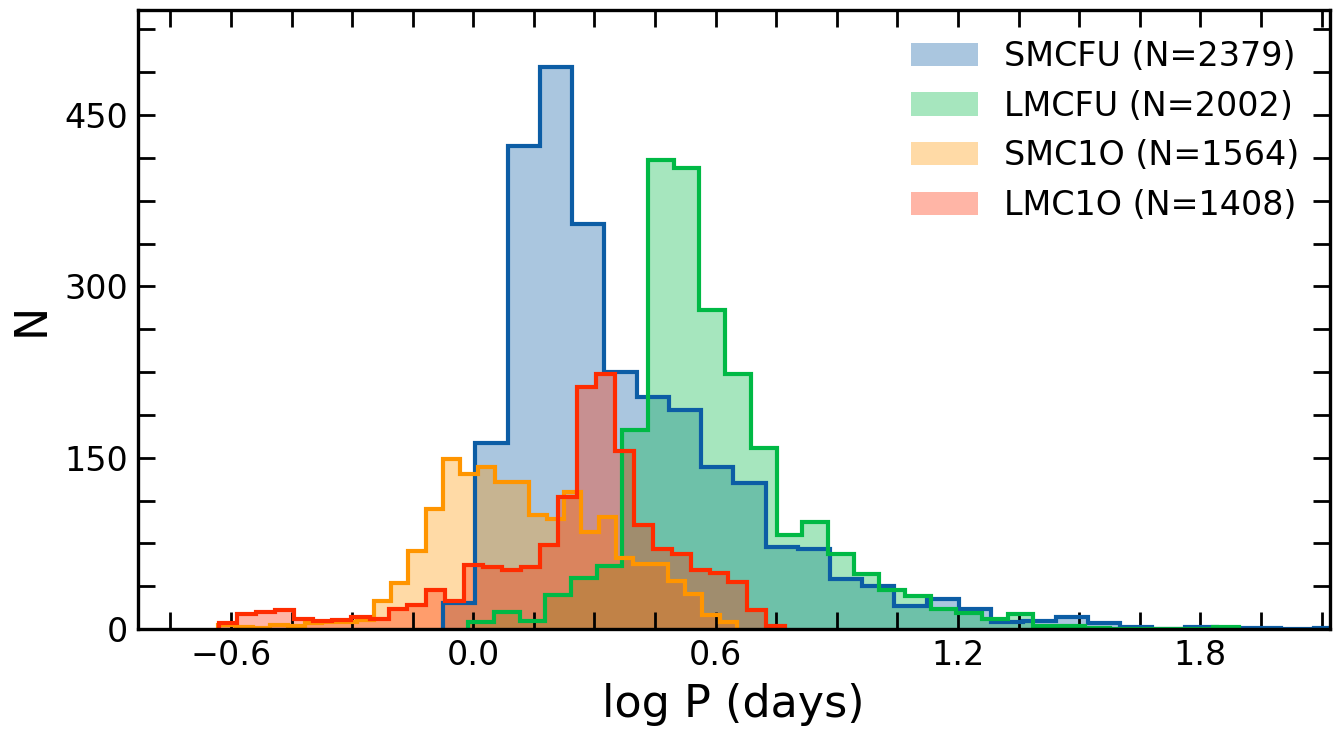}
\caption{Period distribution of FU and 1O Cepheids observed in all 12 S-PLUS bands in the SMC and LMC. The colors distinguish each sample, and the total number of stars is indicated in the legend.}
\label{fig:hist}
\end{figure}

We crossmatched the S-PLUS DR4 sources within a 1\arcsec\ limit with the Optical Gravitational Lensing Experiment data release IV catalog \citep[OGLE-IV;][]{2017_OGLE_CC_MC_updated}. The catalog of \citet{2017_OGLE_CC_MC_updated} is an updated version of \citet{2015_OGLE_CC_MC}, which contains a few newly identified Cepheids and totals 4704 CCs in the LMC and 4945 CCs in the SMC. The 1\arcsec\ matching radius was chosen to ensure a balance between precision and completeness in crossmatching, given the high astrometric accuracy of OGLE-IV ($\sim$0.1--0.3\arcsec\ in empty fields; \citealt{2015_OGLE_CC_MC}) and the slightly higher positional uncertainties ($\lesssim$0.3--0.4\arcsec) reported for the crowded LMC and SMC 
fields in S-PLUS DR4 \citep{Herpich:2024}. The observed separation distribution, sharply peaked below 0.5\arcsec, further supports this choice. This minimizes spurious matches while preserving genuine associations, which is particularly critical for the accurate identification of Cepheid variables. Furthermore, we applied cuts with S/N 
greater than five and uncertainties smaller than 0.1 mag in all 12 filters to ensure the reliability of the photometric measurements. The final selected sample consists of 2671 FU and 1728 1O CCs in the SMC, and 2224 FU and 1578 1O CCs in the LMC. It should be noted that not all CCs were observed in all 12 S-PLUS filters. Table~\ref{tab1:sample} presents the final sample of CCs obtained after crossmatching with the OGLE catalog \citep{2017_OGLE_CC_MC_updated}. For both FU and 1O CCs in the LMC and SMC, we report the number of sources detected in at least one band, in five or more bands, and in all 12 S-PLUS filters. Figure~\ref{fig:hist} illustrates the period distribution of Cepheids crossmatched in the SMC and LMC, taken from the OGLE-IV catalog, that were observed in all 12 S-PLUS filters.

\begin{table}[ht]
    \centering
    \caption{Number of classical Cepheids used in this study.}
    \resizebox{\linewidth}{!}{%
    \begin{tabular}{lcccccc}
\toprule &
 \multicolumn{3}{c}{FU Cepheids} & \multicolumn{3}{c}{1O Cepheids}  \\
\cmidrule(lr){2-4} \cmidrule(lr){5-7}
       & $\geq$1 band & $\geq$5 bands & 12 bands & $\geq$1 band & $\geq$5 bands & 12 bands  \\
\midrule
SMC    & 2671 & 2664 & 2379 & 1728 & 1717 & 1564  \\
LMC    & 2224 & 2217 & 2002 & 1578 & 1573 & 1408  \\
\bottomrule 
\end{tabular}}
\tablefoot{
The sample was obtained by crossmatching S-PLUS DR4 \citep{Herpich:2024} with OGLE-IV
classical Cepheids \citep{2017_OGLE_CC_MC_updated}. For each galaxy and pulsation mode, the number of sources detected in at least one band, in five or more bands, and in all twelve S-PLUS bands are reported.}
\label{tab1:sample}
\end{table} 

\section{Methodology}\label{Sec:Methodology}
We adopt the formalism of \citet{niko2004} to adjust single-epoch measurements through a phase-dependent correction and derive statistical reddening and distance moduli for individual Cepheids.  To improve on this framework, we further introduce a pivot period in the P–L relation, employ simultaneous fitting of multiband distance moduli, and estimate parameters using least squares, maximum likelihood estimation (MLE), 
and Bayesian Markov chain Monte Carlo (MCMC) methods.
\subsection{Theoretical framework}
\label{theoretical}
The Leavitt Law is expressed as
\begin{equation}
M_{\lambda} = \alpha_{\lambda} \left( \log P - \log P_{0} \right) 
+ \beta_{\lambda} 
+ \varepsilon_{\lambda}(M, T_{\mathrm{eff}}, Z, \xi, \xi', \dots) \, ,
\label{eq:leavittlaw}
\end{equation}
where $P$ denotes the pulsation period and $M_{\lambda}$ is the mean absolute magnitude of Cepheid variable stars observed in a photometric band $\lambda$. The coefficients $\alpha_{\lambda}$ and $\beta_{\lambda}$ represent the band-dependent slope and zero point of the P-L relation, respectively.  $P_{0}$ is a pivot period introduced to reduce covariance between $\alpha_{\lambda}$ and $\beta_{\lambda}$ and is chosen as the median period of each sample. In this work, we adopt $\log P_{0} = 0.294$ and $0.103$ for the SMC FU and 1O Cepheids, and $\log P_{0} = 0.553$ and $0.300$ for the LMC FU and 1O Cepheids, respectively. The remaining term, $\ \varepsilon_\lambda(M, \ T_\text{eff}, \ Z, \ \xi, \ \xi', \dots)$, hereafter, $\varepsilon_{\lambda}(.)$, is the unknown function of stellar parameters, such as mass ($M$), effective temperature ($T_\text{eff}$), metallicity ($Z$), or correction parameters ($\xi, \xi', \dots$) due to nonlinearity of P-L relations.  Rewriting Eq.~\ref{eq:leavittlaw} in terms of the observed mean magnitude $\overline{m}_{\lambda}$, we get
\begin{align}
\overline{m}_{\lambda} &= M_{\lambda} + \mu + \mathcal{A}_{\lambda} \notag \\
&= \alpha_{\lambda} \left( \log P - \log P_{0} \right)
+ \beta_{\lambda}
+ \varepsilon_{\lambda}(\cdot)
+ \mu
+ R_{\lambda}\,E(B-V) \, ,
\label{eq:leavittlaw_reddening}
\end{align}
where $\mu$ is the distance modulus and $\mathcal{A}_{\lambda} = R_{\lambda} E(B-V)$ is the extinction in the $\lambda$ band, which depends on the adopted reddening law.

The term $\varepsilon_{\lambda}(\cdot)$ represents the intrinsic dispersion of the P–L relation and is assumed to follow $\mathcal{N}\!\left(0, \sigma^{2}_{0,\lambda}\right)$ \citep{niko2004}. Since any nonzero mean can be absorbed into $\beta{_\lambda}$, this assumption entails no loss of generality. For an individual Cepheid $i$, the reddening and distance modulus can be decomposed as
\begin{subequations}\label{eq:mean_addition}
\begin{align}
E(B-V)_i &= \overline{E(B-V)}_{\mathrm{MC}} + \Delta E(B-V)_i \, , \\
\mu_{i} &= \overline{\mu}_{\mathrm{MC}} + \Delta \mu_{i} \, , 
\end{align}
\end{subequations}
where $\overline{E(B-V)}_{\mathrm{MC}}$ and $\overline{\mu}_{\mathrm{MC}}$ denote the mean reddening and mean distance modulus of the host galaxy (SMC or LMC), and the $\Delta$ terms represent the star-by-star offsets.  Substituting these expressions into Eq.~\ref{eq:leavittlaw_reddening} and absorbing the mean quantities and intrinsic dispersion into a redefined zero point,
\[
\beta'_{\lambda} \;=\; 
\beta_{\lambda} 
+ \overline{\mu}_{\mathrm{MC}}
+ R_{\lambda}\,\overline{E(B-V)}_{\mathrm{MC}},
\]
the final expression for an individual Cepheid $i$ in band $\lambda$ becomes
\begin{equation}
\overline{m}_{\lambda,i} =
\alpha_{\lambda} \left( \log P_i - \log P_{0} \right)
+ \beta'_{\lambda}
+ \Delta \mu_{i}
+ R_{\lambda}\,\Delta E(B-V)_i \, .
\label{eq_m}
\end{equation}

\subsection{Correction for single-epoch measurements}
Since single-epoch S-PLUS magnitudes differ from intensity-averaged means, we introduce a phase-dependent correction term $\Omega_{\lambda,i}(\phi)$, such that
\begin{align}
m_{\lambda,i} 
&= \overline{m}_{\lambda,i} + \Omega_{\lambda,i}(\phi),
\qquad \lambda \in \{\mathrm{S\text{-}PLUS\ bands}\} \notag \\
&= \alpha_{\lambda} \left( \log P_i - \log P_{0} \right)
+ \beta'_{\lambda}
+ \Delta \mu_{i}+ R_{\lambda}\,\Delta E(B-V)_i+\Omega_{\lambda,i}(\phi),
\label{eq_mfinal}
\end{align}
where the pulsation phase ($\phi$) of each Cepheid is computed by folding the S-PLUS single-epoch observation onto the high-quality, well-sampled $I$-band light curve from OGLE-IV \citep{2017_OGLE_CC_MC_updated}. Eq.~\ref{eq_mfinal} parameterizes single-epoch magnitudes in terms of the P–L relation, incorporating reddening and distance modulus offsets relative to the mean values of their host galaxies. The term $\Omega_{\lambda,i}(\phi)$ accounts for the deviation from the mean magnitude and is modeled, following \citet{niko2004}, as a low-order $(N=2)$ Fourier series,
\begin{equation*}
\label{eq_omega}
\Omega_{\lambda,i}(\phi) =
\sum_{j=1}^{N}
\left[
A_{j,\lambda} \cos(2\pi j \phi)
+ B_{j,\lambda} \sin(2\pi j \phi) \, ,
\right]
\end{equation*}
where $\{A,B\}_{j,\lambda}$ are Fourier coefficients that describe the pulsation behavior in each band. 
We used least-squares fitting iteratively to derive the coefficients of Eq. \ref{eq_mfinal}. The steps were as follows:
\begin{enumerate}[label=(\Roman*)]
\item In the first iteration, the coefficients $\alpha_\lambda$, $\beta'_{\lambda}$, and $\{A,B\}_{j,\lambda}$ are initially determined for all S-PLUS bands using an unweighted least-squares fit. At this stage, Eq.~\ref{eq_mfinal} is approximated as
\begin{equation*}
m_\lambda \thickapprox \alpha_\lambda (\log P_i - \log P_{0}) + \beta'_{\lambda} + \Omega_{\lambda}(\phi) \, ,
\end{equation*}
thereby ignoring the offset terms for reddening and distances. Once the coefficients are determined, we compute the wavelength-dependent observed distance modulus offsets for each Cepheid, $\Delta\mu^{\mathrm{obs}}_{\lambda, i}$, and solve for $\Delta\mu_i$ and $\Delta E(B-V)_i$ as given by
\begin{subequations}\label{eq:dist_eqn}
\begin{align}
\Delta\mu^{\mathrm{obs}}_{\lambda, i}(x) &= m_{\lambda, i} - [\alpha_\lambda (\log P_i - \log P_{0}) + \beta'_{\lambda} + \Omega_{\lambda}(\phi)] \, , \\
\Delta\mu^{\mathrm{obs}}_{\lambda, i}(x) &= \Delta\mu_{i} + \left[ a(x) \, R_V + b(x) \right] \times \Delta E(B-V)_i \, ,
\end{align}

\end{subequations}
where $x \equiv \lambda^{-1}$, and $a(x)$ and $b(x)$ are the coefficients of the adopted reddening law. This method relies on the fact that the amount of extinction at a given wavelength is inversely proportional to the wavelength \citep{1985freed, 1988freed}. Thus, $\Delta\mu^{\mathrm{obs}}_{\lambda, i}$ varies with wavelength, and from the simultaneous fitting of the multiband distance moduli as a function of the inverse effective wavelength, assuming the WC23 extinction law, we obtain the $\Delta\mu_i$ and $\Delta E(B-V)_i$ and their uncertainties from the intercept and slope of the fitted relation, respectively \citep{2014rich, 2016Inno}.
\item In the second iteration, each single-epoch S-PLUS magnitude ($m_{\lambda,i}$) is corrected for the $\Delta\mu_{i}$ and $\Delta E(B-V)_i$ calculated in the first iteration and Eq.~\ref{eq_mfinal} is solved using a weighted least-squares fit with weights $w_i = 1/\sigma_{i,\lambda}^2$ to obtain the P-L relation coefficients. The variance $\sigma_{i, \lambda}^2$ is defined as
\begin{equation*}
    \sigma^2_{i, \lambda}  = \sigma^2_{0}+ \sigma_{\text{phot}}^2 + \sigma_{\text{corr}, \lambda}^2 + \sigma_{\alpha, \lambda}^2 (\log P_{i} - \log P_{0}) + \sigma_{\beta,\lambda}^2 \, ,
\end{equation*}
where $\sigma^2_{0}$, $\sigma_{\text{phot}}^2$, $\sigma_{\text{corr},\lambda}^2$, $\sigma_{\alpha, \lambda}^2 (\log P_{i}- \log P_{0})$, and $\sigma_{\beta, \lambda}^2$ represent the intrinsic error, photometric uncertainty, the uncertainty in the correction term, in the P–L slope, and in the zero-point, respectively. Following \citet{niko2004}, we adopt $\sigma_{0} = 0.05$ mag, independent of photometric band. It should be noted that the single-epoch magnitudes in each S-PLUS band are corrected only for differential quantities, i.e., $\Delta\mu_{\lambda, i}$ and $ \Delta E(B-V)_i$, for each individual star relative to the mean values in their host galaxies. The correction does not account for the total line-of-sight reddening. Since these differential quantities vary from star to star, decoupling their effects has enabled us to derive an improved P-L relation in iteration 2 for each passband. 

\end{enumerate}

In addition to the traditional least-squares solution, we performed both an MLE and a Bayesian MCMC analysis in a six-dimensional parameter space to solve Eq.~\ref{eq_mfinal}, assuming Gaussian-distributed residuals. These served as control tests for methodological consistency, complementing the traditional least-squares approach. 

The MLE optimization was carried out using the Nelder--Mead algorithm \citep{1965_nelder, gao:212}. The Bayesian analysis employed the affine-invariant ensemble sampler implemented in \textsc{emcee}\footnote{\url{https://emcee.readthedocs.io/en/stable/}} \citep{2013_emcee}, using 50 walkers, 1000 iterations, and a burn-in of 300 steps. The coefficients computed using the least-squares method were adopted as informative priors, with parameter bounds defined by the MLE estimates and their $\pm3\sigma$ uncertainties derived from the Hessian matrix. These MLE and MCMC analyzes results are consistent.

We derived the solutions separately for the FU and FO Cepheid samples. The distances inferred from the P–L relations are combined with the corresponding equatorial coordinates $(\alpha, \delta)$ to compute the 3D distribution of the Cepheids with respect to the centers of their host galaxies. The reddening maps of the SMC and LMC are constructed from the absolute reddening values of the selected CCs, averaged over spatial grids. The results are discussed in subsequent sections.

\subsection{Extinction}
\label{sec:extinction_subsection}

We considered the extinction law of WC23, which adapts the functional form of \citet[][hereafter C89]{1989cardelli} for environment-specific $R_V$ values derived from Cepheid and red supergiant populations in the LMC and SMC. The C89 was later refined by \citet[][hereafter F99]{1999fitz} using an improved stellar dataset, introducing a spline-based optical prescription and greater flexibility in the UV. Both C89 and F99 derived $R_V \sim 3.1$ and provided a canonical extinction law for the MW. Unlike C89 or F99, WC23 yields $R_V = 3.40 \pm 0.07$ for the LMC and $R_V = 2.53 \pm 0.10$ for the SMC, indicating intrinsic differences in the dust properties relative to the MW. Several large and deep surveys within the Galaxy also report $R_V < 3.1$, consistent with smaller dust grains (e.g., $R_V = 2.50$ for Westerlund 1 at 4 kpc; \citealt{2016damineli}), whereas the higher LMC value suggests a dust population skewed toward larger grains compared to the canonical MW average. The WC23 extinction ratios $A_\lambda/A_V$ are therefore well suited for MC Cepheids and are adopted throughout this work. Nevertheless, we note that foreground MW contamination or mixed stellar populations may introduce additional systematic uncertainties.

The extinction ratios $A_\lambda / A_V$ were computed by interpolating the $R_V$-dependent extinction curve of WC23 at the S-PLUS filter wavelengths, from which we derived $R_\lambda = (A_\lambda/A_V)\,R_V$, where $R_\lambda = A_\lambda/E(B-V)$. Propagating only the uncertainties in $R_V$ yields negligible statistical errors; the dominant contributions arise from systematics, including the adopted extinction law and the spatial dust variations within the MCs \citep{2003gordon,2015nataf}. We therefore adopt conservative wavelength-dependent systematic uncertainties of 8\% ($u$, $J0378$--$J0430$), 5\% ($g$, $J0515$), 4\% ($r$, $J0660$), and 3\% ($i$, $J0861$, $z$), added in quadrature to the statistical uncertainty from $\Delta R_V$. Table~\ref{tab:extinction_coeffs} lists the central wavelengths, extinction ratios $A_\lambda/A_V$, and reddening coefficients $R_\lambda$ for the S-PLUS filters. The F99 values are taken from \citet{Herpich:2024}.

\begin{table}[htbp]
\centering
\caption{Central wavelengths, extinction ratios $A_{\lambda}/A_V$, and reddening coefficients $R_{\lambda}$ for the S-PLUS filters.}
\label{tab:extinction_coeffs}
\resizebox{\linewidth}{!}{%
\begin{tabular}{c c c c c c c}
\toprule
S-PLUS & $\lambda$ &
\multicolumn{2}{c}{$A_\lambda / A_V$ (WC23)} &
 \multicolumn{2}{c}{$R_\lambda$} &
$A_\lambda / A_V$  \\
\cmidrule(lr){3-4} \cmidrule(lr){5-6}
Filters & $(\mathring{A})$ & SMC & LMC & SMC & LMC & (F99) \\
\midrule
u      & 3577 & $1.73 \pm 0.14$ & $1.53 \pm 0.12$ & $4.37 \pm 0.35$ & $5.19 \pm 0.42$ & 1.610 \\
J0378  & 3771 & $1.65 \pm 0.13$ & $1.48 \pm 0.12$ & $4.19 \pm 0.33$ & $5.03 \pm 0.40$ & 1.518 \\
J0395  & 3941 & $1.60 \pm 0.13$ & $1.44 \pm 0.12$ & $4.04 \pm 0.32$ & $4.89 \pm 0.39$ & 1.459 \\
J0410  & 4094 & $1.54 \pm 0.12$ & $1.39 \pm 0.11$ & $3.88 \pm 0.31$ & $4.74 \pm 0.38$ & 1.403 \\
J0430  & 4292 & $1.45 \pm 0.12$ & $1.33 \pm 0.11$ & $3.66 \pm 0.29$ & $4.53 \pm 0.36$ & 1.334 \\
g      & 4774 & $1.24 \pm 0.06$ & $1.18 \pm 0.06$ & $3.14 \pm 0.16$ & $4.02 \pm 0.20$ & 1.199 \\
J0515  & 5133 & $1.10 \pm 0.06$ & $1.08 \pm 0.05$ & $2.79 \pm 0.14$ & $3.67 \pm 0.18$ & 1.098 \\
r      & 6275 & $0.85 \pm 0.03$ & $0.87 \pm 0.04$ & $2.15 \pm 0.09$ & $2.96 \pm 0.12$ & 0.864 \\
J0660  & 6614 & $0.79 \pm 0.03$ & $0.82 \pm 0.03$ & $2.00 \pm 0.08$ & $2.78 \pm 0.11$ & 0.798 \\
i      & 7702 & $0.61 \pm 0.02$ & $0.66 \pm 0.02$ & $1.53 \pm 0.05$ & $2.24 \pm 0.07$ & 0.648 \\
J0861  & 8611 & $0.47 \pm 0.01$ & $0.54 \pm 0.02$ & $1.19 \pm 0.04$ & $1.82 \pm 0.06$ & 0.539 \\
z      & 8882 & $0.45 \pm 0.01$ & $0.51 \pm 0.02$ & $1.13 \pm 0.04$ & $1.74 \pm 0.05$ & 0.512 \\
\bottomrule
\end{tabular}
}
\tablefoot{Values are listed for both the SMC and LMC using the WC23 extinction law, along with corresponding $A_{\lambda}/A_V$ values from F99. Uncertainties for the WC23 values include both propagated statistical errors from $\Delta R_V$ and conservative systematic contributions accounting for spatial variations in dust properties, narrowband sensitivity, and the adopted extinction law for SMC ($R_V = 2.53 \pm 0.10$) and the LMC ($R_V = 3.40 \pm 0.07$).}
\end{table}

\section{Results and discussion}\label{Sec:Results}

In this section, we present the wavelength dependence of the P--L relations across all S-PLUS bands, evaluate their uncertainties. We also compare our results with previous studies, and examine differences between the SMC and LMC populations, including the construction of the reddening map and the 3D distribution of the CC population. 

\subsection{Multiband P--L relations for LMC and SMC CCs}

\begin{figure*}[htb]
    \centering
    \hspace*{0.35cm} 
    \includegraphics[width=0.96\linewidth]{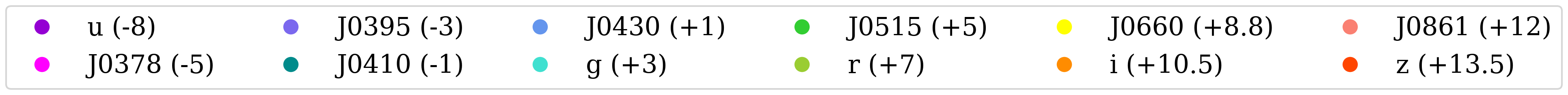} \\[0.5ex]
    
    \includegraphics[width=0.49\linewidth]{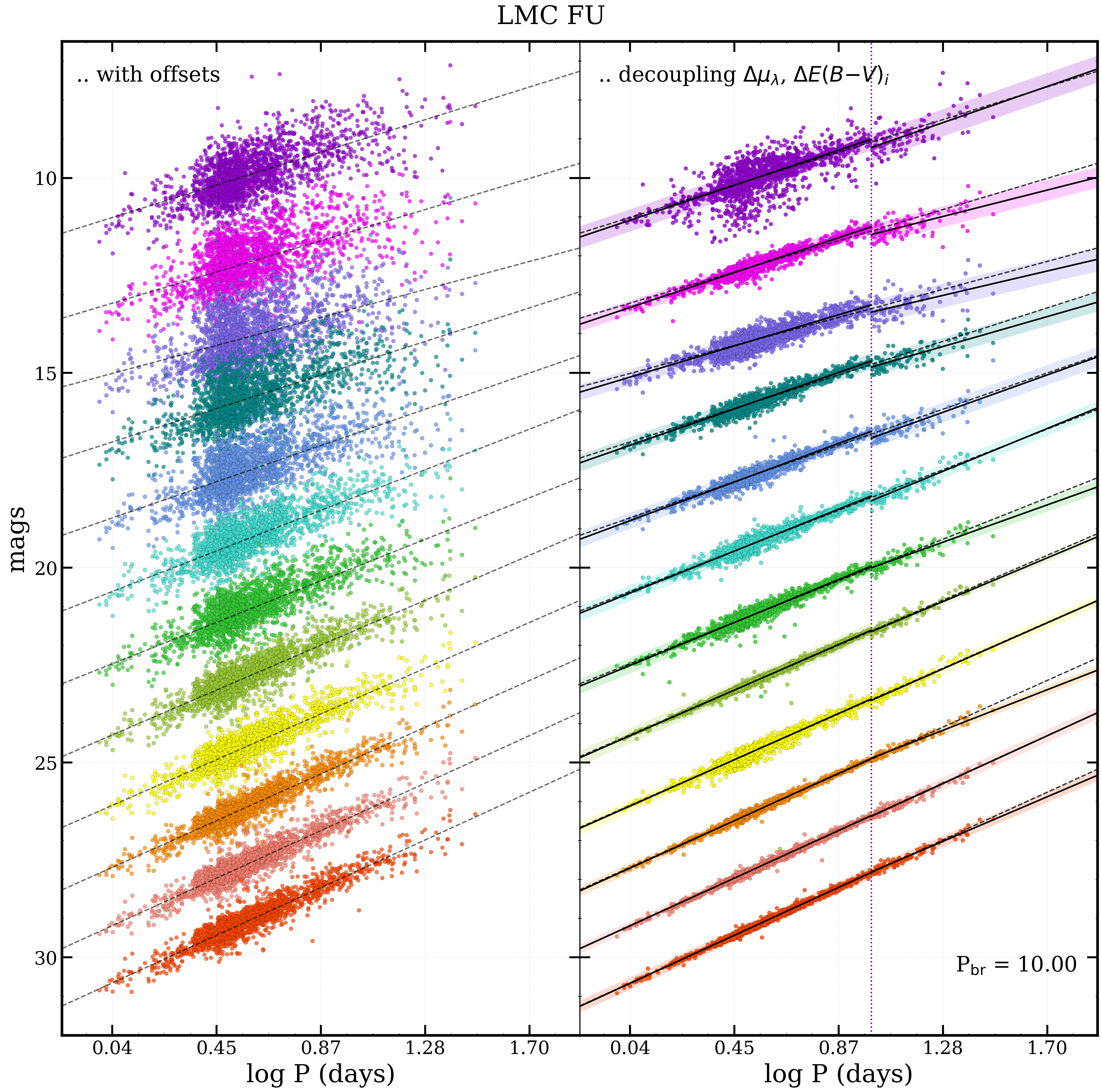}
    \includegraphics[width=0.49\linewidth]{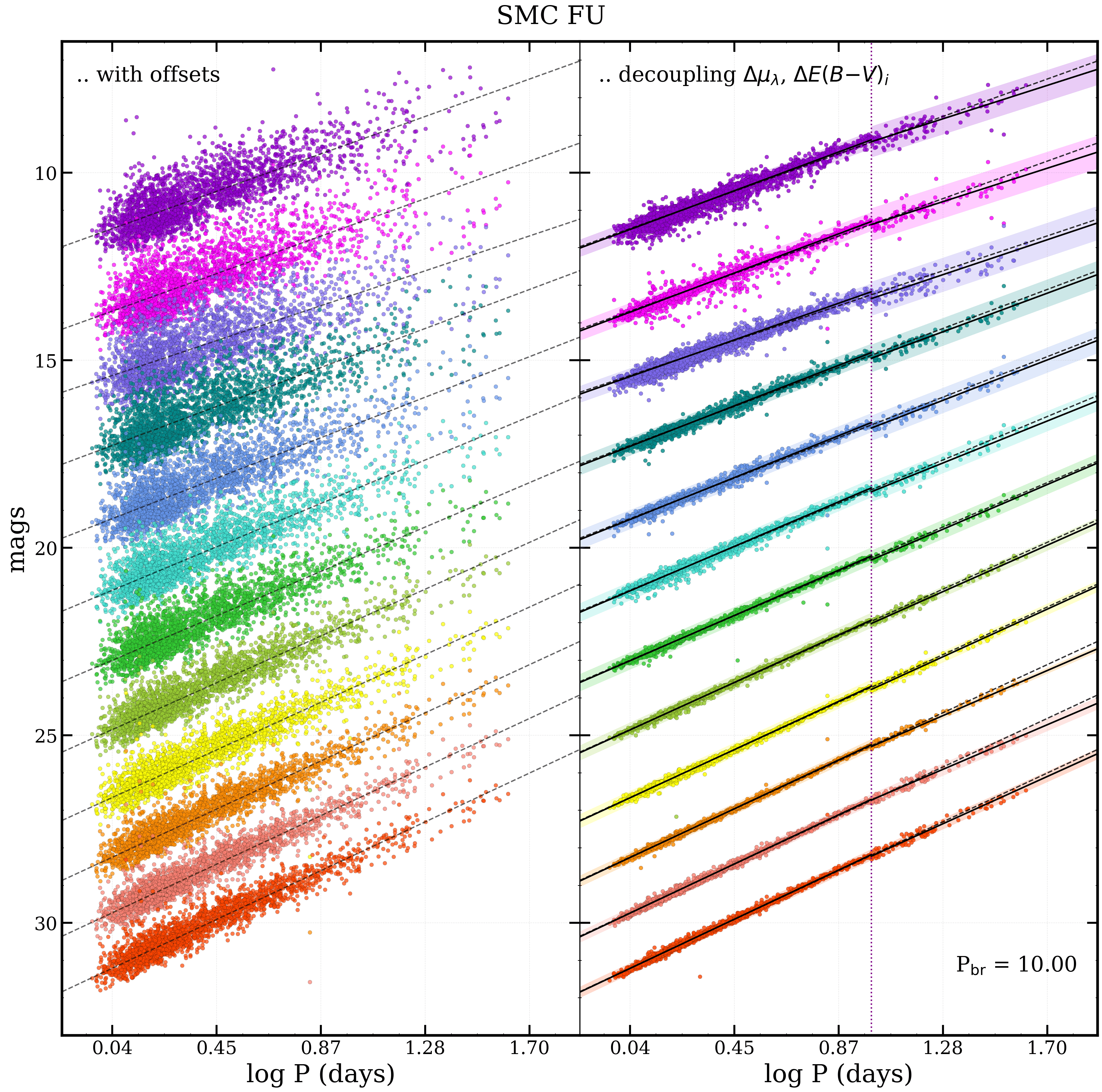} \\
    [0.5ex]
    \includegraphics[width=0.49\linewidth]{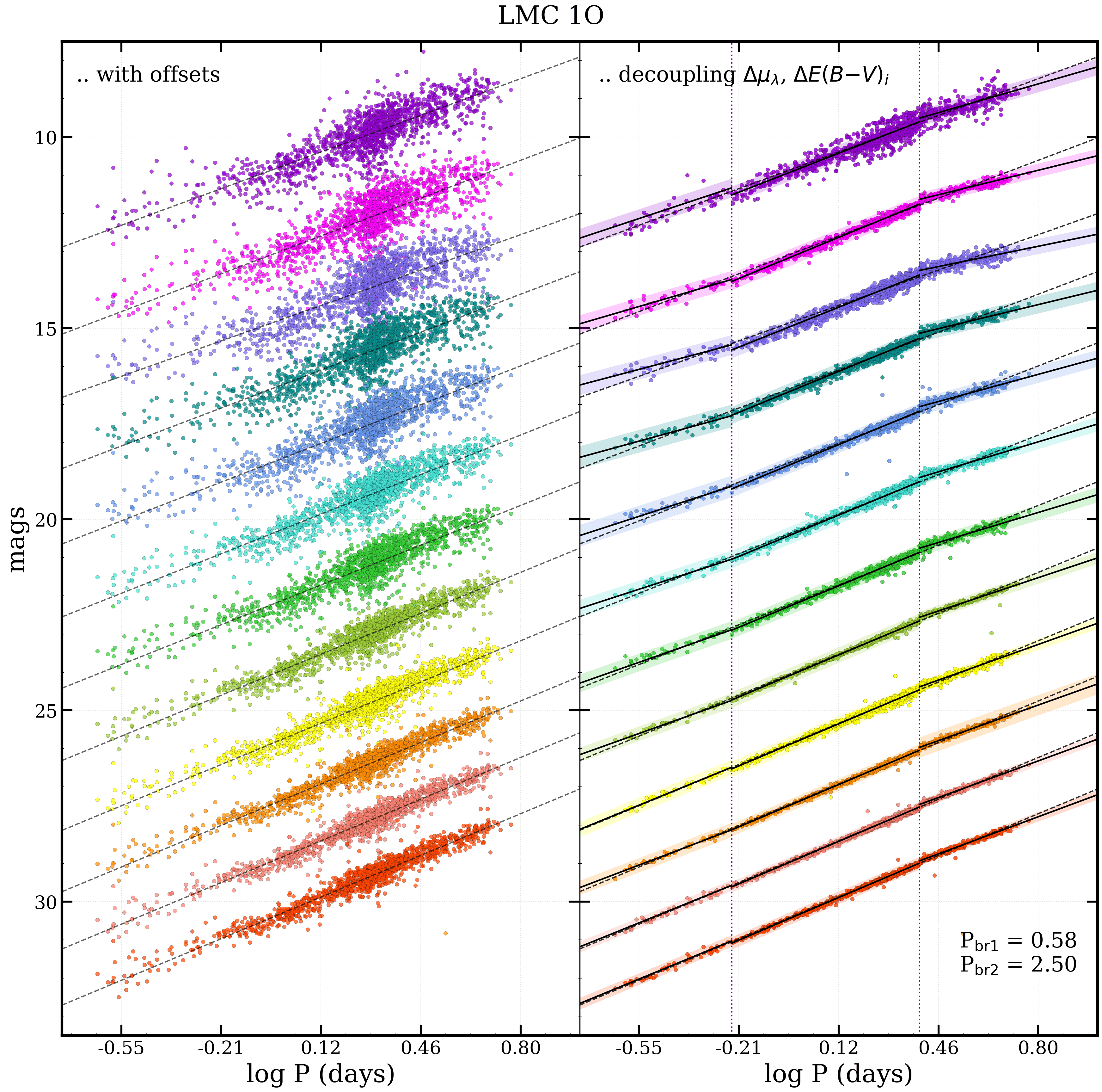}
    \includegraphics[width=0.49\linewidth]{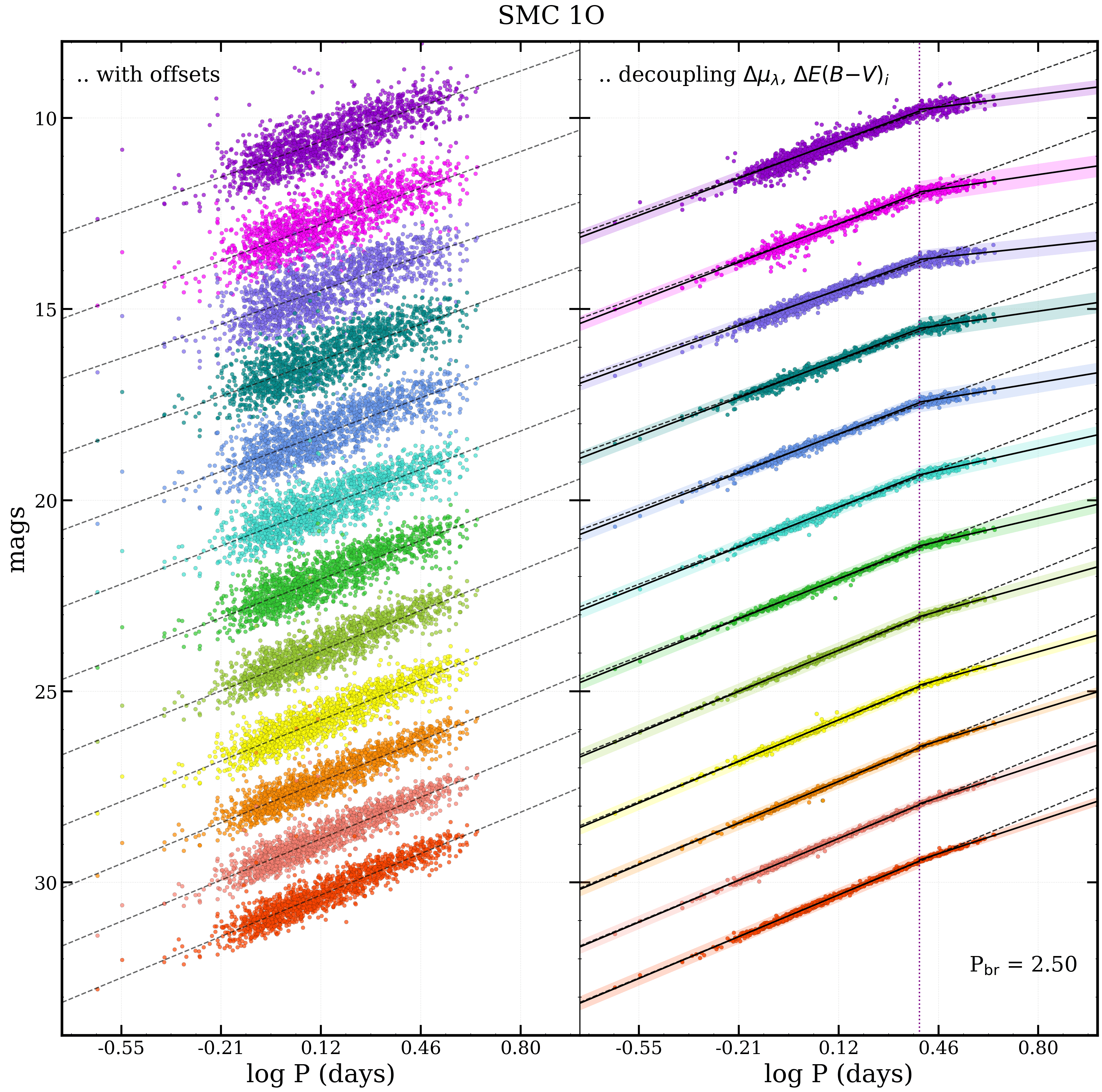}
    
    \caption{Multiband P–L relations for classical Cepheids in the LMC (left panels) and SMC (right panels), with FU (top) and 1O (bottom) pulsators. Each colored sequence corresponds to one of the 12 S-PLUS filters, and the numbers in parentheses in the legend indicate shifts applied for visualization only. Left sub-panels show offset P-L relations, while right sub-panels present results after decoupling $\Delta\mu$ and $\Delta E(B-V)$. Black solid and dashed lines denote best-fit relations with and without breakpoints. Shaded regions indicate uncertainties and vertical dashed lines mark the adopted period breakpoints.}    
    
    \label{fig:PL_FU}  
\end{figure*}

The coefficients of the P--L relations were derived for each of the S-PLUS photometric bands (including seven narrow bands for the first time) following the methodology described in Sect.~\ref{Sec:Methodology}, considering both scenarios: with and without incorporating the period break for both FU and 1O CCs obtained separately. The period break in Cepheids refers to a statistically significant change in the slope of the P--L or period-color (P--C) relation at a specific period, indicating that short-period and long-period Cepheids follow slightly different luminosity or color trends. The break at $P_{\rm br} \sim 10$ days in LMC FU Cepheids was first identified in the P--C and P--L relations by \citet{2002tamm} and statistically confirmed by \citet{2004kanb} using OGLE $V$- and $I$-band light curves at maximum, mean, and minimum light. Subsequent optical and near-IR studies further established this nonlinearity across multiple bands and suggested a possible secondary break near $P_{\rm br} \sim 2.95$ days in the optical \citep{Bhardwaj2016}. In the SMC, \citet{2015subra} identified a break around $\sim 2.95$ days in the optical P--L relation of FU Cepheids, which was later confirmed in the near-IR by \citet{2016ripe}. For 1O Cepheids, evidence of breaks at shorter periods has emerged more recently. \citet{2022ripepi} reported a statistically significant break near $\sim 0.58$ days in LMC 1O pulsators, while other studies have identified nonlinear behavior around $\sim 2.5$--3 days \citep[e.g.,][]{2024breu,2024pilecki}. Collectively, these results demonstrate that the P--L relation is intrinsically nonlinear across different pulsation modes, with break locations that may depend on metallicity, pulsation mode, and wavelength. The physical origin of these breaks has been attributed to changes in stellar structure and pulsation physics, including interactions between the helium ionization front and the photosphere, as well as resonance effects in the pulsation driving mechanism \citep[e.g.,][]{2024anupam_griz}. The results of this study are presented in Table~\ref{tab:lmc_pl_FU} and in Appendix~\ref{app:table} (Tables~\ref{tab:smc_pl_FU}, \ref{tab:lmc_pl_1O}, and \ref{tab:smc_pl_1O}), and are illustrated in Fig.~\ref{fig:PL_FU}. Given the extensive observational evidence from the LMC, SMC, and the Milky Way \citep[e.g.,][]{Narloch_Hajdu_2023,2024anupam_griz,2025madore,2025madore2}, we adopted empirically supported fixed break values of $P_{\rm br} \sim 10$ days for FU Cepheids and $P_{\rm br} \sim 2.5$ days for 1O Cepheids throughout our analysis. Fixing the break at empirically supported values improves the stability and convergence of the P--L relation fits. This approach also facilitates direct comparison across datasets and photometric systems.

The dispersion of the P--L relations decreases significantly after disentangling the effects of individual distance and reddening offsets. The observed scatter in iteration~1 is due to the combined influence of the intrinsic variation of the $\epsilon(.)$ term in Eq.~\ref{eq_mfinal}, line-of-sight depth, interstellar reddening, and photometric uncertainties. Applying individual $\Delta\mu$ and $\Delta E(B-V)$ corrections removes a large part of this contribution and noticeably tightens the relations. Table~\ref{tab:rms_improvement} quantifies the reduction in scatter in the P--L relations after these corrections are applied. The improvement is band dependent, with scatter decreasing by about $30$--$65\%$ across the filters. Both FU and 1O Cepheids show this behavior. The LMC relations, which are already relatively tighter in initial fits than their SMC counterparts, show smaller reductions in comparison; the most modest change occurs for the LMC 1O relation in the $i$ band ($\sim29\%$). A notable exception is the $u$ band for LMC FU Cepheids, where the increased sensitivity to extinction and larger photometric uncertainties at short wavelengths likely dominate the scatter. The number of Cepheids contributing to the fits also varies across bands, and the long-period regime above the adopted break points contains relatively few stars, which increases statistical uncertainties and broadens the dispersion. The uncertainties in the fitted slopes and intercepts follow a different pattern from the dispersions in iteration~1, where the relations are derived before applying individual distance and reddening offset corrections. In several bands, particularly for FU Cepheids, the LMC relations show slightly larger uncertainties in the slope and zero point despite having smaller rms values than those of the SMC. This arises from differences in the period distributions of the Cepheid samples. As shown in Fig.~\ref{fig:hist}, LMC Cepheids are more concentrated at longer periods, whereas the SMC sample contains a larger fraction of short-period variables. Splitting the data at the adopted break periods reduces the number of stars in each subset and narrows the range of $\log P$ values. The effect is strongest above the break points, where only a small number of Cepheids are available, leading to larger formal uncertainties in the fitted slopes and intercepts even when the rms values remain comparable or slightly smaller. For the redder bands, the P--L relations have less dispersion, with the least scatter observed in the $z$ band. This is because the effects of reddening and amplitude are less prominent at longer wavelengths. Additionally, the redder bands have smaller photometric uncertainties, which further contributes to the reduced scatter in their P--L relations \citep[see the calibration process of S-PLUS photometry][]{Almeida-Fernandes-2022,Herpich:2024}. Overall, the P--L relations of 1O Cepheids have less scatter than those of FU Cepheids (see Fig.~\ref{fig:PL_FU}), due to their smaller pulsation amplitudes. This highlights the influence of pulsation amplitude on the scatter in the P--L relations and suggests that accounting for amplitude effects, particularly when light curves in these bands become available, may improve calibration precision \citep[e.g.,][]{niko2004,2025madore}. Moreover, a clear ``belly'' feature is visible in the P--L relations around $\log P \sim 0.3$, which corresponds to the well-known 2:1 resonance between the FU and second overtone modes \citep{1981simon,2009buch,2017smolec}. This resonance alters the light-curve morphology and mean magnitudes, which produces a detectable nonlinearity that is especially pronounced in the metal-poor SMC. Similar short-period breaks have been reported in previous observational and theoretical studies \citep[e.g.,][]{Bono1999,Bhardwaj2016}.

\begin{figure}[!htbp]
    \centering
    \includegraphics[width=\linewidth]{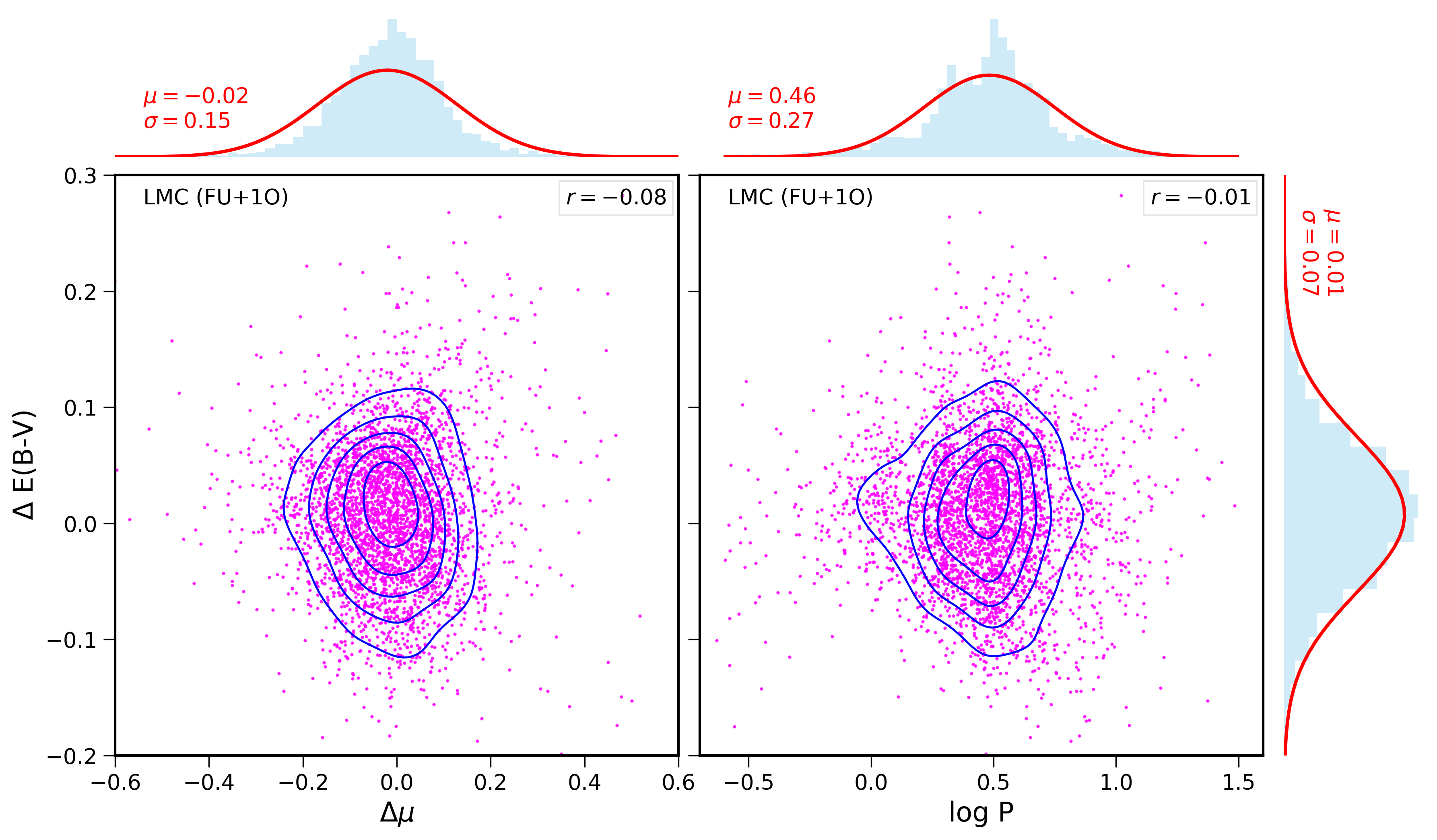}
        
    \caption{Dependence of $\Delta E(B-V)$ on $\Delta\mu$ and $\log P$. Weak correlations (Pearson’s $r=-0.08$ and $-0.01$) indicate independent estimates of reddening and distance modulus, with no systematic trend of reddening with period. Contours show iso-density levels from a Gaussian kernel density estimator; the corresponding SMC distribution is provided in the Appendix~\ref{app:figures}.}
    \label{fig:lmc_ebv_mu}
\end{figure}  

It should be noted that throughout the analysis, we adopted the extinction law of WC23, which is specifically calibrated for the LMC and/or SMC environment (see Section~\ref{sec:extinction_subsection}). As a consistency check, we repeated the P--L analysis using the classical MW extinction law of F99. This results in only small systematic shifts in the zero points, while the derived P--L slopes remain unchanged within the quoted uncertainties. This demonstrates that the results are robust against reasonable variations in the adopted reddening law and supports the use of WC23 as an environment-appropriate prescription for MCs, consistent with previous optical and near-IR studies of Cepheids in these systems \citep{2015_OGLE_CC_MC,2016ripe,2016Inno,2022ripepi,2024anupam_griz}. Figures~\ref{fig:lmc_ebv_mu} and \ref{app:smc_fu_ebv} show the distribution of the offsets $\Delta E(B-V)$ and $\Delta \mu$ derived using the methodology adopted in this work. The lack of correlation between these quantities indicates that the determinations of the reddening and distance modulus are largely independent, and the derived reddening values do not show a systematic trend with period. Fig.~\ref{fig:mu_fit} illustrates the simultaneous fitting of the multiband distance moduli as a function of inverse effective wavelength, assuming the \citet{2023wang} extinction law for two randomly selected stars, shown with (blue squares) and without (red circles) the inclusion of a period break. The slope and zero point of the fit correspond to $\Delta\mu$ and $\Delta E(B-V)$ of the star with respect to the mean values of the host galaxy. 

To assess the reliability of the P--L relation coefficients, we compared results from both frequentist and Bayesian approaches. The close agreement between the MLE-derived parameters and those obtained via least-squares fitting and MCMC sampling for uncertainty estimation demonstrates the consistency and robustness of the parameter estimates. The MCMC analysis, in particular, provided well-constrained posterior distributions, allowing us to quantify credible intervals and assess correlations between parameters, thereby reinforcing the reliability of the adopted values. By utilizing priors adopted from the traditional least-squares solution, we ensured that our MCMC sampling started from a well-informed position, enhancing the efficiency and reliability of the uncertainty estimation. Fig.~\ref{fig:mcmc} illustrates an example of the marginalized posterior distributions and associated uncertainties of the coefficients of the P--L relation for SMC FU Cepheids in two broad bands ($z$, $u$) and two narrow bands ($J0378$, $J0395$).

\begin{figure*}[hb]
    \centering
    \includegraphics[width=0.48\textwidth]{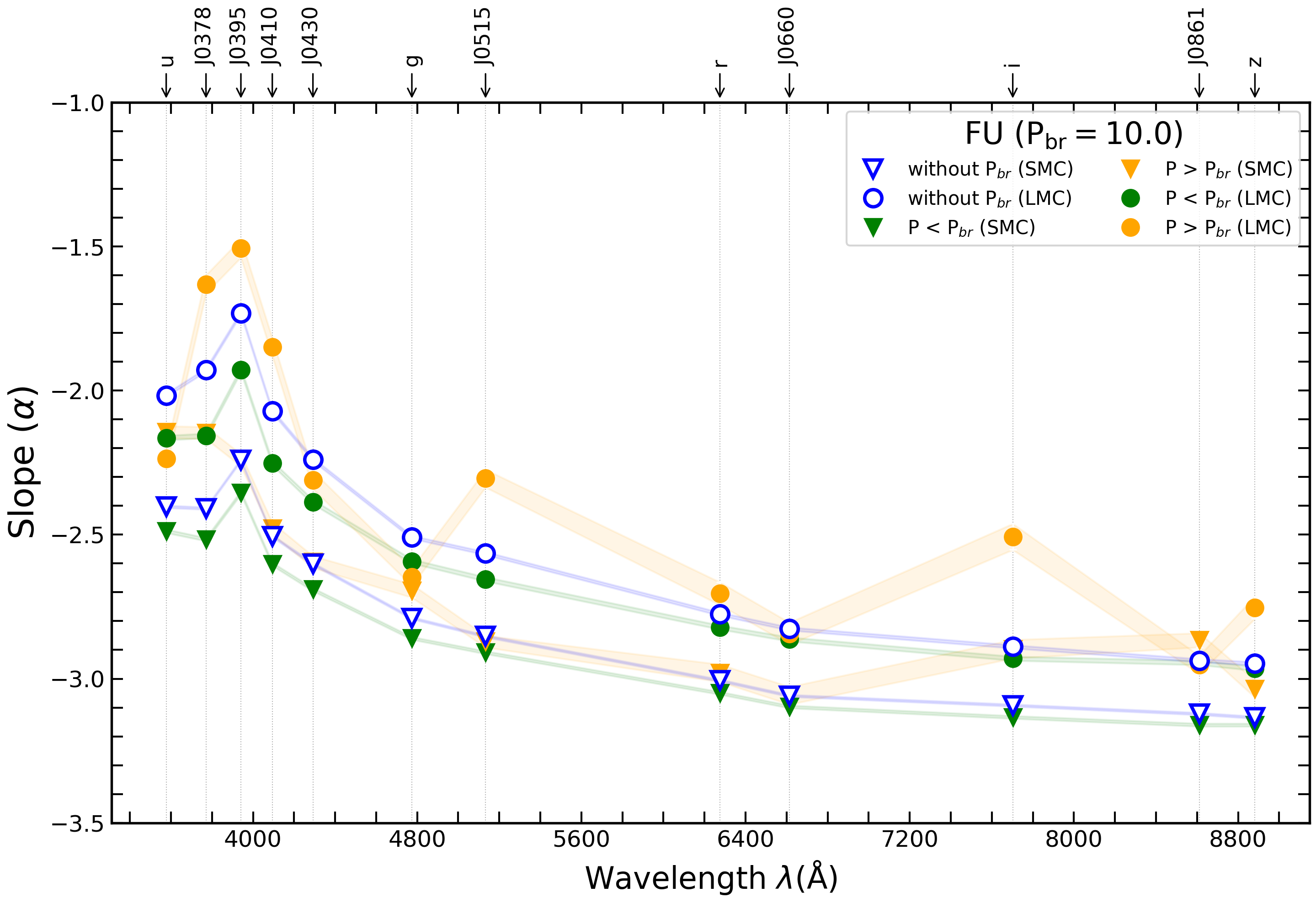}
    \includegraphics[width=0.48\linewidth]{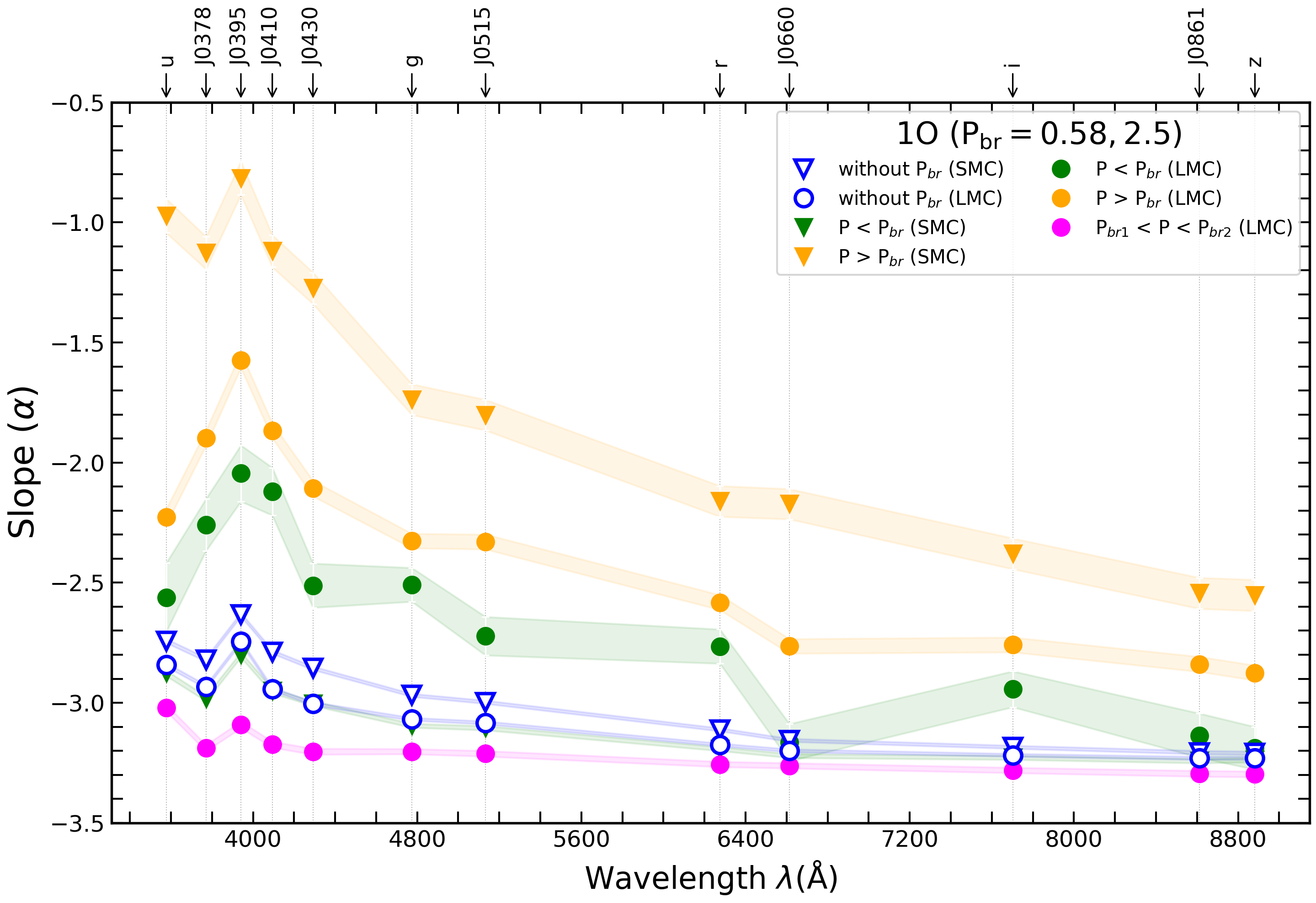}
    \includegraphics[width=0.48\textwidth]{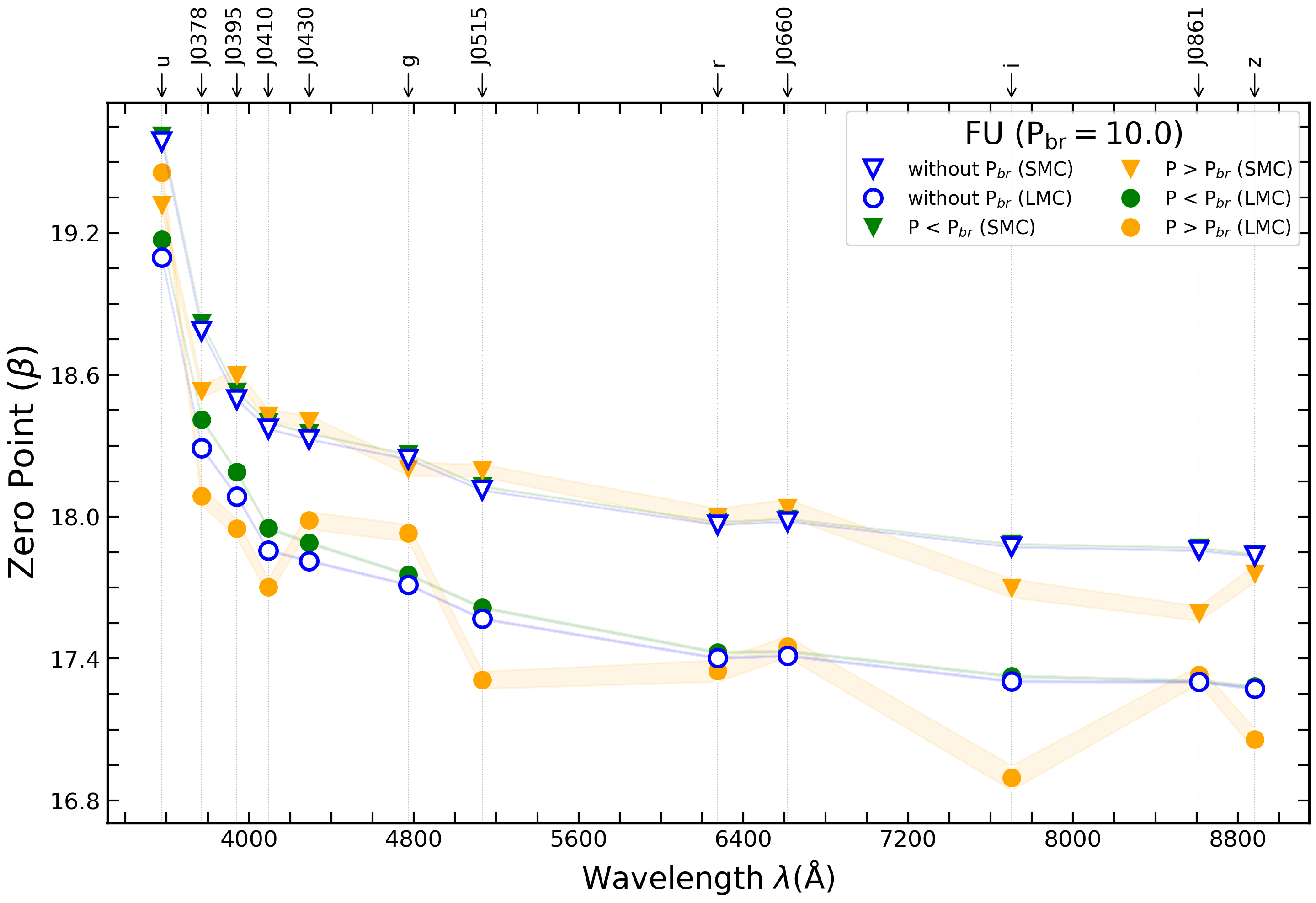}
    \includegraphics[width=0.48\textwidth]{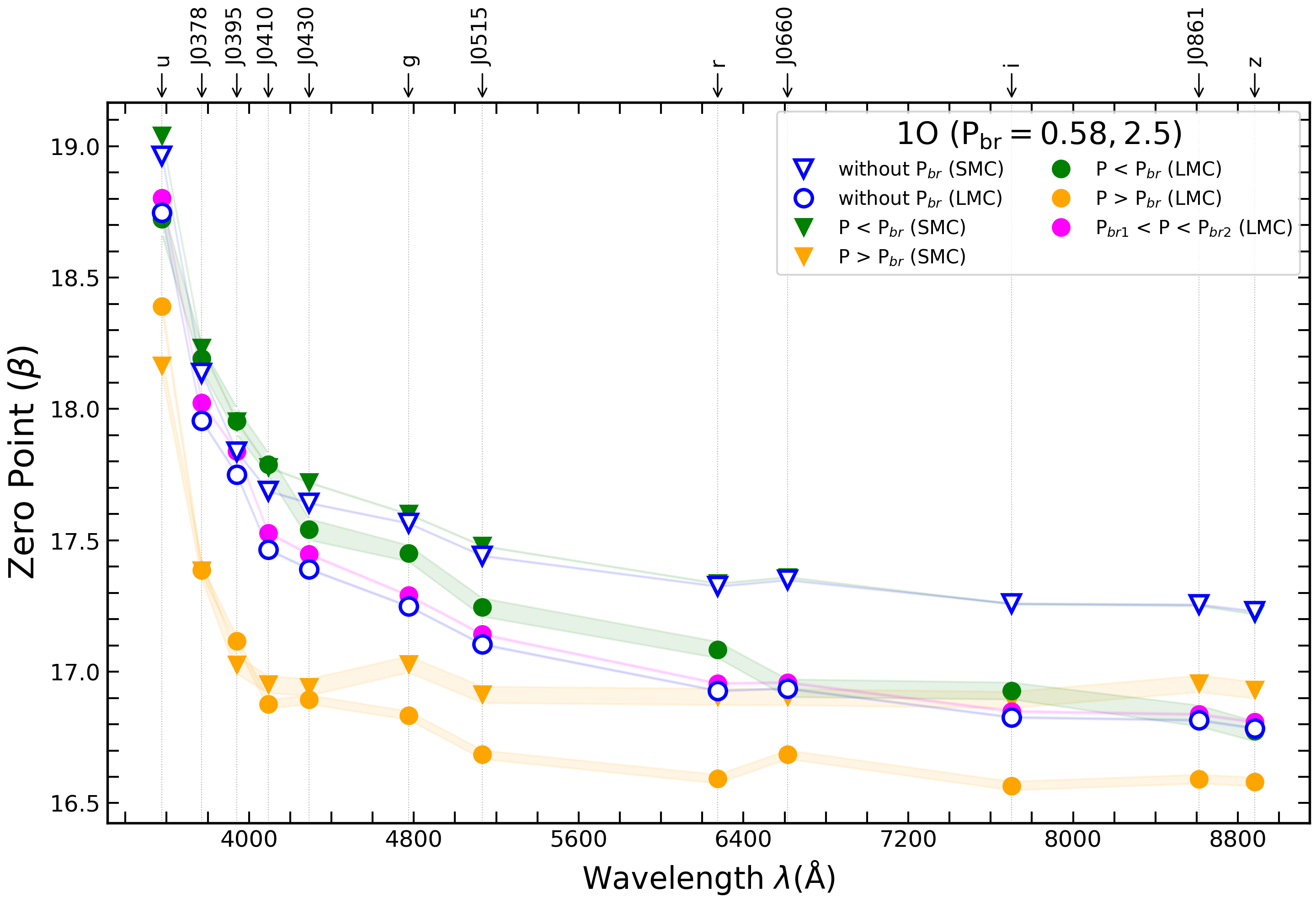}

    \caption{Wavelength dependence of the P-L relation parameters for FU and 1O classical Cepheids across the 12 S-PLUS filters. The top panel shows the variation of the slope ($\alpha$), and the bottom panel shows the zero point ($\beta$). Results are shown for both cases: with and without the inclusion of a period break.}

    \label{fig:slope-zeropt}  
\end{figure*}

Fig.~\ref{fig:omegaplot_lmc} in Appendix~\ref{app:figures} shows the phase dependence of the residual term $\Omega_{\lambda}(\phi)$ for Cepheids in the LMC and SMC across the 12 S-PLUS bands, obtained by modeling the residual magnitudes of the P--L relations. The residual magnitudes show a periodic pattern in every filter, indicating that part of the scatter in the P--L relation arises from the pulsation phase at which the star is observed, in agreement with the formulation introduced by \citet{niko2004}. The amplitude of $\Omega_{\lambda}(\phi)$ generally decreases toward longer wavelengths, consistent with the smaller pulsation amplitudes of CCs in the redder bands. Small departures from this trend are seen in some of the narrow S-PLUS filters, particularly J0395, J0410, and J0430, which probe temperature- and line-sensitive regions of the spectrum (e.g., the Balmer jump, Ca~H+K, and H$\delta$). Because these bands sample spectral features whose strengths change during the pulsation cycle, the phase-dependent residual amplitudes can differ slightly from the simple wavelength trend expected for the broad bands, which are less sensitive to variations affecting individual spectral lines.

\begin{table*}[!ht] 
\centering
\caption{P-L relations for LMC FU Cepheids derived in this study.}

\label{tab:lmc_pl_FU}
\resizebox{0.97\textwidth}{!}{
\begin{tabular}{ccccccccccccc}

\multicolumn{13}{c}{\textbf{LMC FU}} \\
\toprule
\textbf{Filter} & 
\multicolumn{4}{c}{\textbf{without \(P_\text{br}\)}} & 
\multicolumn{4}{c}{\textbf{with \(P \leq P_\text{br}(=10.0)\)}} & 
\multicolumn{4}{c}{\textbf{with \(P > P_\text{br} (=10.0)\)}} \\
\cmidrule(lr){2-5} \cmidrule(lr){6-9} \cmidrule(lr){10-13}
 & \(\alpha\) & \(\beta\) & rms & N & 
\(\alpha\) & \(\beta\) & rms & N & 
\(\alpha\) & \(\beta\) & rms & N \\
\midrule

\multicolumn{13}{c}{\textbf{Iteration 1}}\\
 u & $-2.048 \pm 0.100$ & $19.124 \pm 0.064$ & $0.445$ & 2150 & $-2.241 \pm 0.134$ & $19.226 \pm 0.079$ & $0.437$ & 2033 & $-1.686 \pm 0.599$ & $18.847 \pm 0.705$ & $0.656$ & 117 \\
\rowcolor[HTML]{EFEFEF}
 J0378 & $-1.943 \pm 0.100$ & $18.307 \pm 0.064$ & $0.483$ & 2149 & $-2.181 \pm 0.134$ & $18.433 \pm 0.079$ & $0.472$ & 2031 & $-1.122 \pm 0.605$ & $17.526 \pm 0.714$ & $0.725$ & 118 \\
 J0395 & $-1.763 \pm 0.100$ & $18.126 \pm 0.064$ & $0.510$ & 2151 & $-1.978 \pm 0.133$ & $18.239 \pm 0.079$ & $0.502$ & 2032 & $-0.965 \pm 0.603$ & $17.357 \pm 0.712$ & $0.739$ & 119 \\
\rowcolor[HTML]{EFEFEF}
 J0410 & $-2.100 \pm 0.098$ & $17.882 \pm 0.063$ & $0.464$ & 2194 & $-2.294 \pm 0.132$ & $17.984 \pm 0.078$ & $0.458$ & 2067 & $-1.380 \pm 0.586$ & $17.177 \pm 0.692$ & $0.630$ & 127 \\
 J0430 & $-2.263 \pm 0.099$ & $17.833 \pm 0.063$ & $0.442$ & 2146 & $-2.432 \pm 0.134$ & $17.922 \pm 0.079$ & $0.436$ & 2021 & $-1.843 \pm 0.582$ & $17.455 \pm 0.690$ & $0.612$ & 125 \\
\rowcolor[HTML]{EFEFEF}
 g & $-2.522 \pm 0.098$ & $17.718 \pm 0.063$ & $0.353$ & 2204 & $-2.630 \pm 0.132$ & $17.775 \pm 0.078$ & $0.351$ & 2081 & $-2.326 \pm 0.593$ & $17.571 \pm 0.699$ & $0.447$ & 123 \\
 J0515 & $-2.579 \pm 0.098$ & $17.579 \pm 0.063$ & $0.349$ & 2210 & $-2.689 \pm 0.131$ & $17.638 \pm 0.078$ & $0.348$ & 2084 & $-2.019 \pm 0.589$ & $16.996 \pm 0.694$ & $0.426$ & 126 \\
\rowcolor[HTML]{EFEFEF}
 r & $-2.758 \pm 0.101$ & $17.391 \pm 0.064$ & $0.289$ & 2190 & $-2.807 \pm 0.131$ & $17.417 \pm 0.078$ & $0.290$ & 2073 & $-2.386 \pm 0.755$ & $16.995 \pm 0.875$ & $0.322$ & 117 \\
 J0660 & $-2.836 \pm 0.100$ & $17.415 \pm 0.064$ & $0.258$ & 2204 & $-2.899 \pm 0.132$ & $17.448 \pm 0.078$ & $0.257$ & 2081 & $-2.585 \pm 0.729$ & $17.171 \pm 0.848$ & $0.316$ & 123 \\
\rowcolor[HTML]{EFEFEF}
 i & $-2.891 \pm 0.102$ & $17.300 \pm 0.065$ & $0.215$ & 2177 & $-2.954 \pm 0.132$ & $17.333 \pm 0.078$ & $0.214$ & 2061 & $-2.350 \pm 0.851$ & $16.731 \pm 0.976$ & $0.273$ & 116 \\
 J0861 & $-2.923 \pm 0.098$ & $17.289 \pm 0.063$ & $0.206$ & 2210 & $-2.966 \pm 0.132$ & $17.311 \pm 0.078$ & $0.203$ & 2083 & $-2.912 \pm 0.590$ & $17.310 \pm 0.695$ & $0.286$ & 127 \\
\rowcolor[HTML]{EFEFEF}
 z & $-2.925 \pm 0.100$ & $17.256 \pm 0.064$ & $0.220$ & 2201 & $-2.945 \pm 0.132$ & $17.267 \pm 0.078$ & $0.221$ & 2077 & $-2.719 \pm 0.729$ & $17.032 \pm 0.849$ & $0.259$ & 124 \\

\multicolumn{13}{c}{\textbf{Iteration 2}}\\
 u & $-2.017 \pm 0.006$ & $19.096 \pm 0.004$ & $0.282$ & 2150 & $-2.164 \pm 0.008$ & $19.172 \pm 0.005$ & $0.281$ & 2033 & $-2.236 \pm 0.032$ & $19.456 \pm 0.037$ & $0.347$ & 117 \\
\rowcolor[HTML]{EFEFEF}
 J0378 & $-1.929 \pm 0.006$ & $18.291 \pm 0.004$ & $0.177$ & 2149 & $-2.157 \pm 0.008$ & $18.411 \pm 0.005$ & $0.175$ & 2031 & $-1.632 \pm 0.032$ & $18.086 \pm 0.037$ & $0.264$ & 118 \\
 J0395 & $-1.731 \pm 0.006$ & $18.086 \pm 0.004$ & $0.203$ & 2151 & $-1.928 \pm 0.008$ & $18.189 \pm 0.005$ & $0.200$ & 2032 & $-1.506 \pm 0.032$ & $17.951 \pm 0.038$ & $0.304$ & 119 \\
\rowcolor[HTML]{EFEFEF}
 J0410 & $-2.072 \pm 0.005$ & $17.858 \pm 0.004$ & $0.208$ & 2194 & $-2.252 \pm 0.008$ & $17.952 \pm 0.005$ & $0.210$ & 2067 & $-1.849 \pm 0.031$ & $17.703 \pm 0.036$ & $0.233$ & 127 \\
 J0430 & $-2.239 \pm 0.005$ & $17.813 \pm 0.004$ & $0.206$ & 2146 & $-2.387 \pm 0.008$ & $17.890 \pm 0.005$ & $0.206$ & 2021 & $-2.311 \pm 0.031$ & $17.984 \pm 0.037$ & $0.252$ & 125 \\
\rowcolor[HTML]{EFEFEF}
 g & $-2.509 \pm 0.005$ & $17.712 \pm 0.003$ & $0.198$ & 2204 & $-2.593 \pm 0.007$ & $17.755 \pm 0.004$ & $0.201$ & 2081 & $-2.648 \pm 0.030$ & $17.931 \pm 0.036$ & $0.178$ & 123 \\
 J0515 & $-2.565 \pm 0.005$ & $17.569 \pm 0.003$ & $0.193$ & 2210 & $-2.655 \pm 0.007$ & $17.615 \pm 0.004$ & $0.197$ & 2084 & $-2.305 \pm 0.031$ & $17.309 \pm 0.036$ & $0.164$ & 126 \\
\rowcolor[HTML]{EFEFEF}
 r & $-2.776 \pm 0.005$ & $17.402 \pm 0.003$ & $0.187$ & 2190 & $-2.822 \pm 0.007$ & $17.426 \pm 0.004$ & $0.191$ & 2073 & $-2.705 \pm 0.039$ & $17.348 \pm 0.046$ & $0.088$ & 117 \\
 J0660 & $-2.827 \pm 0.005$ & $17.412 \pm 0.003$ & $0.162$ & 2204 & $-2.863 \pm 0.007$ & $17.431 \pm 0.004$ & $0.164$ & 2081 & $-2.843 \pm 0.038$ & $17.450 \pm 0.044$ & $0.136$ & 123 \\
\rowcolor[HTML]{EFEFEF}
 i & $-2.889 \pm 0.006$ & $17.304 \pm 0.004$ & $0.124$ & 2177 & $-2.930 \pm 0.007$ & $17.326 \pm 0.004$ & $0.126$ & 2061 & $-2.507 \pm 0.045$ & $16.896 \pm 0.052$ & $0.114$ & 116 \\
 J0861 & $-2.937 \pm 0.006$ & $17.302 \pm 0.004$ & $0.122$ & 2210 & $-2.945 \pm 0.008$ & $17.306 \pm 0.005$ & $0.120$ & 2083 & $-2.952 \pm 0.032$ & $17.332 \pm 0.037$ & $0.157$ & 127 \\
\rowcolor[HTML]{EFEFEF}
 z & $-2.947 \pm 0.006$ & $17.273 \pm 0.004$ & $0.121$ & 2201 & $-2.965 \pm 0.007$ & $17.282 \pm 0.004$ & $0.152$ & 2077 & $-2.753 \pm 0.039$ & $17.059 \pm 0.046$ & $0.139$ & 124 \\

\bottomrule
\end{tabular}
}
\tablefoot{Coefficients are listed for fits without a period break and for fits with a period break, for both iteration 1 and iteration 2. Similar tables for LMC 1O, SMC FU, and SMC 1O are included in Appendix~\ref{app:table} (Tables \ref{tab:lmc_pl_1O}, \ref{tab:smc_pl_FU}, \ref{tab:smc_pl_1O})}
\end{table*}

\begin{table}[!ht]
\centering
\small
\caption{Fractional improvement in rms of the P-L relations (without $P_{\rm br}$) after disentangling $\Delta E(B-V)$ and  $\Delta\mu$.}
\label{tab:rms_improvement}
\setlength{\tabcolsep}{7pt}
\renewcommand{\arraystretch}{0.97}
\begin{tabular}{lcccc}
\toprule
Filter & $\Delta R^\mathrm{LMC}_\mathrm{FU}$ & $\Delta R^\mathrm{SMC}_\mathrm{FU}$ & $\Delta R^\mathrm{LMC}_\mathrm{1O}$ & $\Delta R^\mathrm{SMC}_\mathrm{1O}$ \\
\midrule
u     & 0.37 & 0.59 & 0.51 & 0.64 \\
J0378 & 0.63 & 0.60 & 0.57 & 0.62 \\
J0395 & 0.60 & 0.67 & 0.58 & 0.65 \\
J0410 & 0.55 & 0.62 & 0.52 & 0.62 \\
J0430 & 0.53 & 0.58 & 0.53 & 0.60 \\
g     & 0.44 & 0.55 & 0.45 & 0.56 \\
J0515 & 0.45 & 0.56 & 0.44 & 0.57 \\
r     & 0.35 & 0.55 & 0.40 & 0.48 \\
J0660 & 0.37 & 0.56 & 0.36 & 0.53 \\
i     & 0.42 & 0.61 & 0.29 & 0.52 \\
J0861 & 0.41 & 0.59 & 0.42 & 0.50 \\
z     & 0.45 & 0.59 & 0.36 & 0.50 \\
\bottomrule
\end{tabular}
\tablefoot{$\Delta  R = (\mathrm{rms}_{\mathrm{it1}} - \mathrm{rms}_{\mathrm{it2}})/\mathrm{rms}_{\mathrm{it1}}$ quantifies the reduction in dispersion between iteration 1 (it1) and iteration 2 (it2).}
\end{table}

\subsection{Variation of P--L slopes and zero points across the 12 bands}

Fig.~\ref{fig:slope-zeropt} shows the variation of slope ($\alpha$) and zero point ($\beta$) of the P--L relations across the 12 S-PLUS bands. A clear wavelength dependence is observed for both FU and 1O Cepheids in the SMC and LMC. In general, the slopes become progressively steeper toward longer wavelengths, while the dispersion of the relations decreases from the blue to the red filters.

For FU Cepheids, the P--L relations without a break exhibit a systematic steepening with wavelength in both galaxies. In the SMC, the slope varies from $\alpha \approx -2.403$ in the $u$ band to $\alpha \approx -3.134$ in the $z$ band. A similar trend is seen in the LMC, where the slope changes from $\alpha \approx -2.017$ in the $u$ band to $\alpha \approx -2.947$ in the $z$ band. Small deviations occur in the narrowband filters in the blue part of the spectrum, particularly around $J0378$ and $J0395$, but the overall monotonic increase in the absolute value of the slope with wavelength is preserved. When the canonical break at $P_{\rm br}=10$ days is introduced, the short-period relations ($P \leq P_{\rm br}$) are generally slightly steeper than the single-slope solutions, whereas the long-period relations show larger uncertainties because of the smaller sample sizes. Nevertheless, both regimes follow the same wavelength-dependent behavior.

A comparable trend is also observed for the 1O Cepheids. For the LMC 1O sample, the slopes of the single-slope relations vary from $\alpha \approx -2.842$ in the $u$ band to $\alpha \approx -3.230$ in the $z$ band, while the SMC slopes range from $\alpha \approx -2.742$ to $\alpha \approx -3.209$ in the same wavelength range. Introducing the two period breaks for LMC 1O Cepheids ($P_{\rm br1}=0.58$ and $P_{\rm br2}=2.5$ days) reveals that the intermediate-period regime ($P_{\rm br1}<P<P_{\rm br2}$) produces the steepest slopes across most bands, reaching $\alpha \approx -3.30$ in the reddest filters, whereas the longest-period regime yields comparatively shallower slopes.

The zero points of the P--L relations show a systematic decrease with wavelength for both pulsation modes and galaxies. For example, the zero point for LMC FU Cepheids decreases from $\beta \approx 19.10$ in the $u$ band to $\beta \approx 17.27$ in the $z$ band, while in the SMC FU it changes from $\beta \approx 19.59$ to $\beta \approx 17.84$. Similar trends are present for the 1O Cepheids. At the same time, the dispersion of the relations decreases significantly toward longer wavelengths. In the LMC FU sample, for example, the rms decreases from about $0.28$ mag in the $u$ band to about $0.12$ mag in the $z$ band, while in the SMC it decreases from $\sim0.23$ to $\sim0.14$ mag across the same wavelength range. A similar reduction is observed for the 1O Cepheids, where the rms values decrease from $\sim0.17$--$0.18$ mag in the bluest filters to $\sim0.10$--$0.12$ mag in the reddest bands (see Appendix~\ref{app:figures}, Fig.~\ref{fig:rms_improvements}).

The observed wavelength dependence of both slope and dispersion in this study is consistent with the physical framework discussed by \citet{2012madore}. In this interpretation, the finite width of the P--L relation reflects a combination of radius and temperature variations across the instability strip. At shorter wavelengths, the flux is more sensitive to temperature variations, which broadens the instability strip in magnitude and produces shallower P--L slopes with larger intrinsic scatter. As the wavelength increases, the sensitivity to temperature variations decreases, and the luminosity becomes increasingly dominated by the geometric period-radius relation, leading to steeper slopes and a simultaneous reduction in the width of the relation. This coupling between the slope and dispersion naturally explains the monotonic steepening of the P--L relations and the progressive tightening of the scatter toward longer wavelengths in this study. A similar phenomenon is seen in the case of RR Lyrae stars \citep{Catelan2004}.

The systematic offsets between the slopes and the zero points obtained for the SMC and LMC probably reflect differences in the chemical composition of the two systems. The mean metallicities of the MCs differ substantially. Measurements based on APOGEE DR17 spectroscopy suggest typical values of [Fe/H] $\approx -0.67$ dex for the LMC and [Fe/H] $\approx -1.04$ dex for the SMC \citep{2022guilherme}. However, these values primarily trace the metallicities of older stellar populations and may not be fully representative of the young stellar populations to which CCs belong. High-resolution spectroscopic studies specifically targeting Cepheids provide more appropriate estimates. For the SMC, \citet{2008roman_SMC} derived a mean metallicity of [Fe/H] $\approx -0.75$ dex from direct abundance measurements of Cepheid variables, and a more recent analysis of LMC Cepheids by \citet{2022roman_lmc} reports a mean value of [Fe/H] $\approx -0.41$ dex. These Cepheid-based measurements confirm that the LMC is systematically more metal-rich than the SMC, supporting the interpretation that metallicity differences may contribute to the observed offsets in the derived P--L relations. Metallicity has long been recognized as an important factor influencing the Cepheid P--L relation, affecting both its slope and zero point and thereby impacting the calibration of extragalactic distance scales \citep[e.g.,][]{2017wielg, 2018gieren, 2023hocd, 2025ripepi, 2025wang, 2025madore, 2025madore2}. Although the differences observed here between the two galaxies are consistent with this expectation, a quantitative assessment of metallicity effects on the P--L relation is beyond the scope of the present analysis given the absence of multi-epoch S-PLUS light curves for MC Cepheids in these bands. Future time-series photometry and spectroscopic observations in the S-PLUS filter system will enable a more robust interpretation of metallicity effects on the P--L relation by reducing systematic uncertainties associated with single-epoch measurements.

Previous investigations of MC Cepheids have derived P--L and period--Wesenheit index relations using several optical and near-IR photometric systems \citep[e.g.][]{2016ripe, Bhardwaj2016, 2018deb_lmc, 2019deb_smc, 2022ripepi}. Direct comparisons of the P--L coefficients with those studies are not straightforward because of differences in filter sets and in the adopted formulations of the relations. Despite these differences, the general behavior of the relations derived in this work remains consistent with earlier findings, particularly the tendency for the slopes to become steeper and the intrinsic scatter to decrease toward longer wavelengths. The results presented here therefore complement previous studies by extending multiband P--L relations to the S-PLUS photometric system and providing the first such characterization for CCs in this filter set, including seven narrow bands.

\begin{figure*}[ht]
    \centering
    \includegraphics[width=0.43\linewidth]{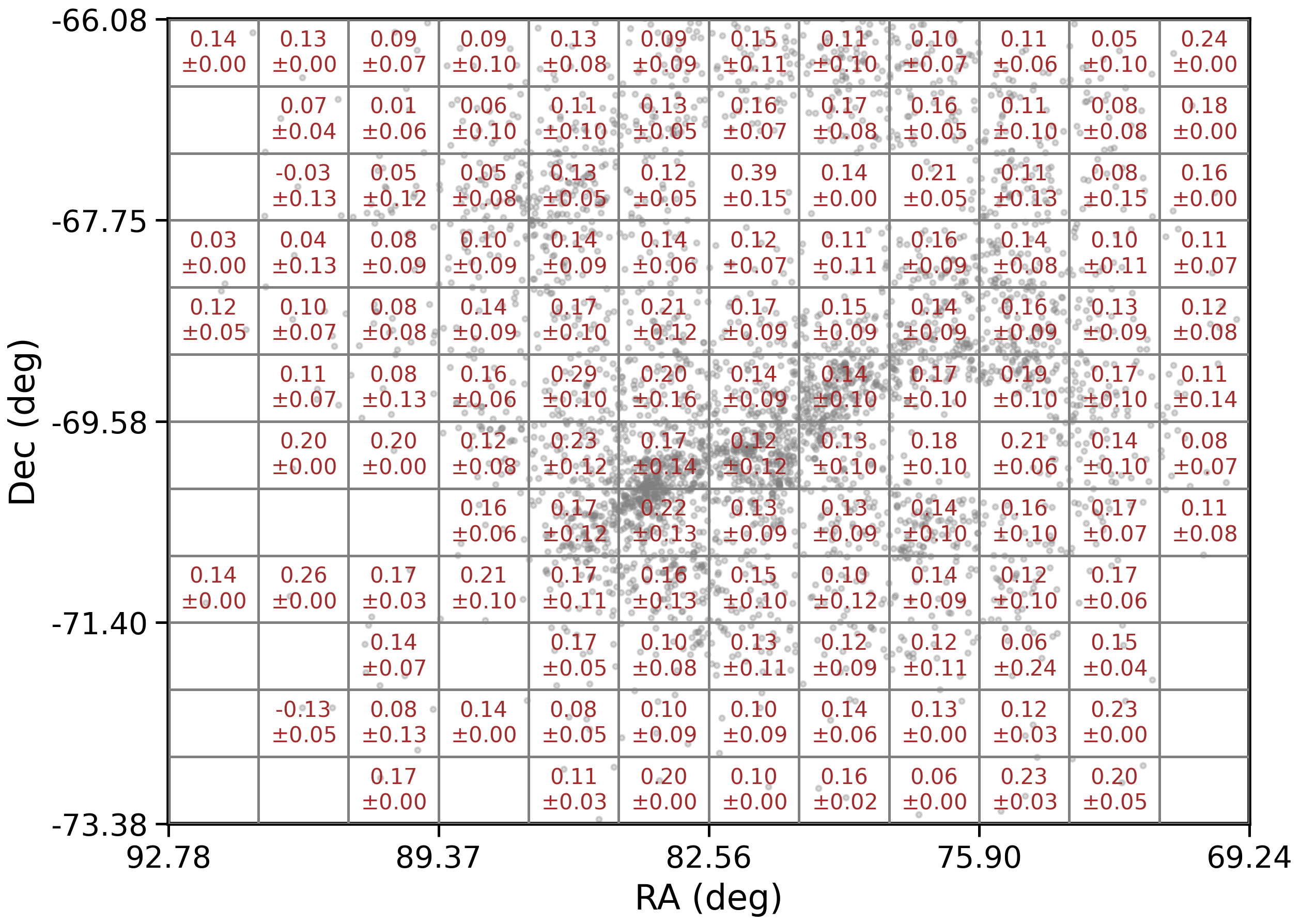}
    \includegraphics[width=0.43\linewidth]{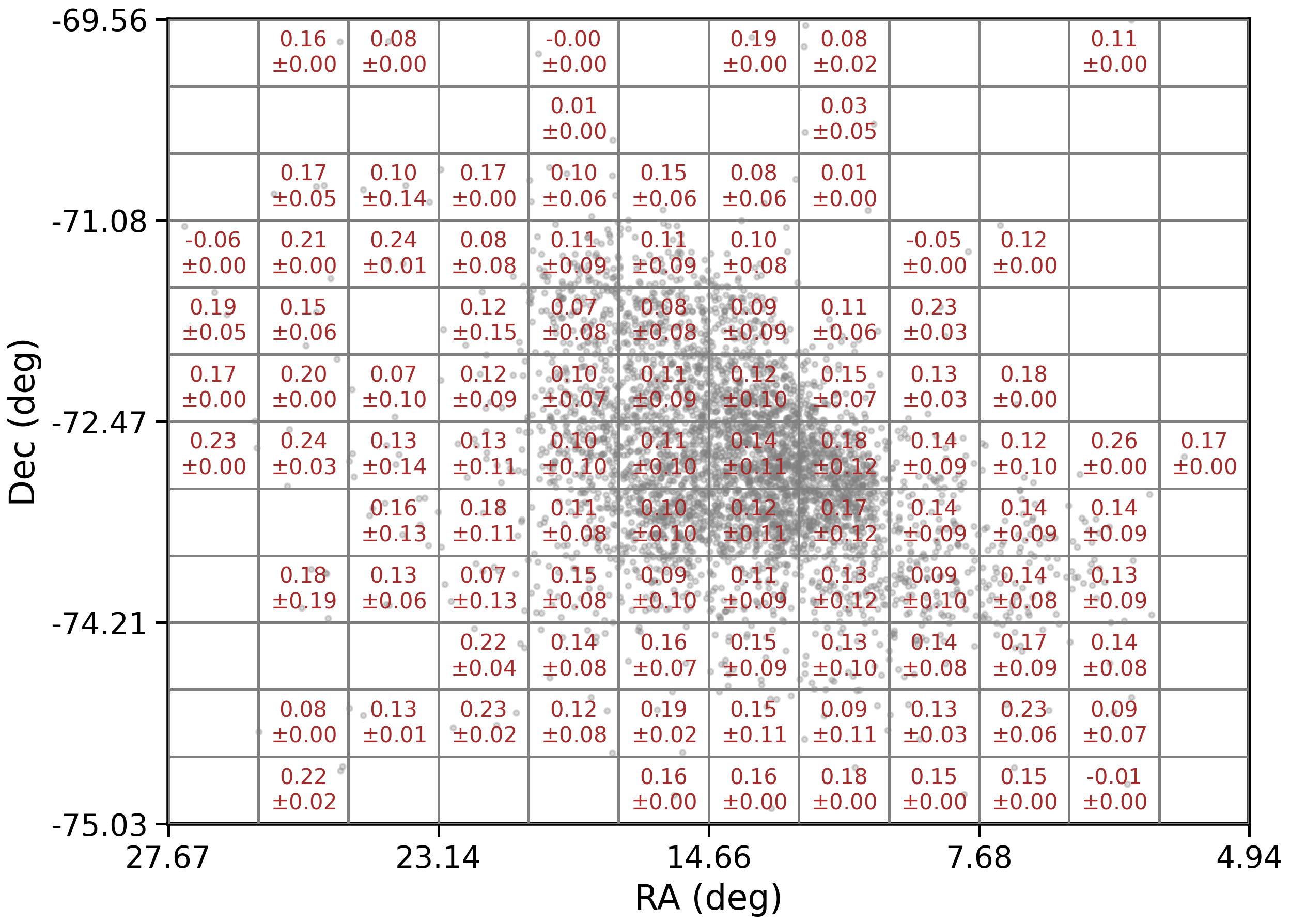}
    \caption{Gridded reddening maps of the LMC (left) and SMC (right) from P-L relations of CCs without breaks. Each cell shows the weighted mean $E(B-V)$ in equatorial spatial bins, with uncertainties representing the standard deviation within each bin. Only bins with $>3$ stars are shown for statistical reliability and visual clarity.}
    \label{fig:gridmap}
\end{figure*}

\begin{figure*}[ht]
    \centering 
    \includegraphics[width=0.32\linewidth]{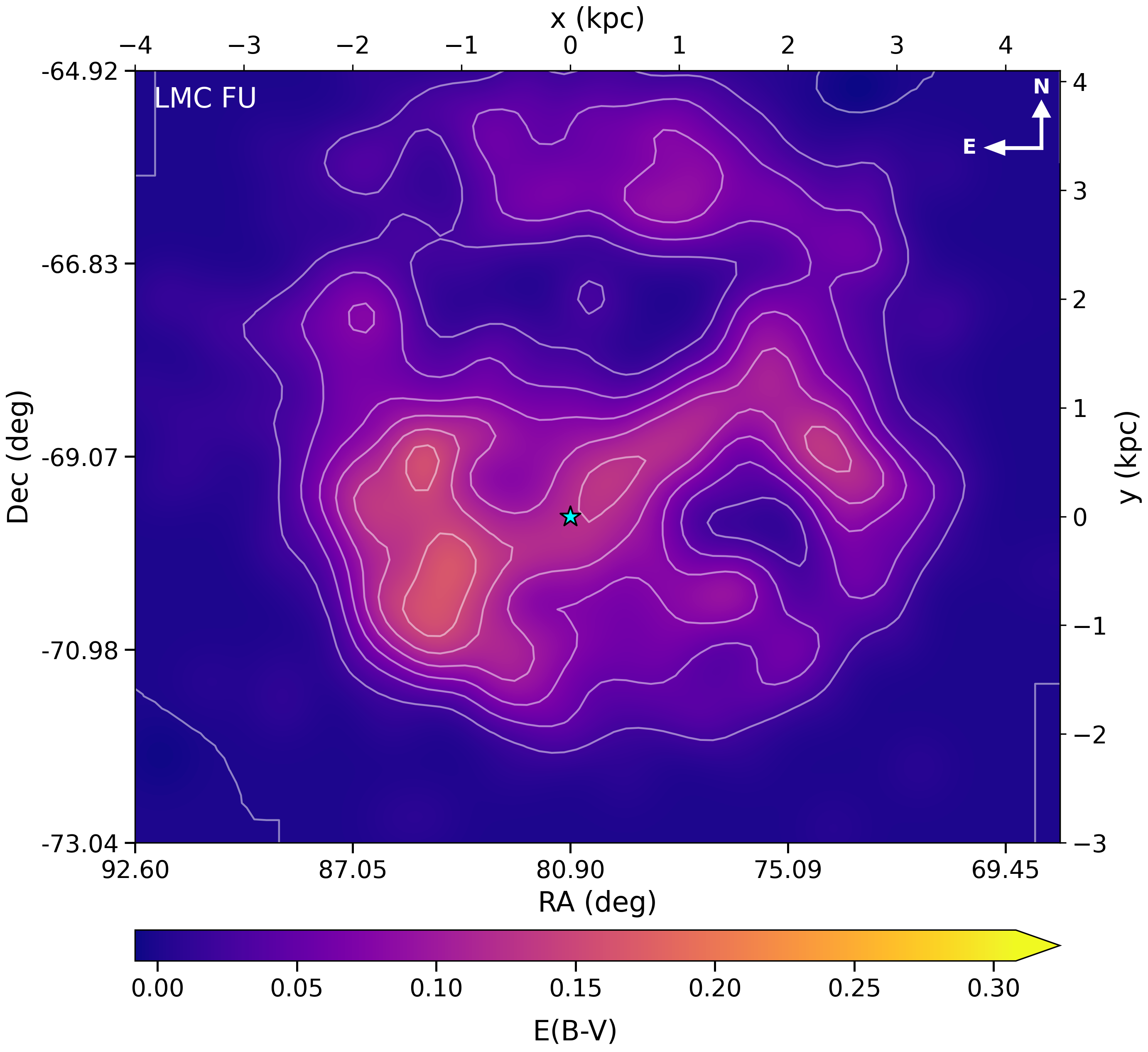}
    \includegraphics[width=0.32\linewidth]{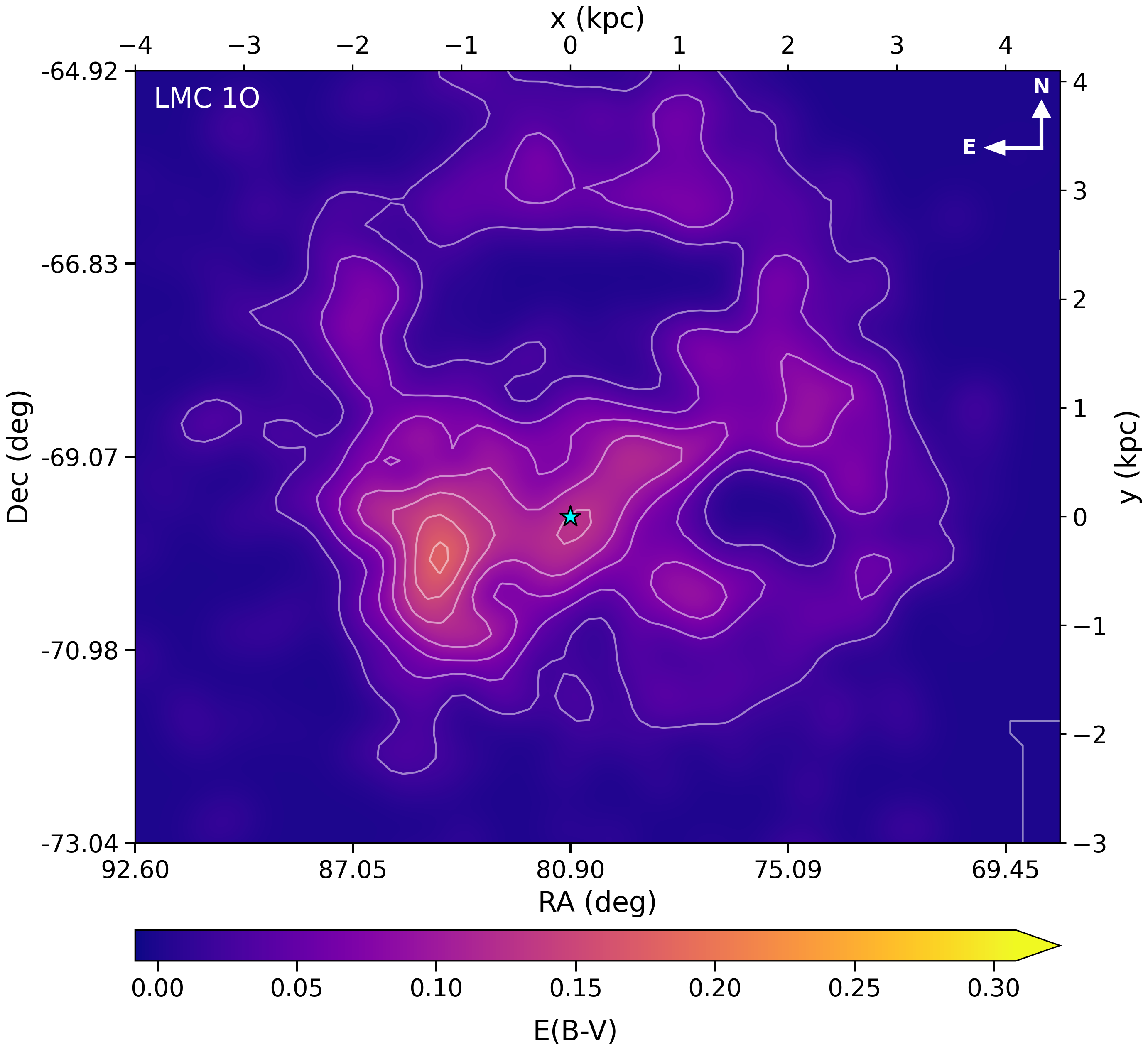}
    \includegraphics[width=0.32\linewidth]{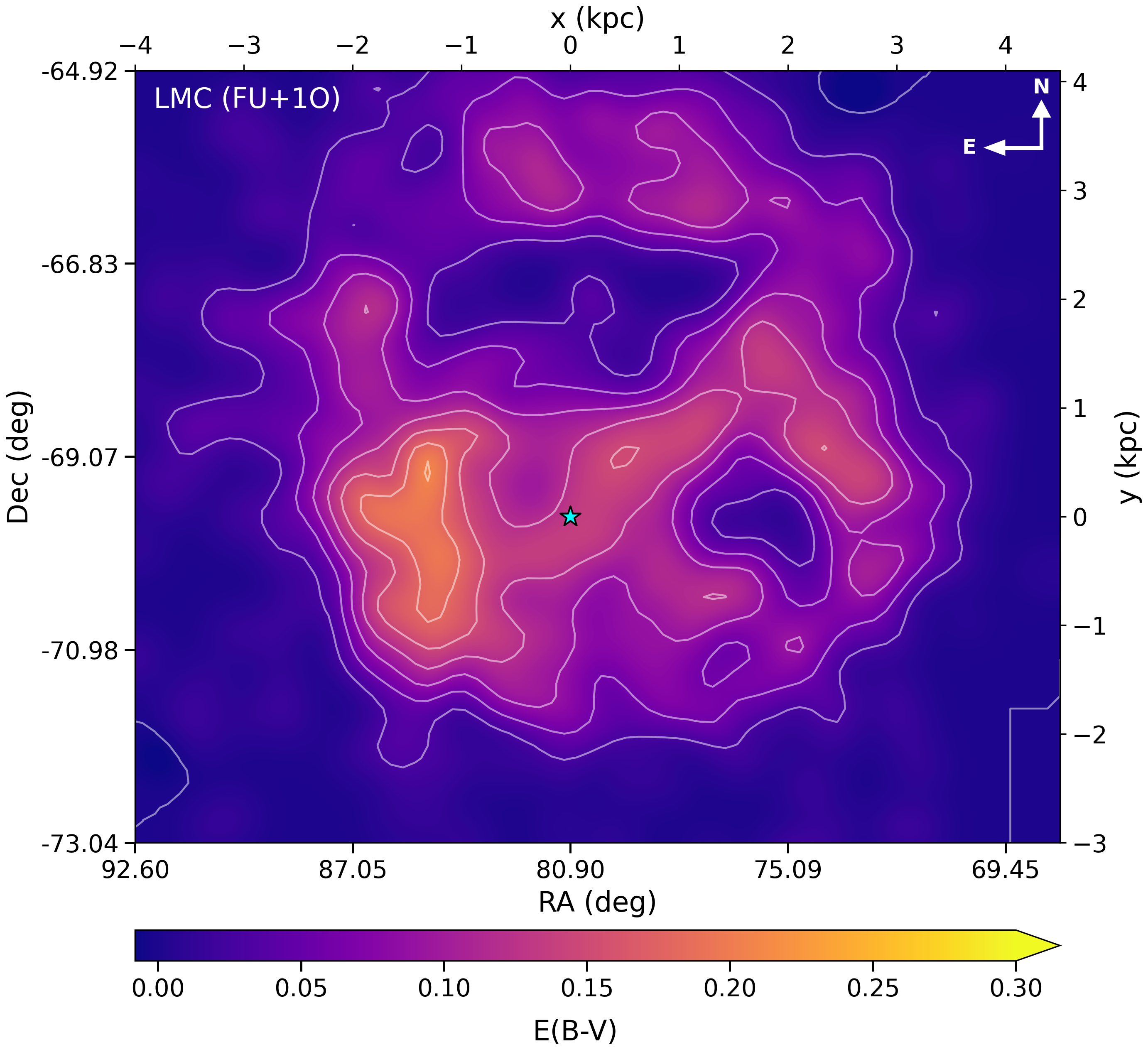} \\
    
    \includegraphics[width=0.325\linewidth]{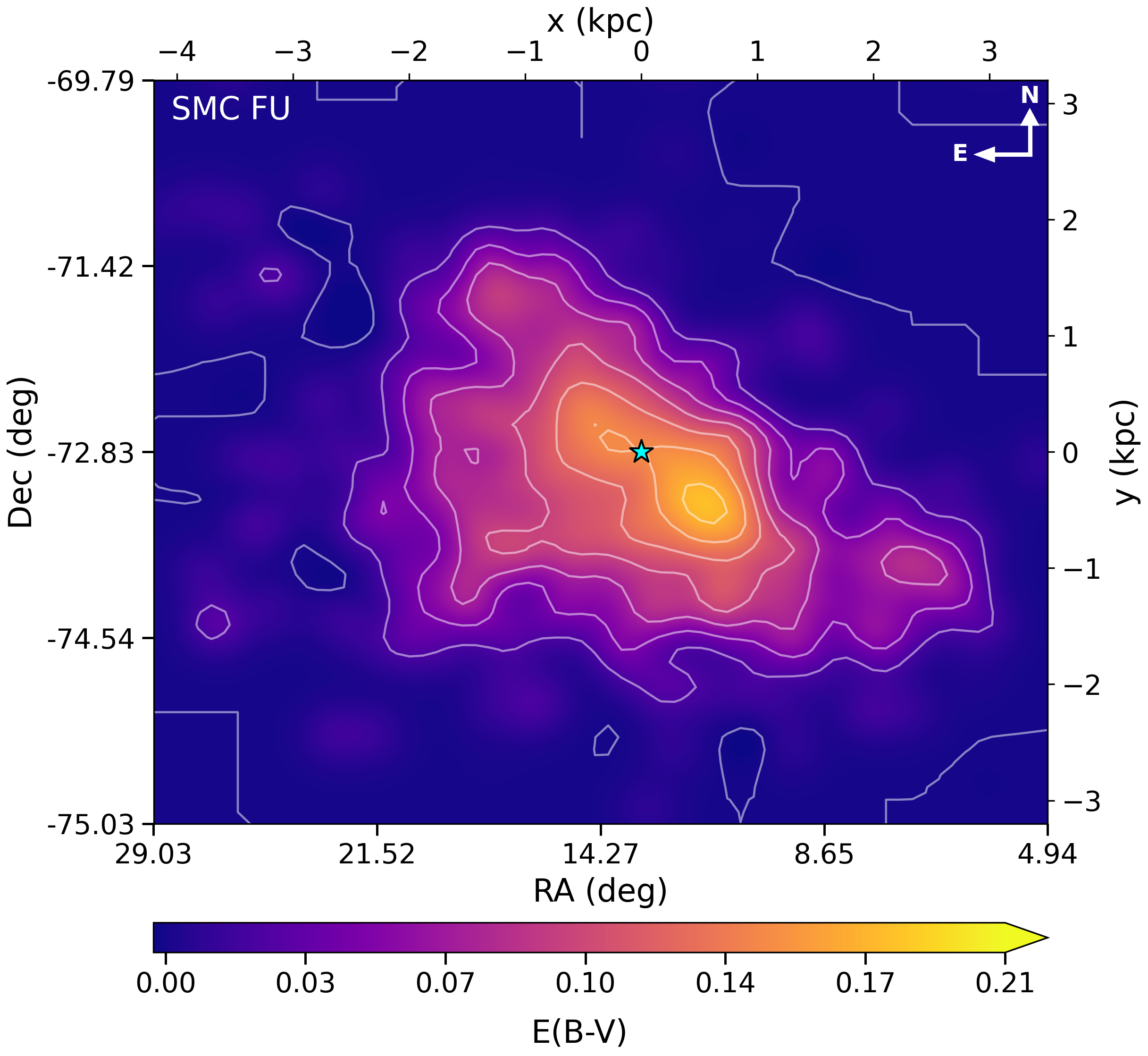}
    \includegraphics[width=0.325\linewidth]{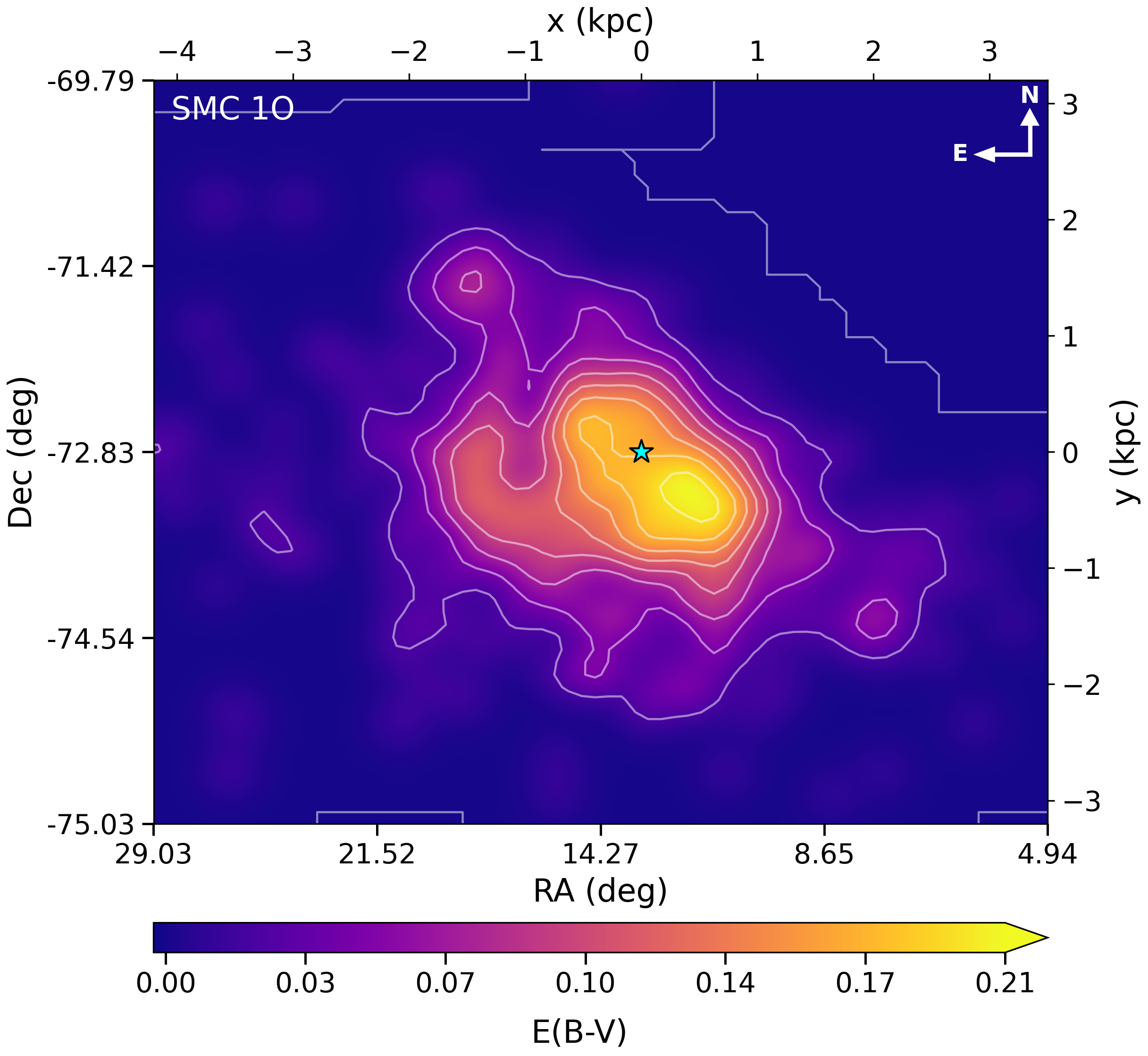}
    \includegraphics[width=0.329\linewidth]{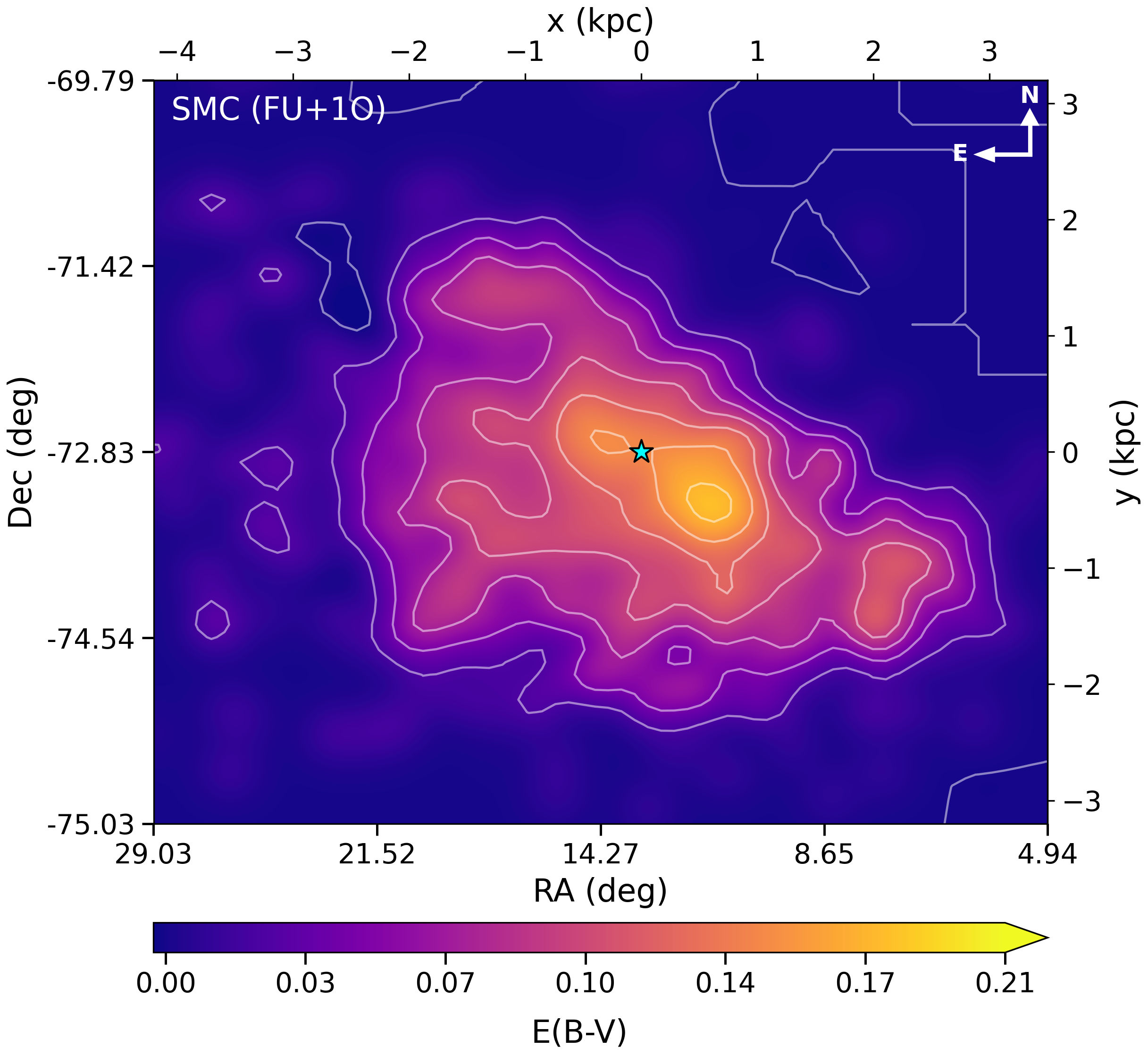} 

    \caption{Reddening maps of the LMC (top panels) and SMC (bottom panels) constructed from Cepheids without considering period breaks in the P--L relations. The maps show the distribution of the mean color excess, $E(B-V)$, derived for individual stars and averaged over a fine spatial grid in equatorial coordinates. A Gaussian smoothing was applied to emphasize large-scale structures, while the contours indicate iso-reddening levels and highlight regions of enhanced extinction. The cyan star marks the adopted center of each galaxy, taken from the literature and used for the coordinate transformation. The panels correspond to different Cepheid subsamples (FU, 1O, and combined). Similar reddening maps constructed by considering the period breaks in the P--L relations are presented in Appendix~\ref{app:figures} (Fig.~\ref{fig:redd_map_lmc_br} and Fig.~\ref{fig:redd_map22})}

    \label{fig:redd_map}
\end{figure*}

\begin{figure}
    \centering
    \includegraphics[width=\linewidth]{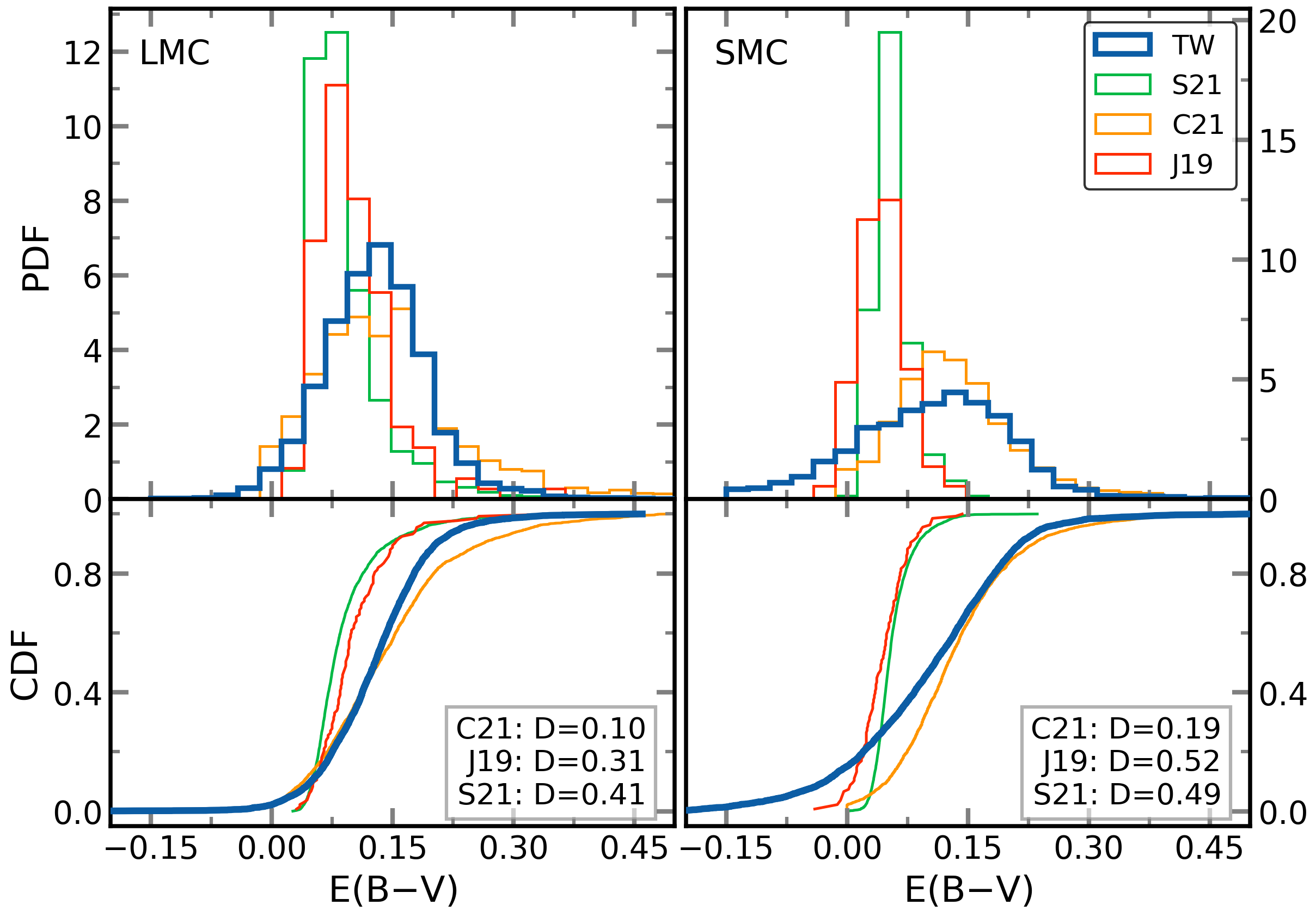}
    \caption{Comparison of the reddening distribution $E(B-V)$ derived in this work (TW; blue) against literature values from \citet{2021chow} (C21; orange), \citet{2021skowron} (S21; green), and \citet{2019joshi} (J19; red) for the LMC (left) and SMC (right). The top panels show the probability density functions (PDFs), while the bottom panels display the cumulative distribution functions (CDFs). Inset boxes provide K-S statistics ($D$) relative to TW, indicating the highest statistical consistency with C21 ($D \approx 0.10$–$0.19$).}
    \label{fig:reddcomp}
\end{figure}

\subsection{LMC and SMC reddening maps}

Once the offset values $\Delta E(B-V)_i$ and $\Delta \mu_i$ for individual CCs, with respect to the mean values of their host galaxies, were determined from the simultaneous fitting of the multiband distance moduli, the corresponding true reddening and distance modulus values were obtained by adopting representative mean values from the literature. Specifically, we adopted $\overline{E(B-V)}_{\rm LMC}=0.14\pm0.02$ mag \citep{niko2004} and $\overline{E(B-V)}_{\rm SMC}=0.125\pm0.02$ mag \citep{1997sasse} for the mean reddening, together with $\overline{\mu}_{\rm LMC}=18.477\pm0.004\,({\rm stat.})\pm0.026\,({\rm sys.})$ \citep{2019piet} and $\overline{\mu}_{\rm SMC}=18.977\pm0.016\,(\mathrm{stat.})\pm0.028\,(\mathrm{sys.})$ \citep{2020grac} for the mean distance modulus. These mean values were combined with the offsets derived to obtain the individual true reddening $E(B-V)_i$ and the distance modulus $\mu_i$ following Eq.~\ref{eq:mean_addition} for each CC. For individual CCs, we report in Appendix~\ref{app:table} (Table~\ref{table:catalog}), the offset quantities $\Delta E(B-V)_i$ and $\Delta \mu_i$ along with the corrected S-PLUS magnitudes in 12 bands.

In the present analysis, we adopt the mean values listed above because they ensure physically meaningful true reddening values for the majority of stars in our sample. Several alternative estimates of the mean reddening for the MCs are available in the literature. For example, OGLE-IV measurements based on red clump stars reported $\overline{E(V-I)}_\mathrm{LMC}=0.100\pm0.043$ and $\overline{E(V-I)}_\mathrm{SMC}=0.047\pm0.025$, corresponding to $\overline{E(B-V)}$ values of $\sim0.13$ and $\sim0.08$, respectively \citep{2021skowron}. Using a different approach, \citet{2019joshi} derived $\overline{E(B-V)}_\mathrm{LMC}=0.091\pm0.050$ and $\overline{E(B-V)}_\mathrm{SMC}=0.038\pm0.053$, while empirical averages of $\overline{E(B-V)}=0.127$ and $0.084$ mag for the LMC and SMC using red clump stars were reported by \citet{2020gorski}. These differences in the reported mean reddening values for the LMC and SMC may arise from variations in the tracers and from the spatial coverage of the samples. Reddening estimates derived from early-type stars may be systematically overestimated because these stars are preferentially located in dusty star-forming regions \citep{1999zari}. Additionally, interstellar dust in the LMC is highly inhomogeneous, showing clumpy structures on arcsecond scales \citep{1997harri}. A similar irregular dust distribution in the SMC suggests that reddening values derived from young stellar populations may likewise be biased toward higher extinction \citep{2002zari}. Adopting smaller mean reddening values in our analysis, particularly for the SMC, would produce negative reddening estimates for a statistically significant number of CCs, which is unphysical. Such negative reddening values have been previously reported in CC- and RR Lyrae-based studies of the MCs. Using OGLE $V$-band data for 1529 RR Lyrae stars in the SMC and 13,490 in the LMC, \citet{Haschke2011} found negative reddening values for about $5\%$ of the sample in the LMC bar region and for approximately $7\%$ of the SMC sample. Similarly, \citet{2019joshi} reported that 9 spatial segments ($6.6\%$ of their sample) in the SMC showed unphysical negative reddening values in Cepheid-based analyzes using OGLE bands, while the multiband CC study of \citet{2018deb_lmc} found such values for less than $1\%$ of their sample. In the present work, after adding the mean values as described above, we identify 115 CCs in the LMC ($\sim3.03\%$) and 292 CCs in the SMC ($\sim6.6\%$) with negative reddening values. The origin of these negative reddening estimates remains unclear both in previous studies and in our analysis, and may require improved multiband light-curve data and further investigation of spatial extinction variations across the MCs.

To validate our reddening estimates, we compared the $E(B-V)$ distributions for LMC and SMC against the maps of \citet{2019joshi}, \citet{2021chow}, and \citet{2021skowron}. The $E(V-I)$ values of \citet{2021skowron} were converted to $E(B-V)$ using $E(B-V)_{\mathrm{LMC}} = E(V-I) / 1.35$ and $E(B-V)_{\mathrm{SMC}} = E(V-I) / 1.35$, adopting the WC23 extinction law at $\lambda = 798$~nm for the $I$ band. Figure~\ref{fig:reddcomp} illustrates the PDFs and CDFs for these samples. Statistical consistency was evaluated using the Kolmogorov--Smirnov (K-S) test, where $D$ denotes the maximum difference between the CDFs of this work and those reported in the literature. The smallest K-S deviations for both LMC and SMC ($D_{\mathrm{LMC}} = 0.10$, $D_{\mathrm{SMC}} = 0.19$) are found with \citet{2021chow}, who employed a mid-IR P--L relation using CCs, indicating the closest agreement with our results. The slight shift in the peak of the PDF of the \citet{2021skowron} distribution is expected, as their map is derived from red clump stars.

We constructed reddening maps for both LMC and SMC using true reddening values for the combined sample of FU and 1O CCs without considering the negative values. To generate the maps, the observed regions of the LMC and the SMC were first divided into a $(12 \times 12)$ spatial grid based on their $\mathrm{RA}$ and $\mathrm{Dec}$ coordinates. For each grid cell, the weighted mean reddening of all CCs within that bin was taken as the representative reddening value, while the standard deviation of the measurements was adopted as the statistical uncertainty. The resulting gridded reddening maps for LMC and SMC for the combined sample of FU and 1O CCs are shown in Fig.~\ref{fig:gridmap}. In order to better visualize the large-scale reddening structure, we also constructed reddening maps by averaging the reddening values of the combined sample of FU+1O CCs on a finer $(80 \times 80)$ spatial grid within the $\mathrm{RA}$ and $\mathrm{Dec}$ range of the LMC and SMC and applying a Gaussian smoothing with $\sigma = 1.6$ pixels to those grids that have at least 3 stars. The smoothed reddening maps with overlaid contour levels are presented in Fig.~\ref{fig:redd_map} with a resolution of $\sim0.4$ kpc for Cepheids in the SMC and LMC, separated by pulsation mode: FU, 1O, and their combined sample (FU+1O) from P--L relations without period breaks. Each panel shows the mean reddening values in the spatial bins of RA, Dec, and the Cartesian coordinate system $(x, y)$ centered on the host galaxy. The details of the coordinate transformation are described in Appendix~\ref{app:coord}. Similar maps constructed from the P--L relations with period breaks are shown in Appendix~\ref{app:figures} (Fig.~\ref{fig:redd_map_lmc_br} and Fig.~\ref{fig:redd_map22}).

The reddening map of the LMC derived in this study, constructed with and without incorporating the break in the P-L relations, shows a clear concentration of high extinction in the region bounded by $\alpha = 82.56^\circ$--$89.37^\circ$ and $\delta = -70.74^\circ$ to $-67.75^\circ$. The most prominent peak is located near $\alpha \sim 85.5^\circ$, $\delta \sim -70.0^\circ$, coinciding with the Tarantula Nebula (30 Doradus) region in the LMC bar, one of the most active star-forming H\,{\sc II} regions in the Local Group \citep{niko2004, 2016Inno, 2013tatto}. Another higher reddening zone is seen around $\alpha \sim 74^\circ$, $\delta \sim -69^\circ$, which has previously been identified as the LMC {\sc HI} supergiant shell SGS 12 (LMC3), a region that hosts young stellar clusters \citep{2010glat}. The LMC reddening maps also reveal a more complex and extended structure, with an apparent ring-like feature and significant variations in $E(B-V)$ across the disk. The ring-like extinction feature in the LMC has previously been identified in OGLE-based maps using red clump stars and Cepheids \citep[e.g.,][]{Haschke2011, 2021skowron}, as well as in near-IR surveys such as VISTA \citep{Tatton2021}. The enhanced extinction toward the southeastern part of the LMC is consistent with previous studies using various tracers and reflects the internal dust structure of the galaxy \citep{Haschke2011, 2016Inno, 2019joshi, 2020bell, 2021skowron, 2022bell}. Although the reddening maps constructed with and without considering P-L breaks show small star-to-star differences, both approaches consistently identify the 30 Doradus region as the location of the highest reddening in the LMC.

The reddening maps of the SMC constructed in this study exhibit a relatively compact central region of higher extinction, particularly toward the southwestern part of the bar, with the strongest peak around $\alpha \sim 12.9^\circ$ and $\delta \sim -73.0^\circ$, while comparatively lower reddening is found toward the northeast bar and the wing. The higher extinction toward the southwestern part of the bar is consistent across FU and 1O populations. Similar trends have been reported in previous studies using different tracers, including red clump stars, red giants, and CCs \citep{2012subra, 2015subra, 2015rube, 2019joshi, 2019deb_smc}. The reddening pattern observed in this study agrees well with earlier extinction maps derived from CCs \citep{2015subra} and young stellar populations \citep{2002zari}, which also report increasing extinction along the bar from northeast to southwest. Studies based on star clusters, on the other hand, report comparatively larger reddening values in the MCs. \citet{2016nayak} analyzed 1072 star clusters in the LMC using OGLE--III data and derived $E(V-I)$ values between 0.05 and 0.50 mag, with most clusters lying in the range 0.1--0.3 mag. Using a similar approach, \citet{2018nayak} studied 179 open clusters in the SMC and also obtained reddening values generally higher than those derived in the present work.

\begin{figure*}[ht]
    \centering
    \includegraphics[width=0.49\linewidth]{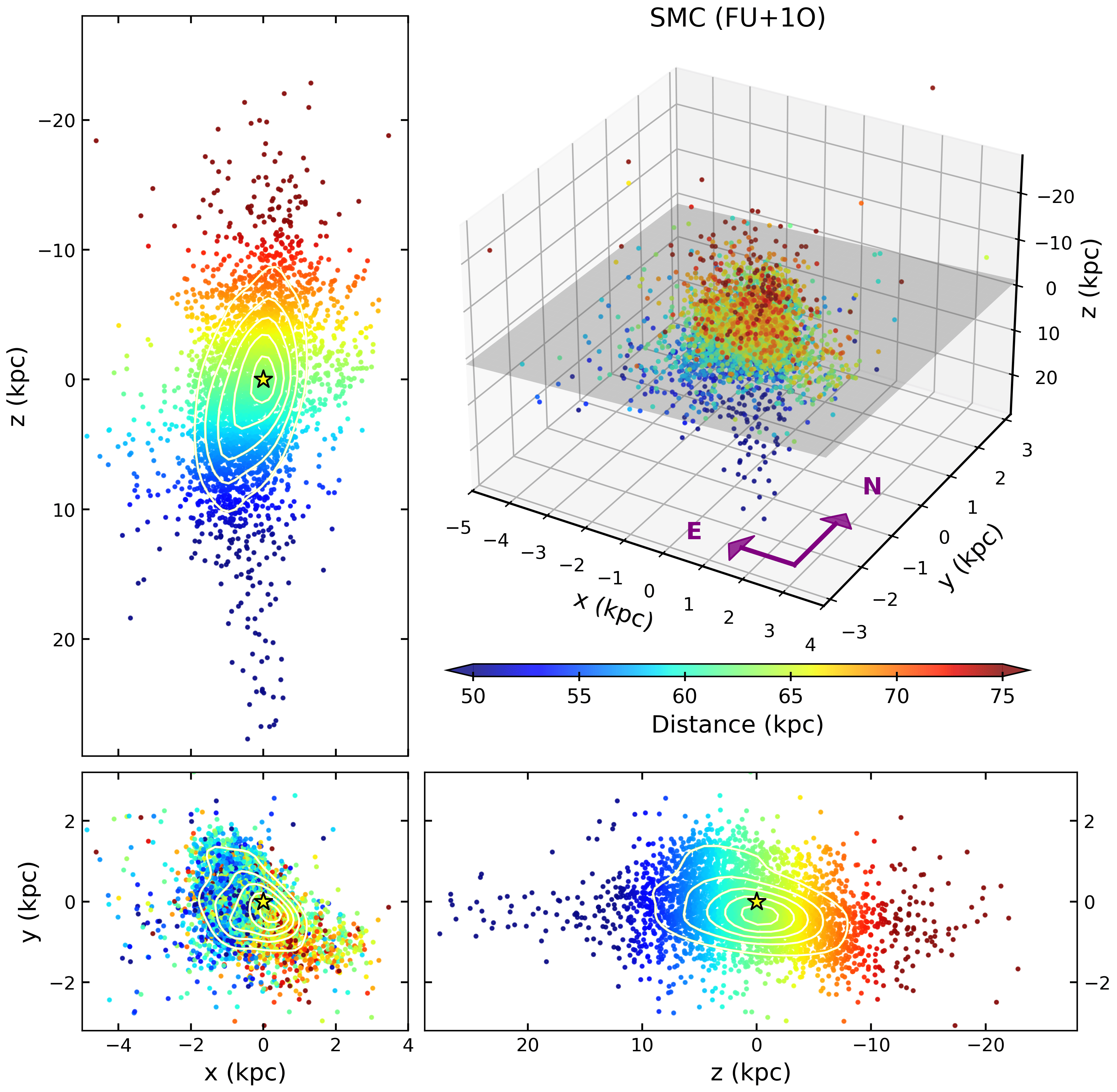}
    \includegraphics[width=0.49\linewidth]{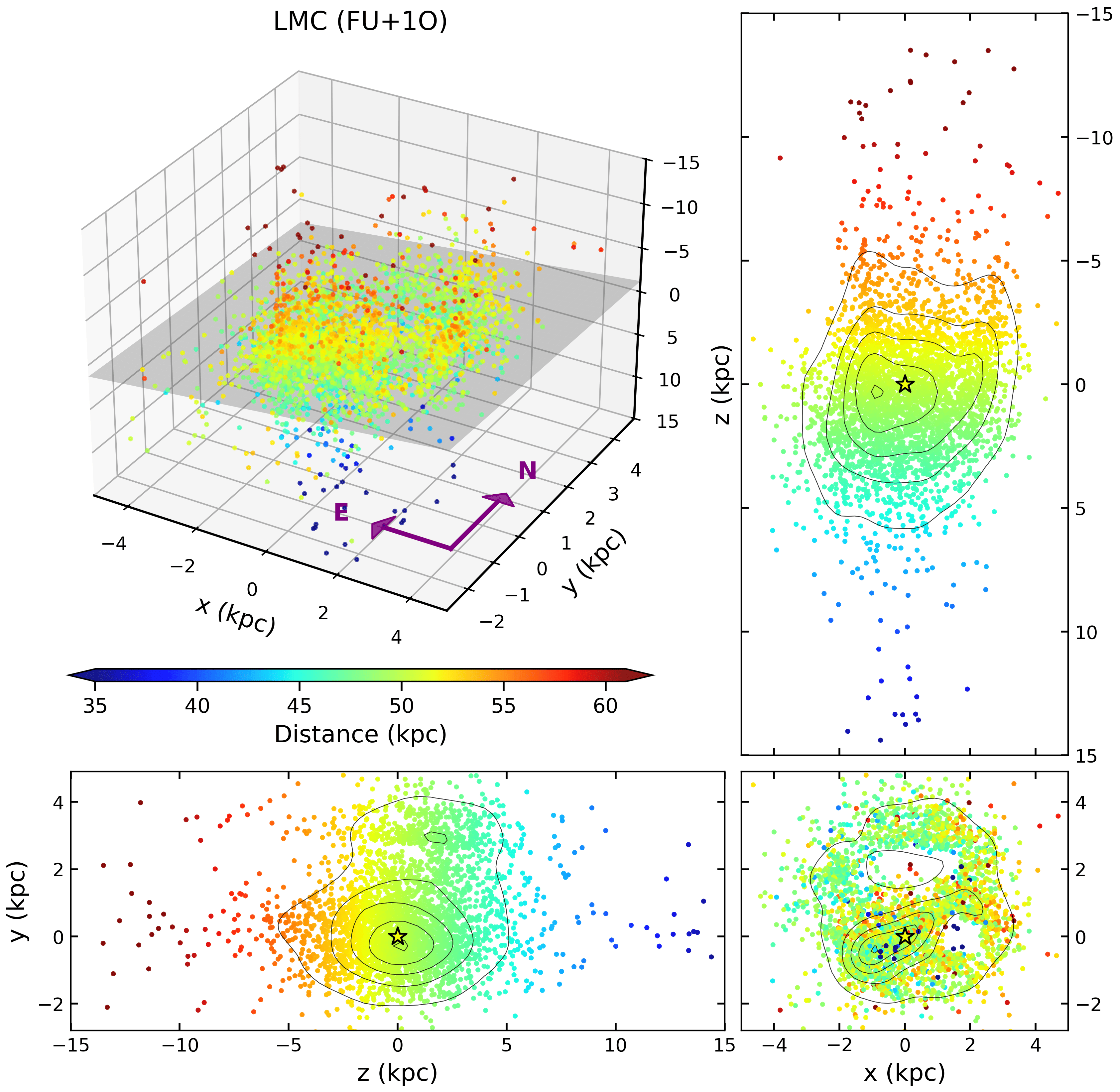}
    \caption{    3D spatial distribution of CCs in the SMC (left panel) and LMC (right panel), shown in a Cartesian coordinate system $(x, y, z)$ centered on each galaxy. The orientation of the sky is indicated by the north (N) and east (E) directions. The points are color-coded according to their line-of-sight distances. The shaded surface represents the best-fit plane of the form $z = Ax + By + C$, obtained from a weighted least-squares fit to the stellar distribution (See text for details).}

\end{figure*}

Differences between reddening maps derived from different tracers likely reflect population-dependent extinction. Young stars such as Cepheids are more closely associated with dusty star-forming regions than intermediate-age tracers like red clump stars \citep{2004stan, Haschke2011}. These comparisons suggest that reddening estimates traced by younger stellar populations are systematically higher than those obtained from older populations across the MCs. In this regard, within the capabilities of S-PLUS multiband photometry to map spatially resolved extinction with high fidelity, using standard candles as tracers provides the advantage of leveraging multiple intermediate and narrow bands, including the Ca {\sc II} H\&K-sensitive $u$ and $J0378$ filters, offering enhanced sensitivity to metallicity and extinction. The $J0660$ filter may also enable the separation of nebular contamination from the stellar continuum, providing additional constraints on $T_\mathrm{eff}$ through H$\alpha$ absorption profiles \citep[][]{2025luis}. The spatial consistency between the FU and 1O Cepheid samples, and the smoothness of the combined maps, also suggest that systematic effects from the P-L relation calibration or period sampling are minimal. These extinction maps not only support the internal consistency of our reddening estimates but also provide valuable constraints for future 3D dust modeling and studies of the Magellanic interstellar medium. Looking ahead, while accurate estimation of photometric metallicity in these narrow bands requires well-sampled light curves and is beyond the scope of this study, the future availability of such data will open exciting opportunities to derive metallicities and further characterize the stellar populations of the MCs.

\subsection{3D distribution of Cepheids in the MCs}
\label{subsec:3D}

To examine the spatial distribution of CCs in the LMC and SMC, the equatorial coordinates $(\alpha,\delta)$ together with the distances $(D)$ were transformed into a Cartesian coordinate system $(x,y,z)$ following \citet{2001marel} and \citet{2001wein}. The distance to each star was obtained using the standard relation $D = 10^{0.2(\Delta\mu + \mu_{\rm mean}) - 2.0}$, where $\mu_{\rm mean}$ represents the mean distance modulus of the host galaxy and $\Delta\mu$ is derived from the simultaneous fitting of the multiband distance moduli as a function of the inverse effective wavelength, assuming the WC23 extinction law. In this coordinate framework, the $z$-axis is directed toward the observer, the $x$-axis is antiparallel to the $\alpha$-axis, and the $y$-axis is aligned with the $\delta$-axis. The adopted centers of the LMC and SMC were taken from the NASA/IPAC Extragalactic Database (NED)\footnote{\url{https://ned.ipac.caltech.edu/}} as $(\alpha_{0}, \delta_{0})=(80.894^\circ, -69.756^\circ)$ for the LMC and $(\alpha_{0}, \delta_{0})=(13.187^\circ, -72.829^\circ)$ for the SMC. The distances to the centers of the two galaxies were adopted as $(D_{0}, \mu_{0})=(49.59\,{\rm kpc}, 18.477)$ for the LMC \citep[from][]{2019piet} and $(D_{0},\mu_{0})=(62.447\,{\rm kpc}, 18.977)$ for the SMC \citep[from][]{2020grac}.

The position angle ($\theta_{\rm lon}$) represents the orientation of the line of nodes, measured from north toward east following the standard astronomical convention. The inclination ($i$) describes the tilt of the galaxy plane $(x',y')$ relative to the sky plane $(x,y)$. These parameters are obtained by fitting a plane of the form $z=f(x,y)$ to the stellar distribution in Cartesian coordinates $(x,y,z)$ and determining its orientation with respect to the sky plane. We performed a weighted least-squares fit in order to account for the different uncertainties in the individual stellar distances. Each star was assigned a weight proportional to the inverse variance of its line-of-sight distance (i.e., $w_i = 1/\sigma_{z,i}^{2}$), so that stars with smaller uncertainties contribute more strongly to the determination of the plane parameters. The best-fitting coefficients were obtained by minimizing the weighted residuals between the observed and modeled values of  $z$, and then $i$ and $\theta_{\rm lon}$ were derived from the fitted coefficients. The uncertainties in $\theta_{\rm lon}$ and $i$ were estimated by standard propagation of errors from the uncertainties in the fitted coefficients. The details of the coordinate transformation, along with the equations for calculating $\theta_{\rm lon}$ and $i$, are included in the Appendix \ref{app:coord}.

The values of $i$ and $\theta_{\rm lon}$ obtained in this study are: LMC: $(i, \theta_{\rm lon}) = (24.3 \pm 1.2^\circ,\ 154.5 \pm 1.8^\circ)$ and SMC: $(i, \theta_{\rm lon}) = (57.9 \pm 1.7^\circ,\ 149.3 \pm 3.2^\circ)$. The viewing angle parameters obtained in the present analysis are broadly consistent with some previous CC-based investigations.

For the LMC, \citet{2018deb_lmc} determined the geometry of the galaxy using multiwavelength photometry of more than 3500 Cepheids and obtained viewing angles of $i = 25.11^\circ \pm 0.37^\circ$ and $\theta_{\rm lon} = 154.70^\circ \pm 1.38^\circ$ from a plane fit to the 3D distribution of the Cepheids. On the other hand, \citet{2016Inno} derived $i = 25.05^\circ \pm 0.55^\circ$ and $\theta_{\rm lon} = 150.76^\circ \pm 0.07^\circ$ using multiwavelength period-Wesenheit index relations for a large sample of CCs. Comparable results have also been reported by \citet{Haschke2011}, who found $i \sim 32^\circ$ and $\theta_{\rm lon} \sim 151^\circ$, and by \citet{2024Bhuyan}, who obtained $i \sim 23^\circ$ and $\theta_{\rm lon} \sim 155^\circ$, which further support a relatively well-defined disk geometry for the LMC. These values agree well with our estimates for the LMC within the uncertainties.

However, for the SMC, a larger spread in the reported viewing angles is commonly found in the literature. For example, using OGLE Cepheids and a weighted plane-fitting method, \citet{2015subra} obtained $i = 64.4^\circ \pm 0.7^\circ$ and $\theta_{\rm lon} = 155.3^\circ \pm 6.3^\circ$. Using the same procedure, \citet{2015rube} calculated $i = 39.3^\circ \pm 5.5^\circ$ and $\theta_{\rm lon} = 179.3^\circ \pm 2.1^\circ$. Similarly, \citet{2016Jacy} reported $i \sim 65^\circ$ and $\theta_{\rm lon} \sim 150^\circ$, while \citet{2019deb_smc} obtained $i = 3.57^\circ \pm 0.04^\circ$ and $\theta_{\rm lon} = 63.09^\circ \pm 0.12^\circ$, depending on the tracer and methodology adopted. The presence of such a wide range of values mainly reflects the intrinsically complex and nonplanar structure of the SMC, which is known to possess a large line-of-sight depth and significant structural subcomponents.

\section{Summary and conclusions}\label{Sec:Conclusions}

We present a homogeneous multiband analysis of CC P-L relations in the MCs using single-epoch S-PLUS DR4 data. This work extends multiband Cepheid analysis to the 12-filter S-PLUS photometric system, combining five broad bands with, for the first time, seven narrow bands. Our main results and conclusions are summarized as follows:

\begin{itemize}

\item We implemented a phase-dependent correction framework tailored for single-epoch photometry that enables statistical recovery of mean magnitudes for a sample of $>8000$ known CCs spanning a period range of $0.23$ to $41.21$ days in LMC and SMC. This approach successfully disentangles the effects of relative distance and reddening offsets to construct consistent multiband P-L relations.

\item The multiband P-L relation coefficients were derived iteratively for both FU and 1O Cepheids in the LMC and SMC. The solutions were validated using least-squares fitting, MLE, and Bayesian MCMC analysis. After correcting for star-by-star reddening and distance offsets, the dispersion of the P-L relations decreases by $\sim 30$--$65\%$ across the filters. This highlights the effectiveness of the multiband approach in reducing extinction, line-of-sight depth, and photometric scatter.

\item The P-L slopes exhibit a clear wavelength dependence, steepening from $\alpha \sim -2.0$ in the bluest to $\alpha \sim -3.1$ in the reddest filters. The zero points decrease systematically toward longer wavelengths.

\item The narrowband filters centered on key spectral features (e.g., Ca H+K, H$\delta$, Mg\,b, H$\alpha$, Ca triplet) enable the extension of Cepheid P-L relations, for the first time, to line-sensitive spectral regions. The results are consistent with broadband behavior.

\item Individual reddening and distances were calculated from a simultaneous fitting of the multiband distance moduli as a function of inverse effective wavelength, assuming the WC23 extinction law. Reddening and distances are found to be uncorrelated for our CC sample in each galaxy. There is no dependence of reddening on the pulsation period for either the LMC or the SMC samples, validating the robustness of the method.

\item From the derived reddening values, we constructed spatially resolved reddening maps of the LMC and SMC that reproduce known extinction structures (e.g., 30 Doradus in the LMC) and are consistent with the literature.

\item From the calculated distances of the CCs, coupled with their sky positions, 3D distributions of the CCs in each galaxy are computed. From these distributions, we infer structural parameters as follows: for the LMC,  $(i, \theta_{\rm lon}) = (24.3 \pm 1.2^\circ,\ 154.5 \pm 1.8^\circ)$, and for the SMC, $(i, \theta_{\rm lon}) = (57.9 \pm 1.7^\circ,\ 149.3 \pm 3.2^\circ)$. The LMC values are consistent with previous Cepheid-based determinations, indicating a well-defined disk geometry. In contrast, the SMC parameters should be interpreted with caution, as the wide range of reported viewing angles depends on the adopted tracers and methods and reflects its complex, nonplanar structure.

\end{itemize}

This work demonstrates that single-epoch photometry, combined with a multiband framework, can yield precise Cepheid P-L relations and reliable reddening and distance estimates. This approach provides a powerful pathway for future large-scale surveys to map the 3D structure of nearby resolved galaxies with high precision.

\section*{Data availability} The single-epoch multiband PSF photometry used in this study is publicly available through the S-PLUS cloud server (\url{https://splus.cloud/}, which adheres to Virtual Observatory standards for data accessibility. Table~\ref{table:catalog} (Appendix~\ref{app:table}) is only available in electronic form at the CDS via anonymous ftp to cdsarc.u-strasbg.fr (130.79.128.5) or via \url{http://cdsweb.u-strasbg.fr/cgi-bin/qcat?J/A+A/}.

\begin{acknowledgements}
The S-PLUS project, including the T80-South robotic telescope and the S-PLUS scientific survey, was founded as a partnership between the Fundação de Amparo à Pesquisa do Estado de São Paulo (FAPESP), the Observatório Nacional (ON), the Federal University of Sergipe (UFS), and the Federal University of Santa Catarina (UFSC), with important financial and practical contributions from other collaborating institutes in Brazil, Chile (Universidad de La Serena), and Spain (Centro de Estudios de Física del Cosmos de Aragón, CEFCA). Thanks to the S-PLUS team for their efforts in data reduction, precise calibration, and photometry. We acknowledge the referee for the valuable comments that have improved the quality of the manuscript. DH acknowledges the fruitful discussion with Claudia Vilega Rodrigues, INPE, Brazil, regarding the adopted reddening law for the SMC and LMC, during his visit to INPE. M.C., C.E.F.L and D.H. acknowledges ANID/FONDECYT Regular grant 1231637 and by ANID/Basal (CATA) grant FB21003. D.H. acknowledges the support provided by ANID through doctoral fellowship grant 21232262 for pursuing Ph.D. P.K.H. gratefully acknowledges the Fundação de Amparo à Pesquisa do Estado de São Paulo (FAPESP) for the support grant 2023/14272-4. C.E.F.L acknowledges support from DIUDA 88231R11; LSST Discovery Alliance grant; and GEMINI/ANID grant 32240028. M.C. acknowledges additional support from ANID's Basal project FB210003.
\end{acknowledgements}

\bibliographystyle{aa}
\bibliography{references}


\begin{appendix}

\section{Coordinate transformation}
\label{app:coord}
The equatorial coordinates $(\alpha, \delta)$ and distances $D$ were transformed into a Cartesian coordinate system $(x, y, z)$ using the following relations from \citet{2001marel} and \citet{2001wein}:
\begin{equation}
\begin{aligned}
x &= -D \sin(\alpha-\alpha_0)\cos\delta, \\
y &= D\sin\delta\cos\delta_0 - D\sin\delta_0\cos(\alpha-\alpha_0)\cos\delta, \\
z &= D_0 - D\sin\delta\sin\delta_0 - D\cos\delta_0\cos(\alpha-\alpha_0)\cos\delta .
\end{aligned}
\label{eq:xyz_transform}
\end{equation}

\noindent To describe the orientation of the host galaxy, a rotated coordinate system $(x', y', z')$ is defined. This system is obtained from $(x, y, z)$ by a counterclockwise rotation through an angle $\theta$ about the $z$-axis, followed by a clockwise rotation through an inclination angle $i$ about the resulting $x$-axis. The transformation is given by:
\[
\begin{bmatrix}
x' \\
y' \\
z'
\end{bmatrix}
=
\begin{bmatrix}
\cos\theta & \sin\theta & 0 \\
-\sin\theta \cos i & \cos\theta \cos i & -\sin i \\
-\sin\theta \sin i & \cos\theta \sin i & \cos i
\end{bmatrix}
\begin{bmatrix}
x \\
y \\
z
\end{bmatrix}.
\]

\noindent The inclination ($i$) and the position angle of the line of nodes ($\theta_{\rm lon}$) are derived by approximating the 3D stellar distribution with a planar model of the form $z = Ax + By + C $, where the coefficients $A$ and $B$ describe the orientation of the stellar distribution with respect to sky plane. Then the $\theta_{lon}$ and $i$ are given by:
\begin{equation}
\theta_{\rm lon}  = \arctan\left(\frac{-A}{B}\right) + \mathrm{sign}(B)\,\frac{\pi}{2}, 
i = \arccos\left(\frac{1}{\sqrt{1 + A^{2} + B^{2}}}\right)
\end{equation}
and their corresponding uncertainties from the propagation of errors in the fitted coefficients $A$ and $B$ are given by:
\begin{equation}
\begin{aligned}
\sigma_{\theta_{\rm lon}} & =
\frac{1}{\sqrt{A^{2}+B^{2}}}
\sqrt{A^{2}\sigma_{B}^{2} + B^{2}\sigma_{A}^{2}} \, , \\
\sigma_{i} & =
\frac{1}{(A^{2}+B^{2}+1)\sqrt{A^{2}+B^{2}}}
\sqrt{A^{2}\sigma_{A}^{2} + B^{2}\sigma_{B}^{2}} \, .
\end{aligned}
\end{equation}

\section{Supplementary figures}
\label{app:figures}

\begin{figure}[ht]
    \centering
    \includegraphics[width=\linewidth]{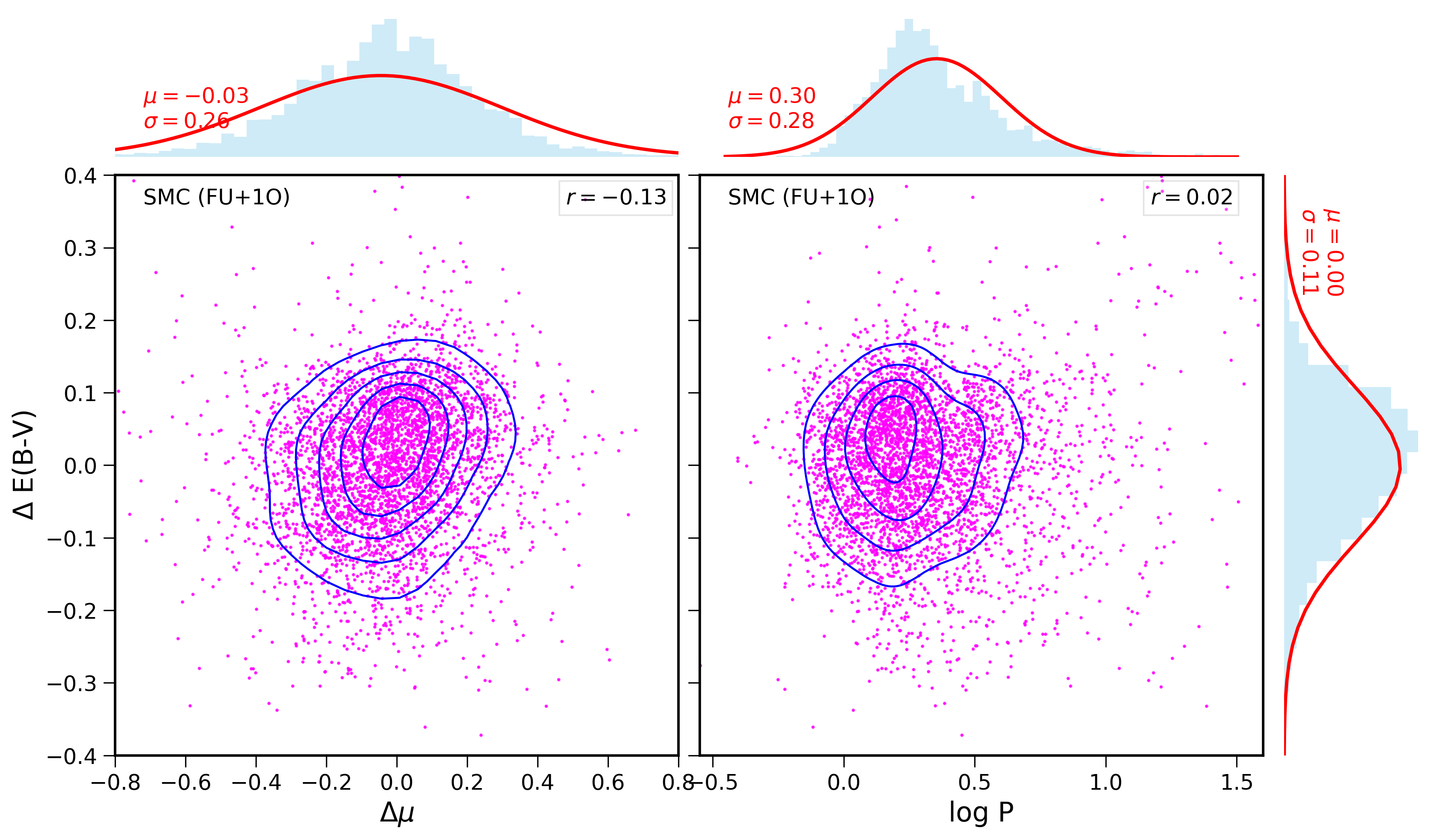}
    \caption{Same as Figure \ref{fig:lmc_ebv_mu} but for the SMC. }
    \label{app:smc_fu_ebv}
\end{figure}

\begin{figure}[ht]
    \centering
    \includegraphics[width=0.95\linewidth]{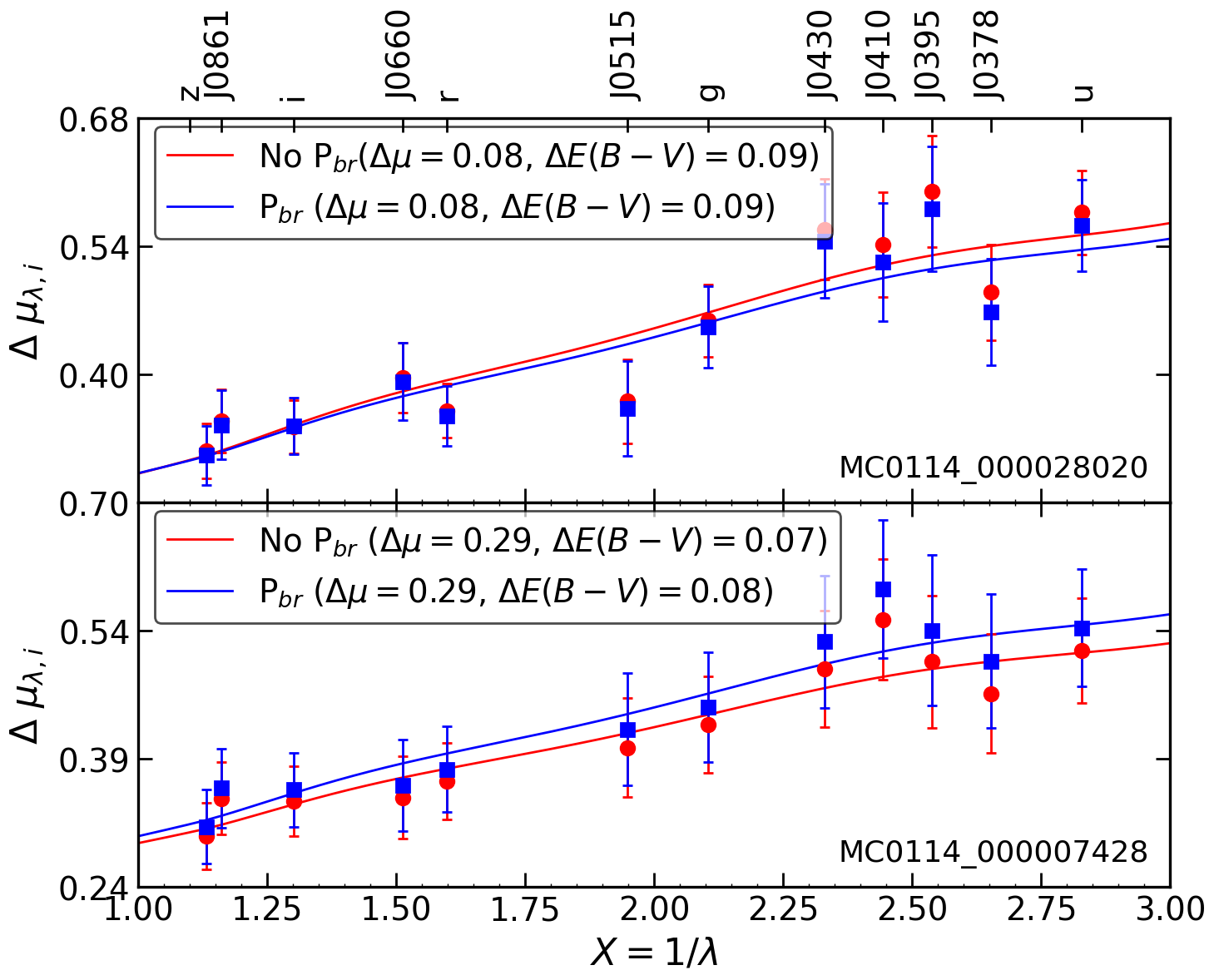}
    \caption{Determination of $\Delta \mu_0$ for two representative Cepheids (S-PLUS IDs indicated). Solid curves show simultaneous multi-band fits (12 bands) to distance moduli using the WC23 extinction law, adopting $R_V=2.53$ (SMC) and $R_V=3.40$ (LMC). Blue squares and red circles represent $\Delta \mu_{\lambda,i}$ values including and excluding period-break cases, respectively.}
    \label{fig:mu_fit} 
\end{figure} 

\begin{figure}[ht]
    \centering
    \includegraphics[width=0.95\linewidth]{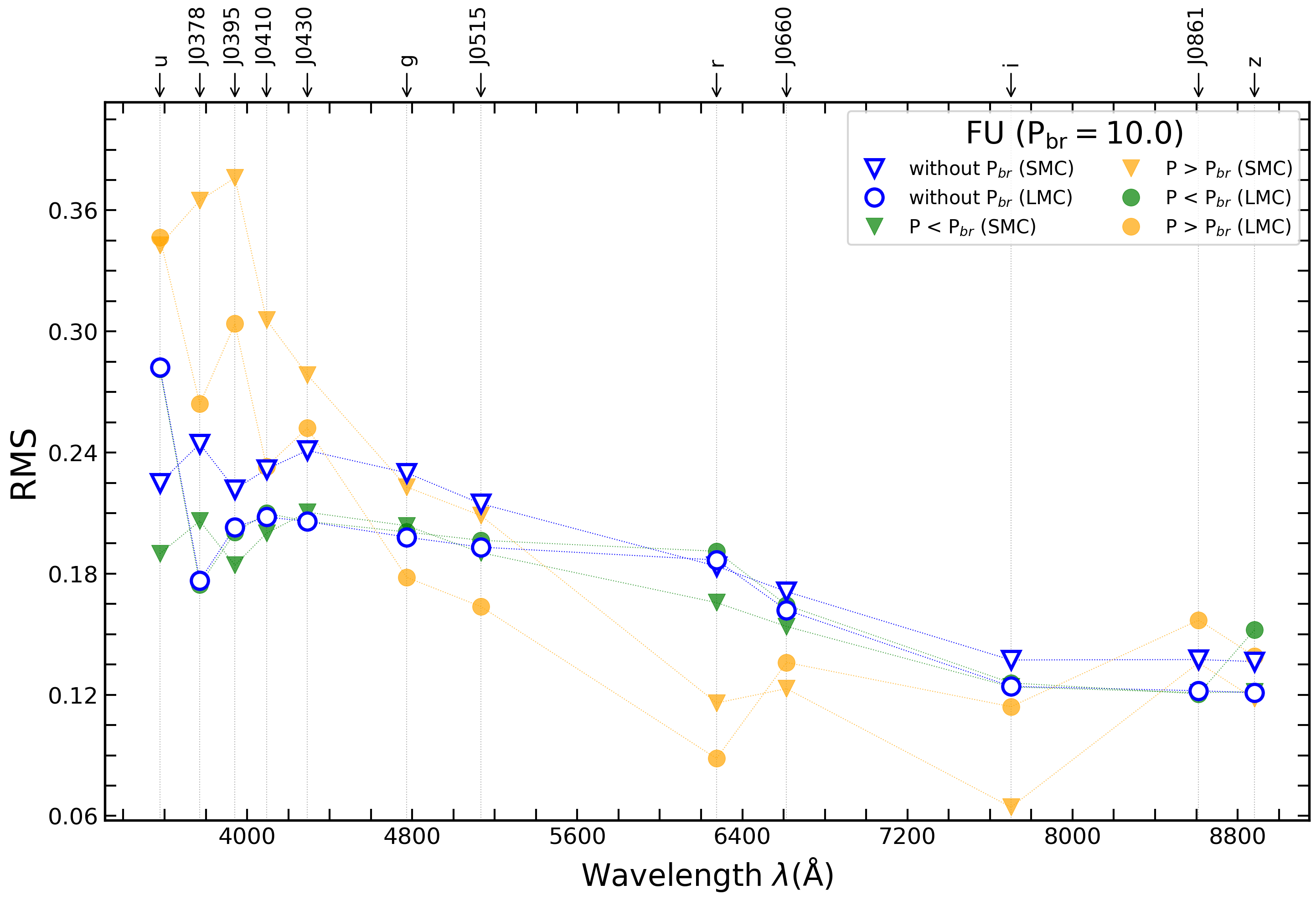}
    \includegraphics[width=0.95\linewidth]{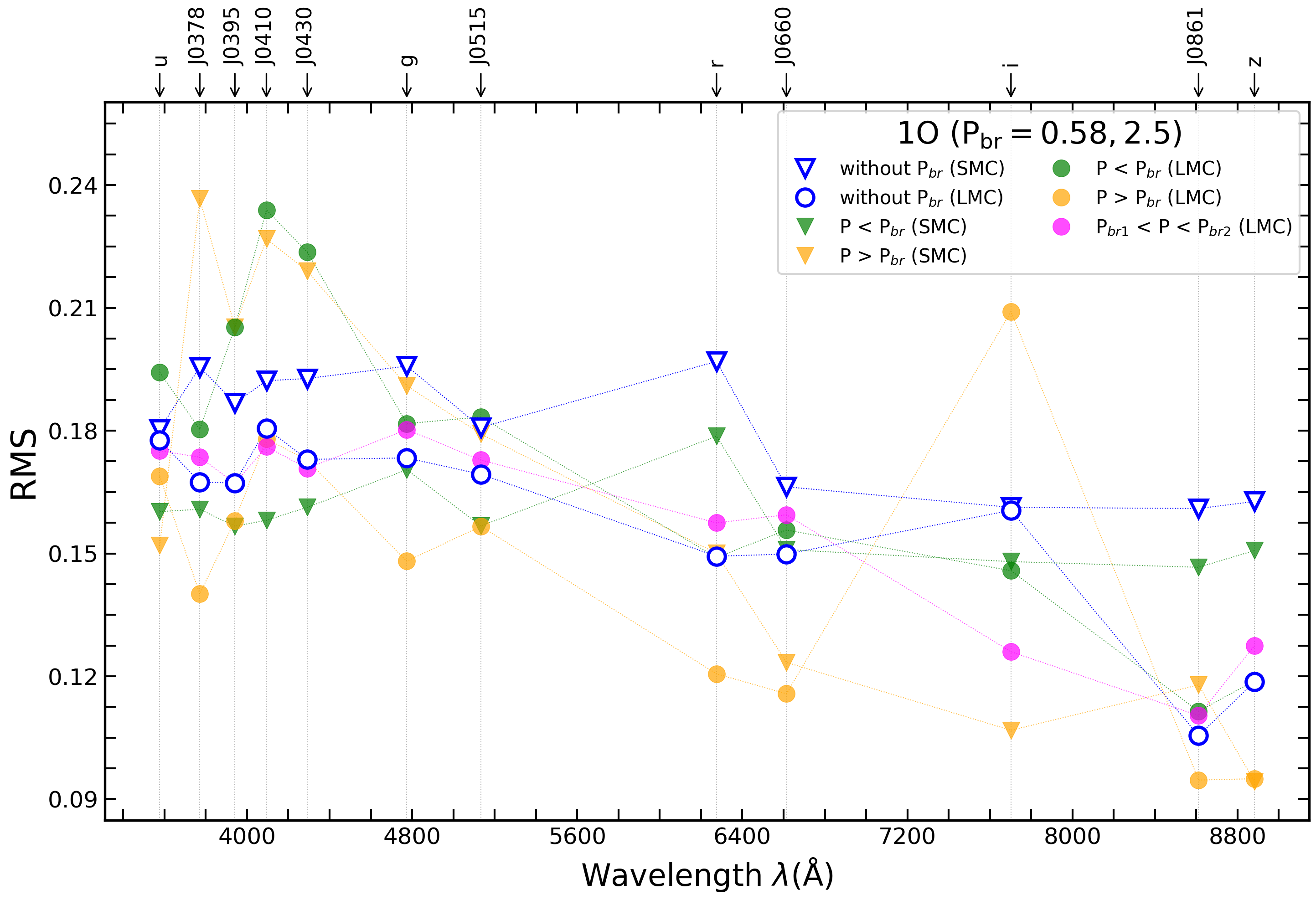}    
    \caption{Variation of the rms of the P-L relation fits across S-PLUS bands, for both broken and single-slope P-L relations. The top and bottom panels show FU and 1O Cepheids, respectively.}
    \label{fig:rms_improvements}
\end{figure}

\begin{figure*}[!]
    \centering 
    \includegraphics[width=0.32\linewidth]{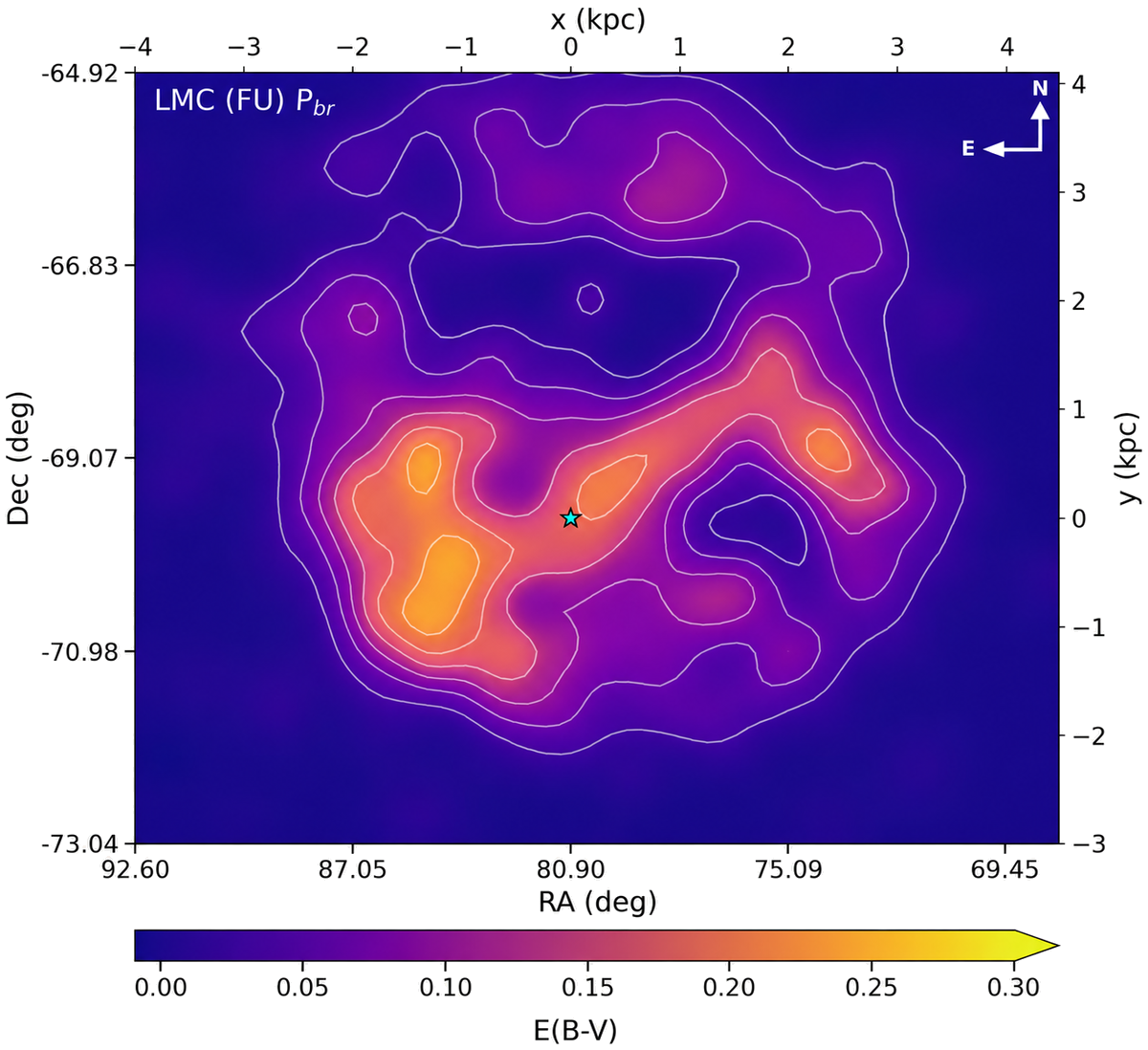}
    \includegraphics[width=0.32\linewidth]{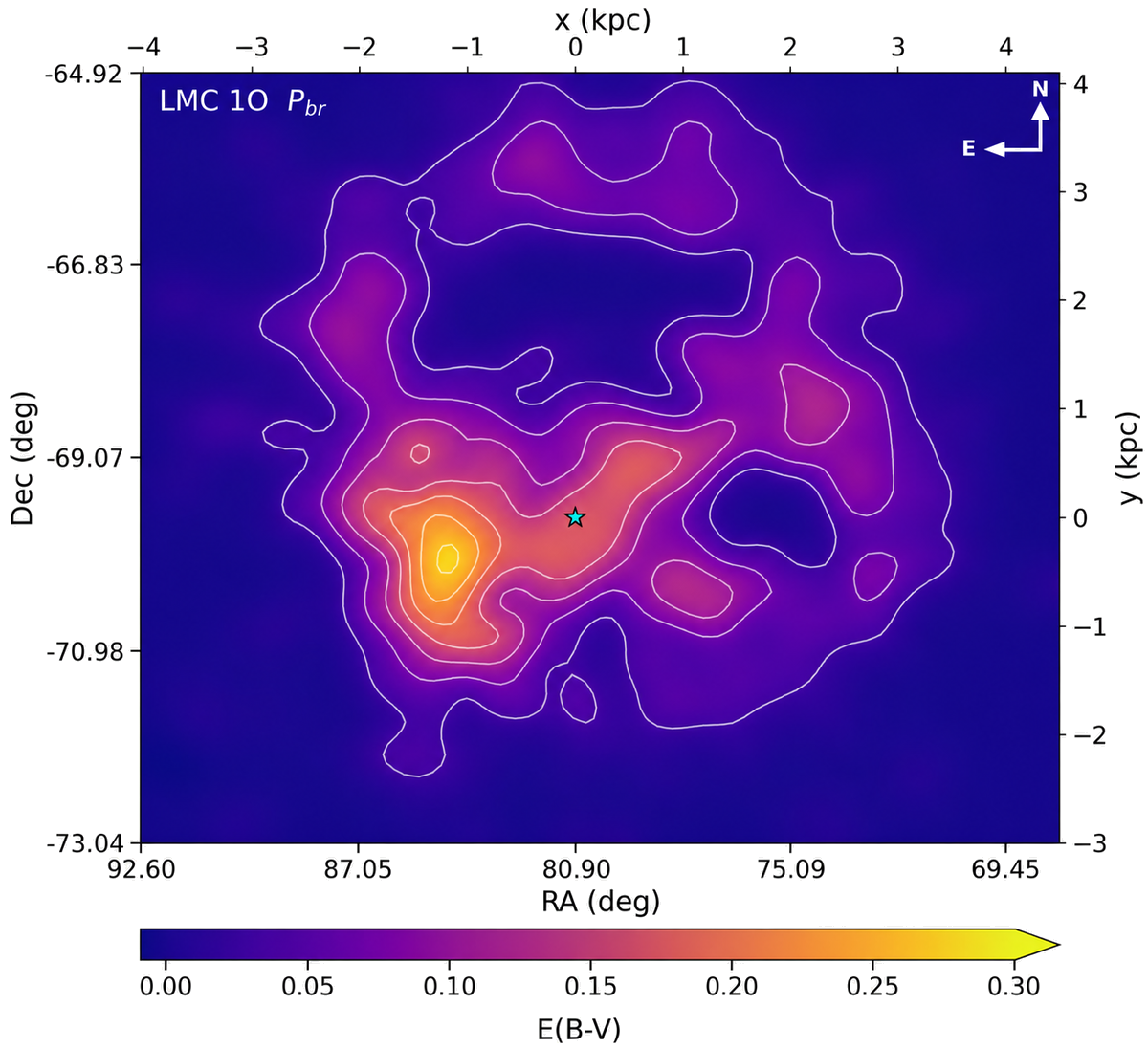}
    \includegraphics[width=0.32\linewidth]{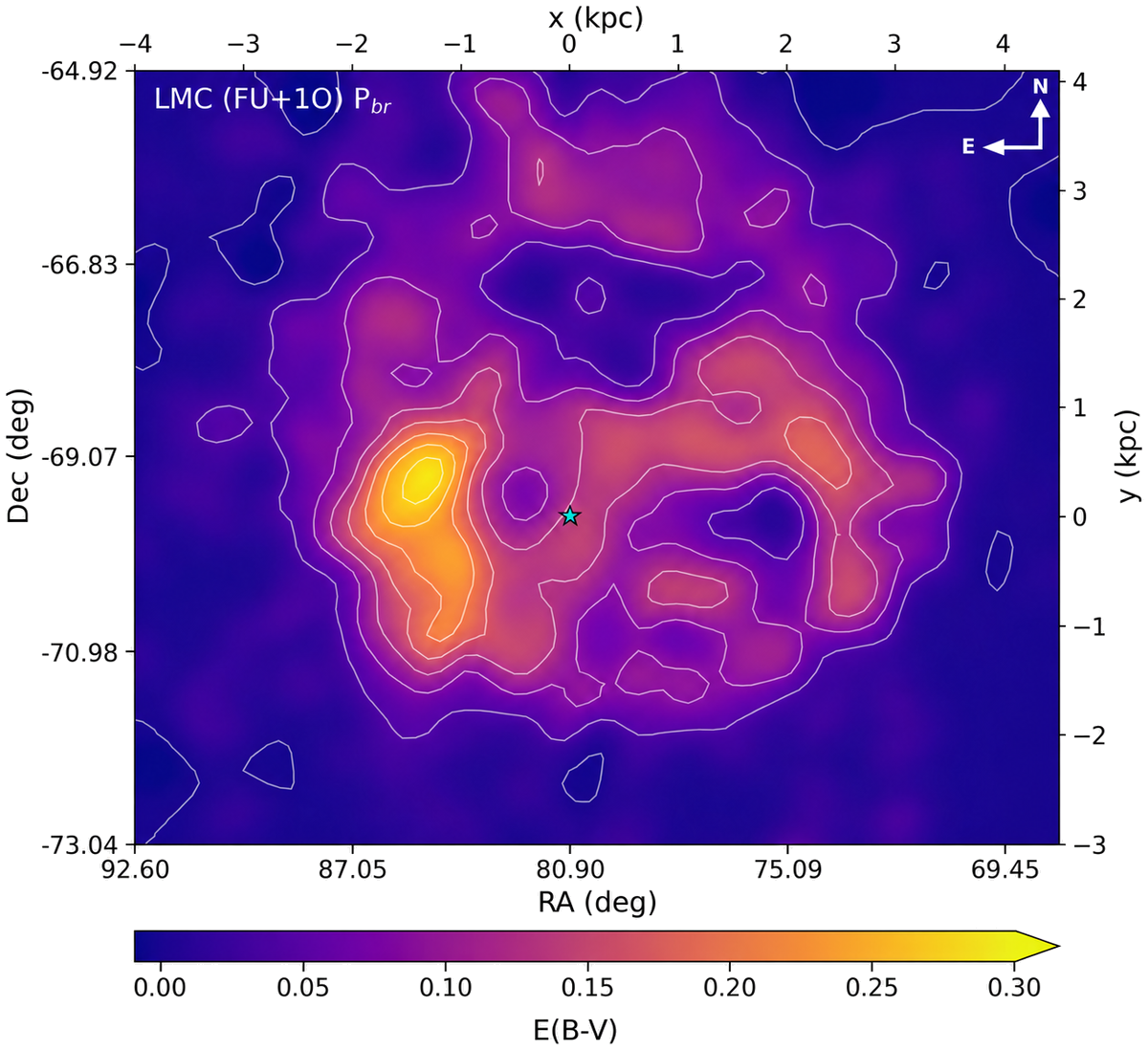}

    \caption{Same as Fig.~\ref{fig:redd_map}, but for the LMC considering break in the P-L relations.}
    \label{fig:redd_map_lmc_br}
\end{figure*}

\begin{figure}[ht]
    \centering
    \includegraphics[width=0.95\linewidth]{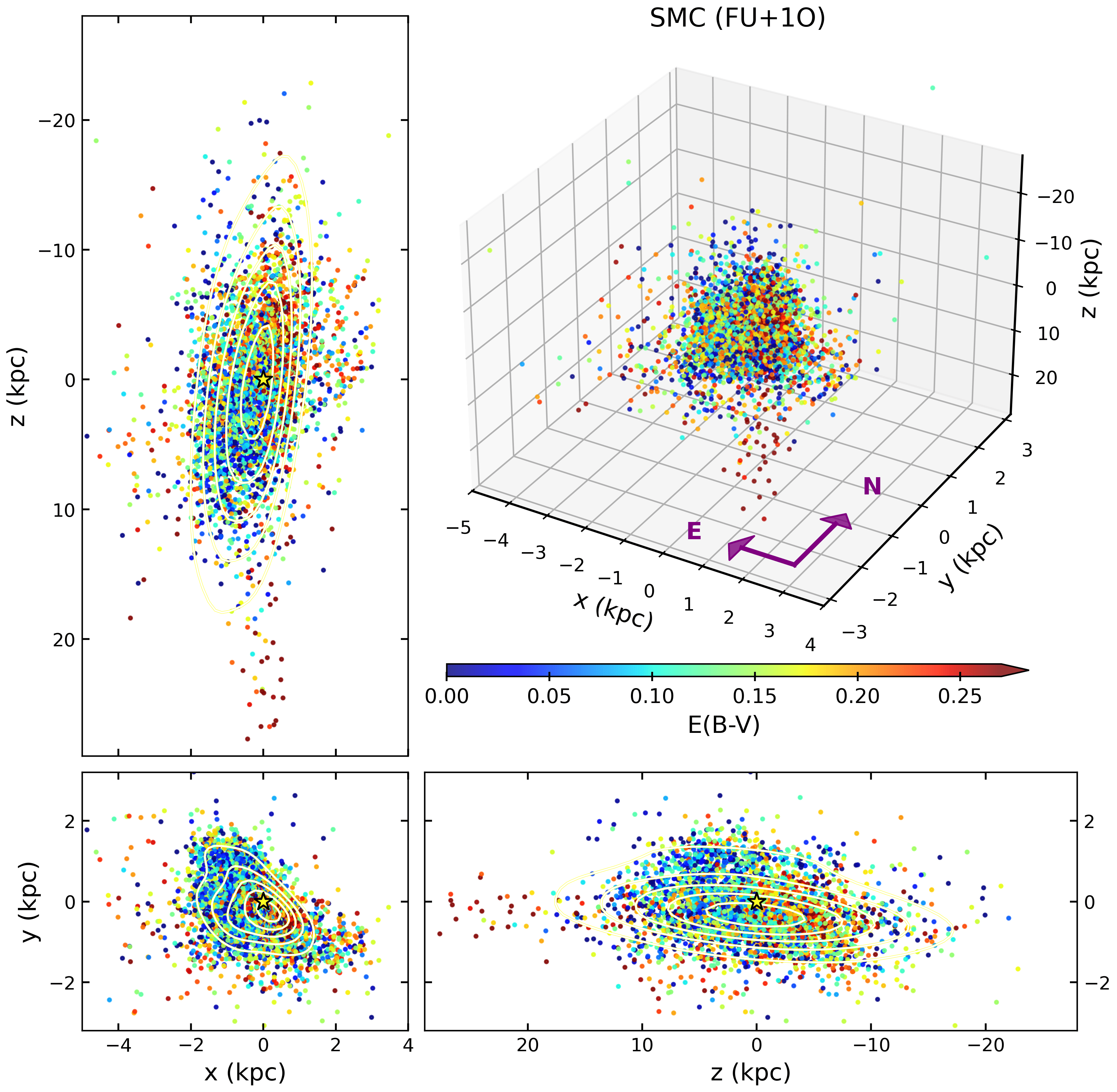}
    \includegraphics[width=0.95\linewidth]{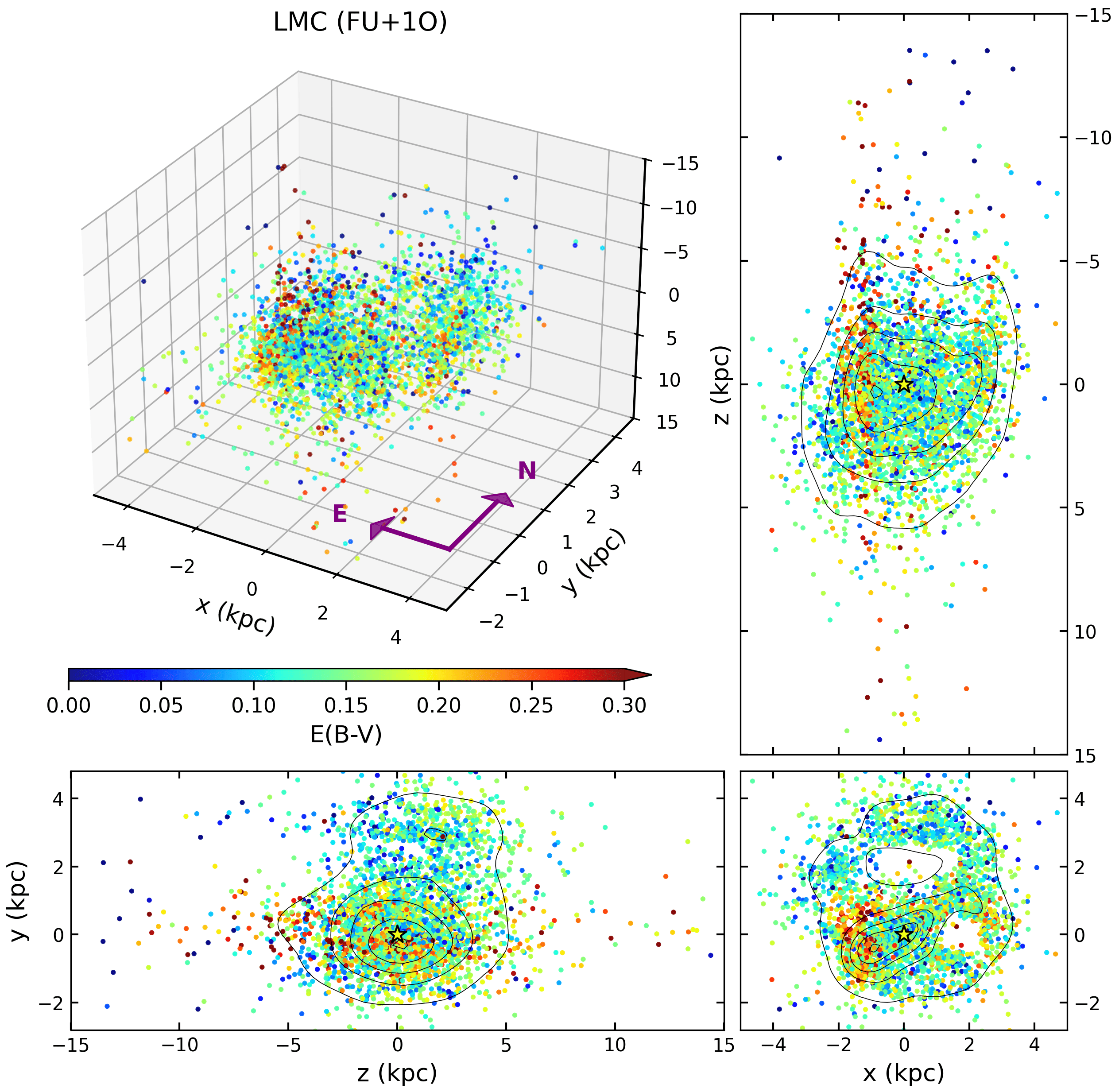}
    \caption{3D spatial distribution of CCs in SMC (top) and LMC (bottom). The points are color-coded according to their $E(B-V)$.}
    \label{fig:3D_EBV}
\end{figure}

\begin{figure}[!ht]
    \centering
    \includegraphics[width=0.95\linewidth]{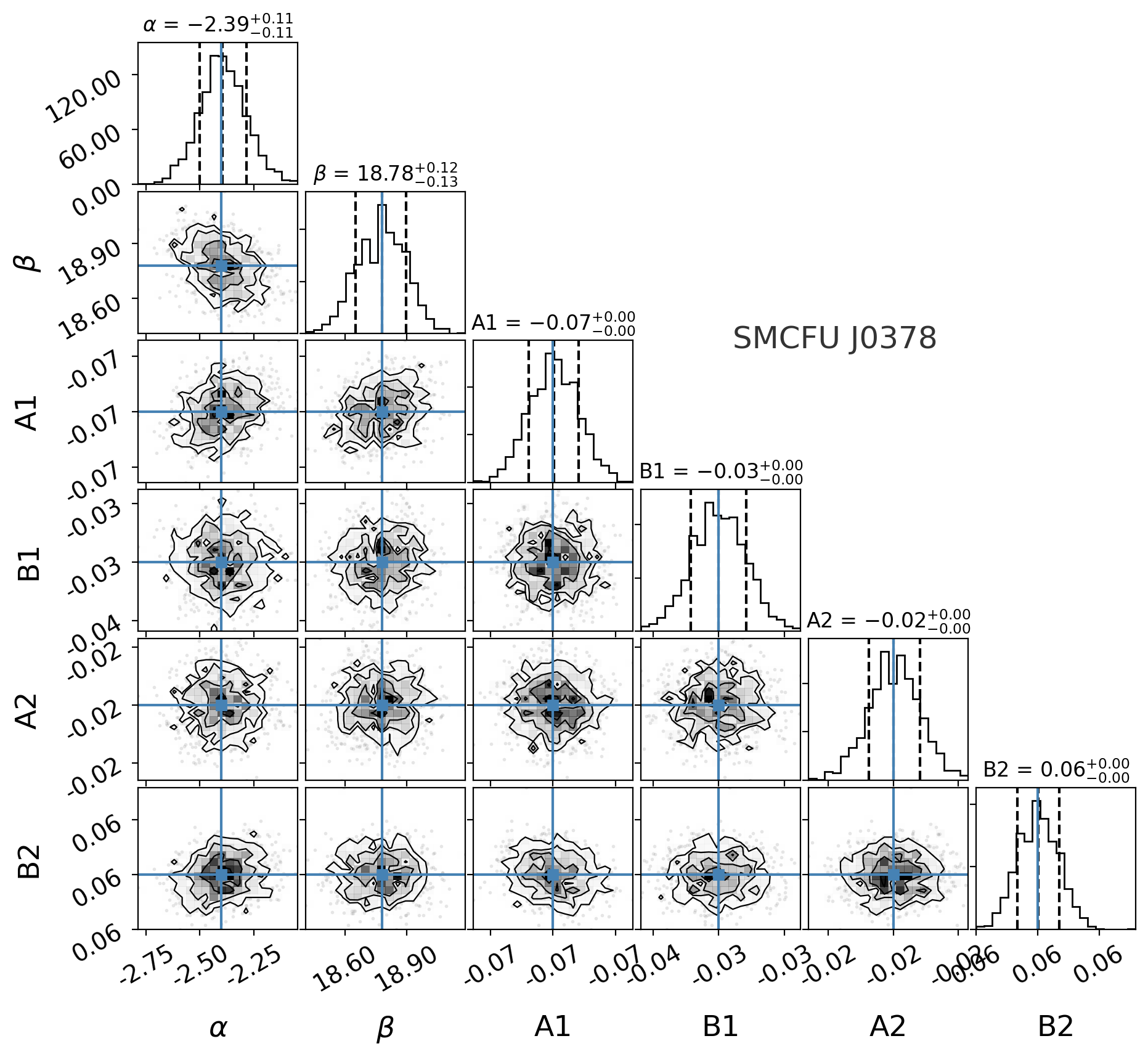}
    \includegraphics[width=0.95\linewidth]{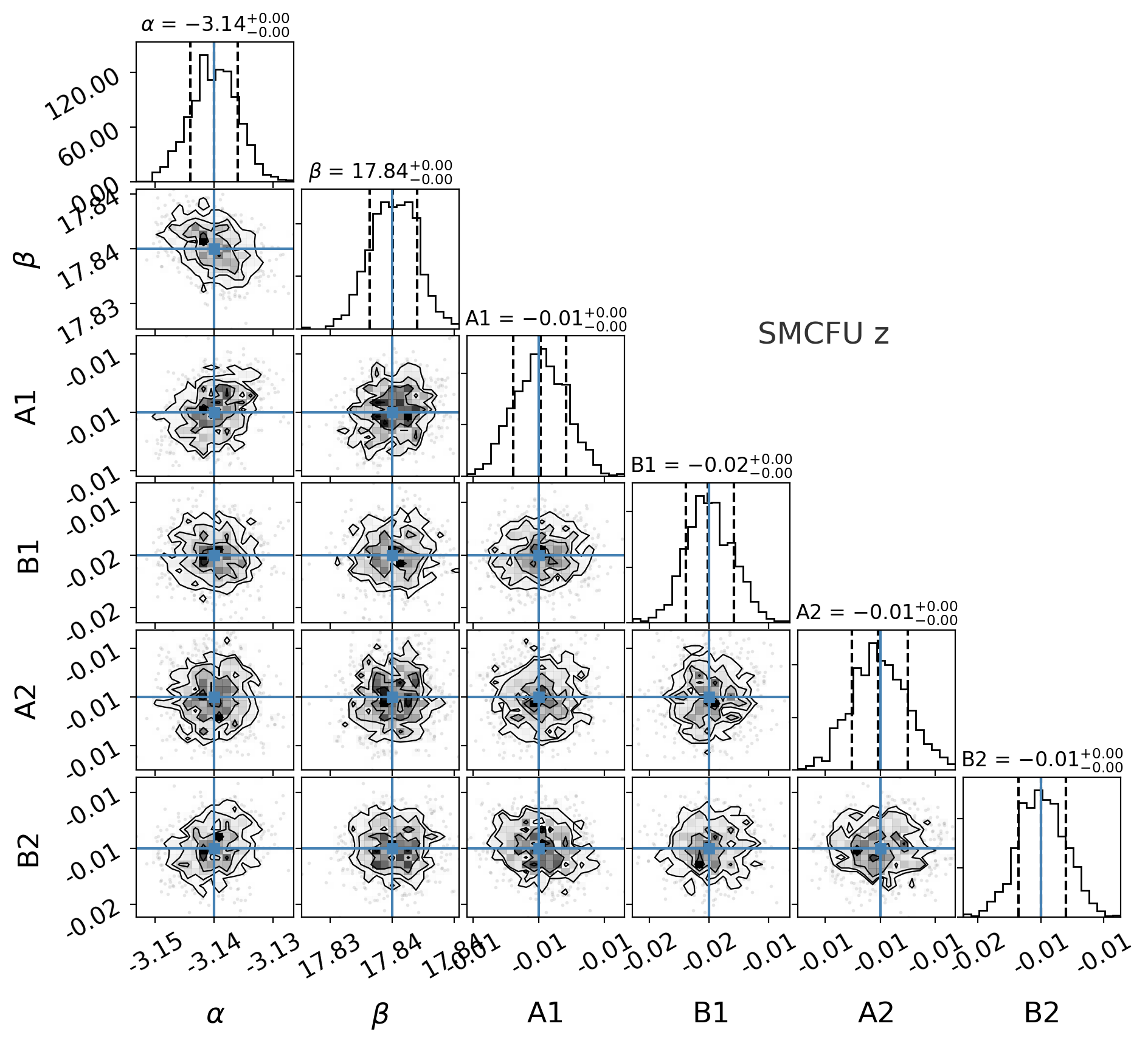}
    
    \caption{Marginalized posterior distributions of the P-L parameters for SMC FU Cepheids from the MCMC analysis, for one narrow ($J0378$) and one broad ($z$) band, without considering break in the P-L relations.}

    \label{fig:mcmc} 
\end{figure}

\begin{figure*}[!]
    \centering 
    \includegraphics[width=0.32\linewidth]{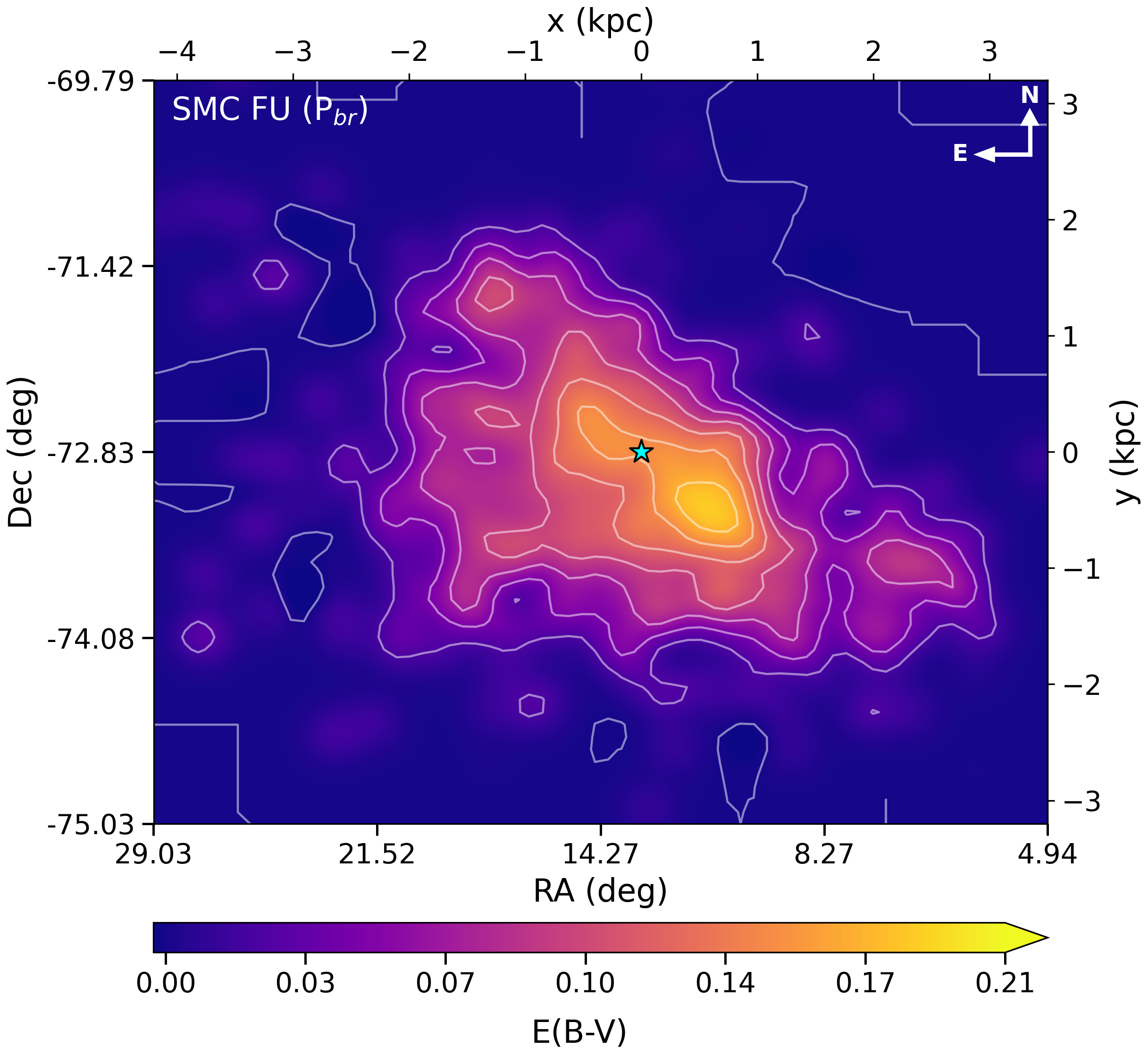}
    \includegraphics[width=0.32\linewidth]{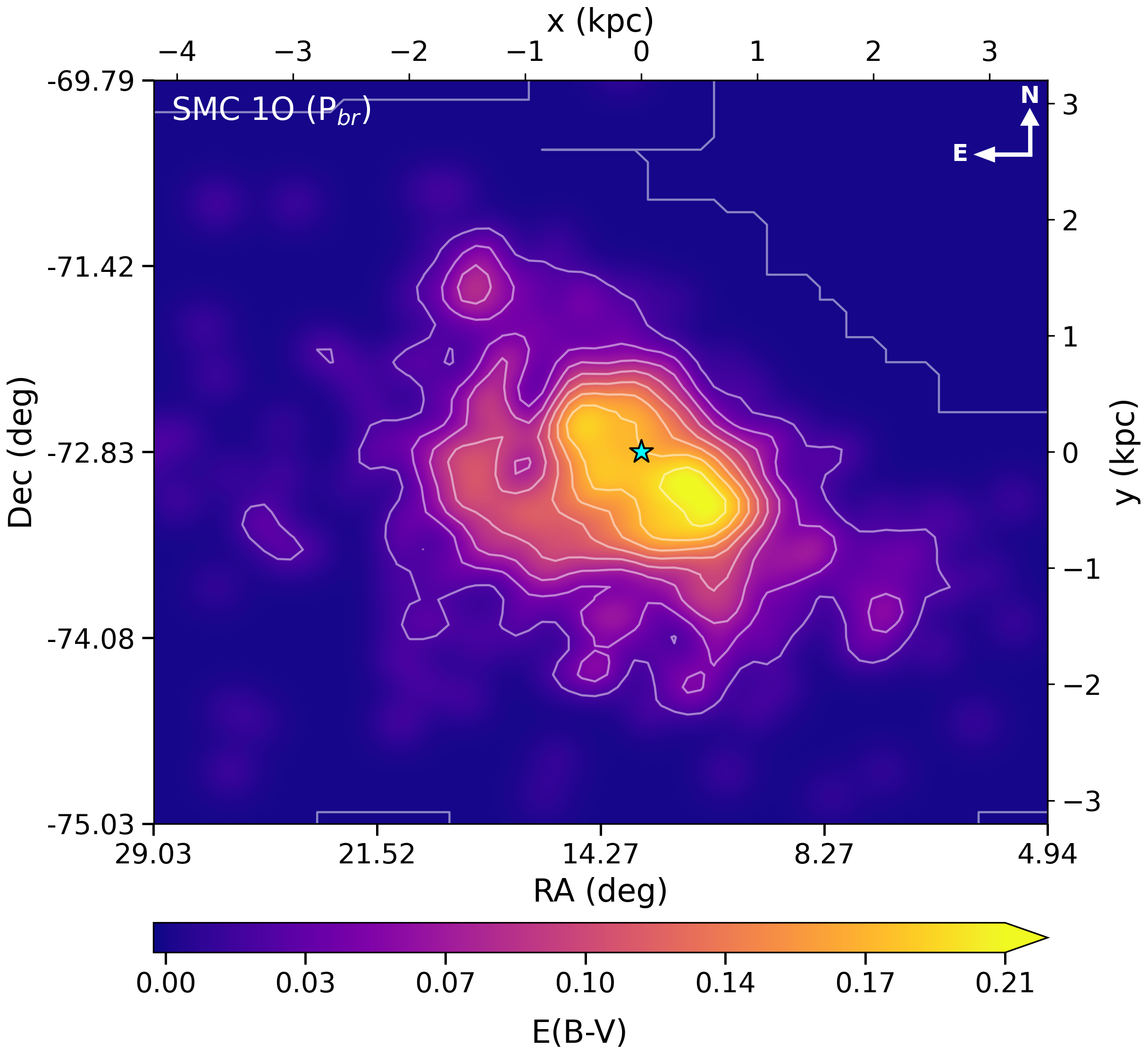}
    \includegraphics[width=0.32\linewidth]{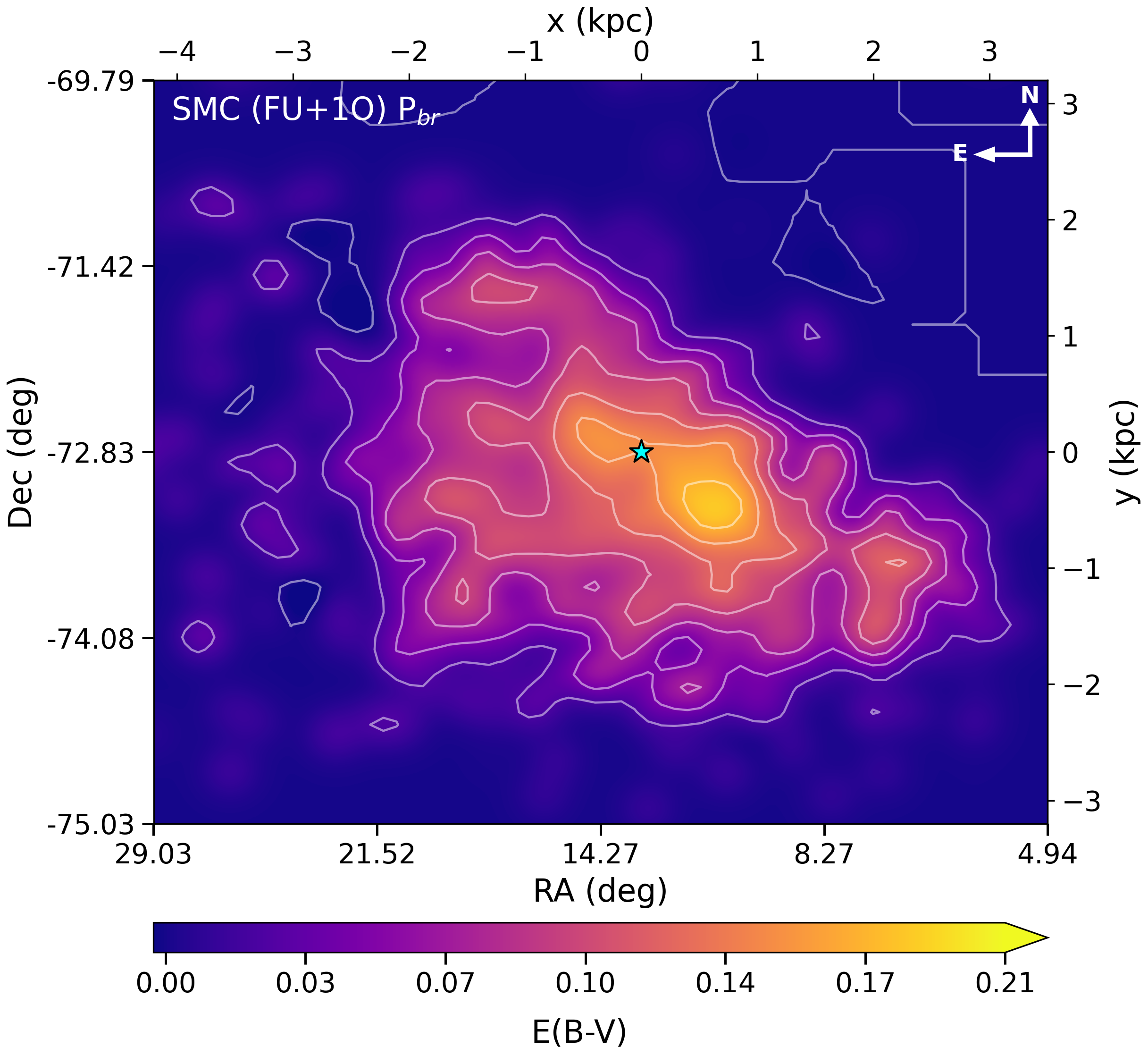}

    \caption{Same as Fig.~\ref{fig:redd_map}, but for the SMC considering break in the P-L relations.}
    \label{fig:redd_map22}
\end{figure*}

\begin{figure*}[ht]
    \centering
    \includegraphics[width=0.246\linewidth]{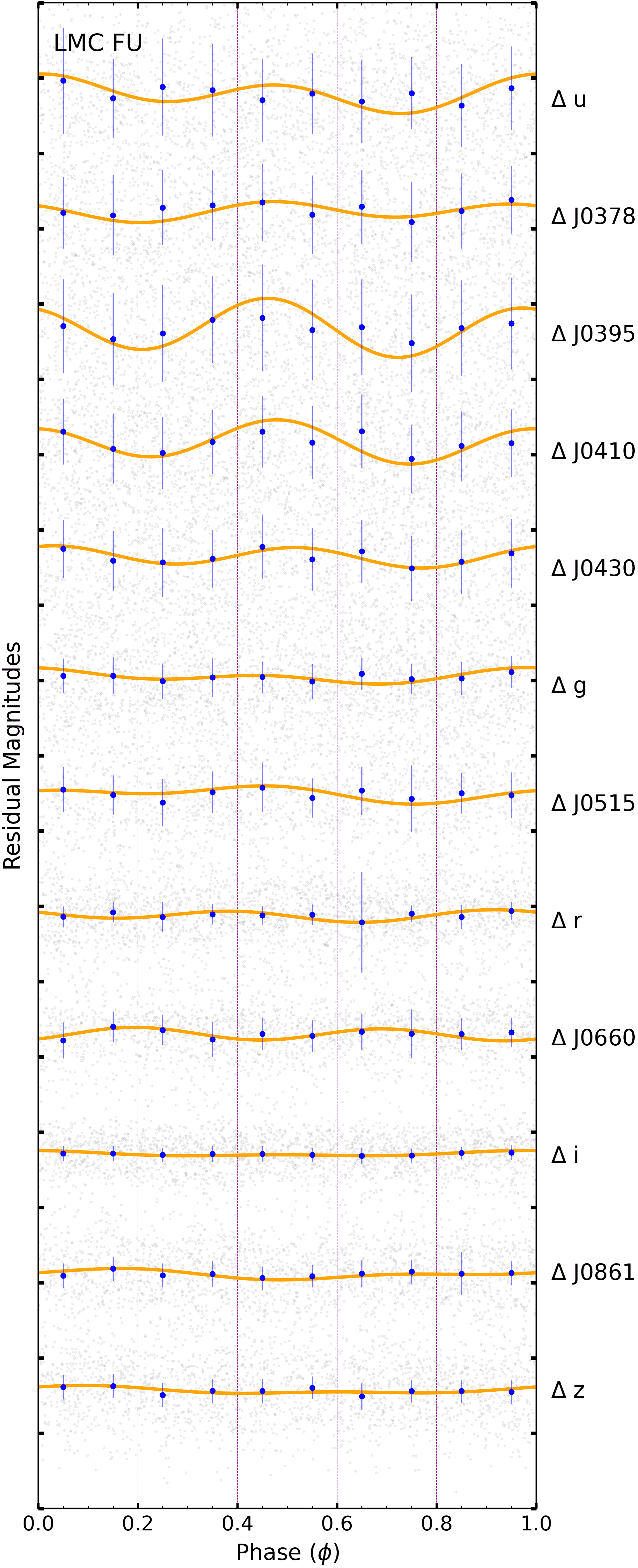}
    \includegraphics[width=0.246\linewidth]{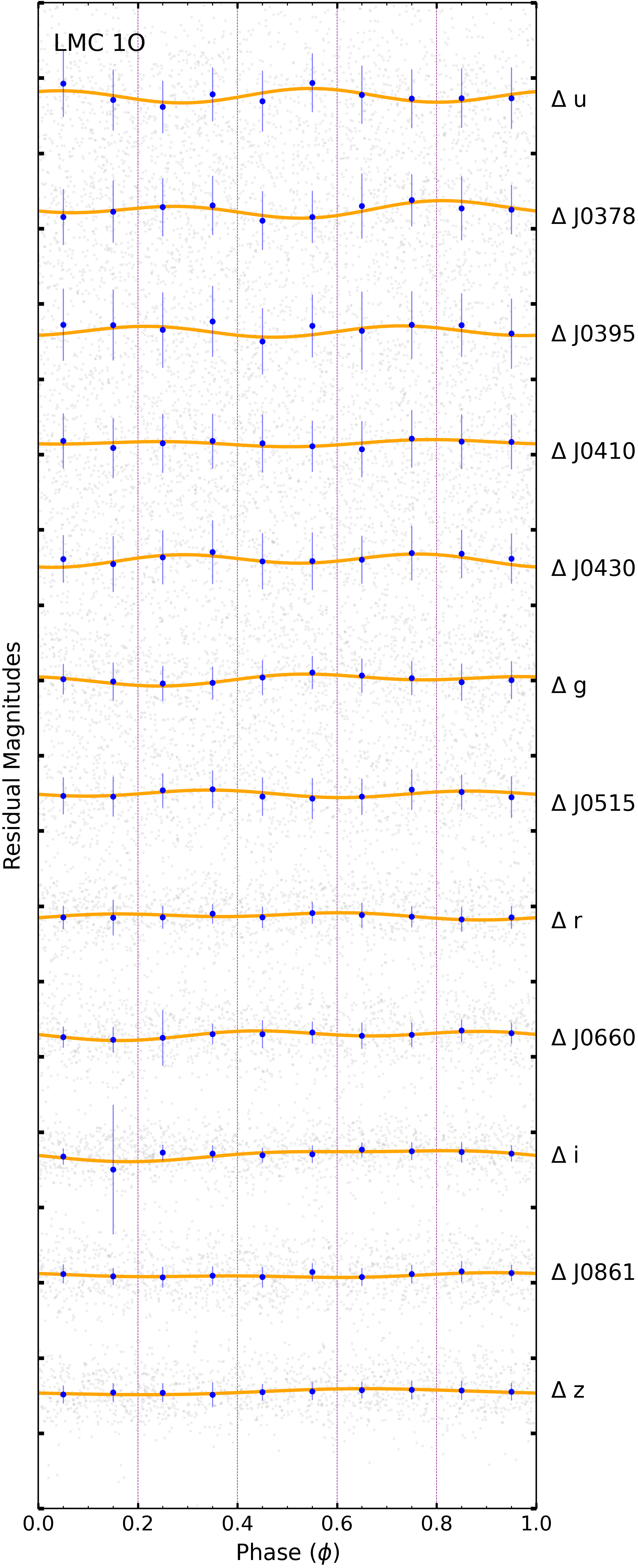}
    \includegraphics[width=0.246\linewidth]{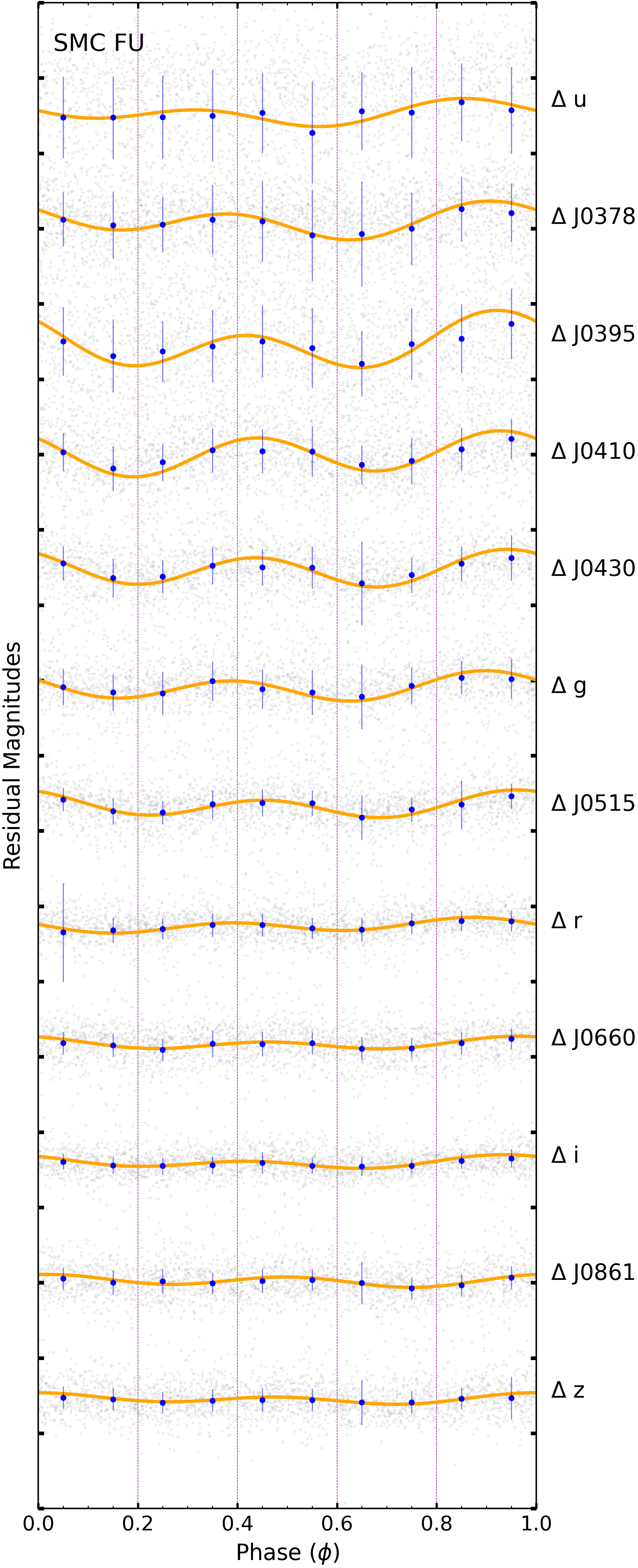}
    \includegraphics[width=0.246\linewidth]{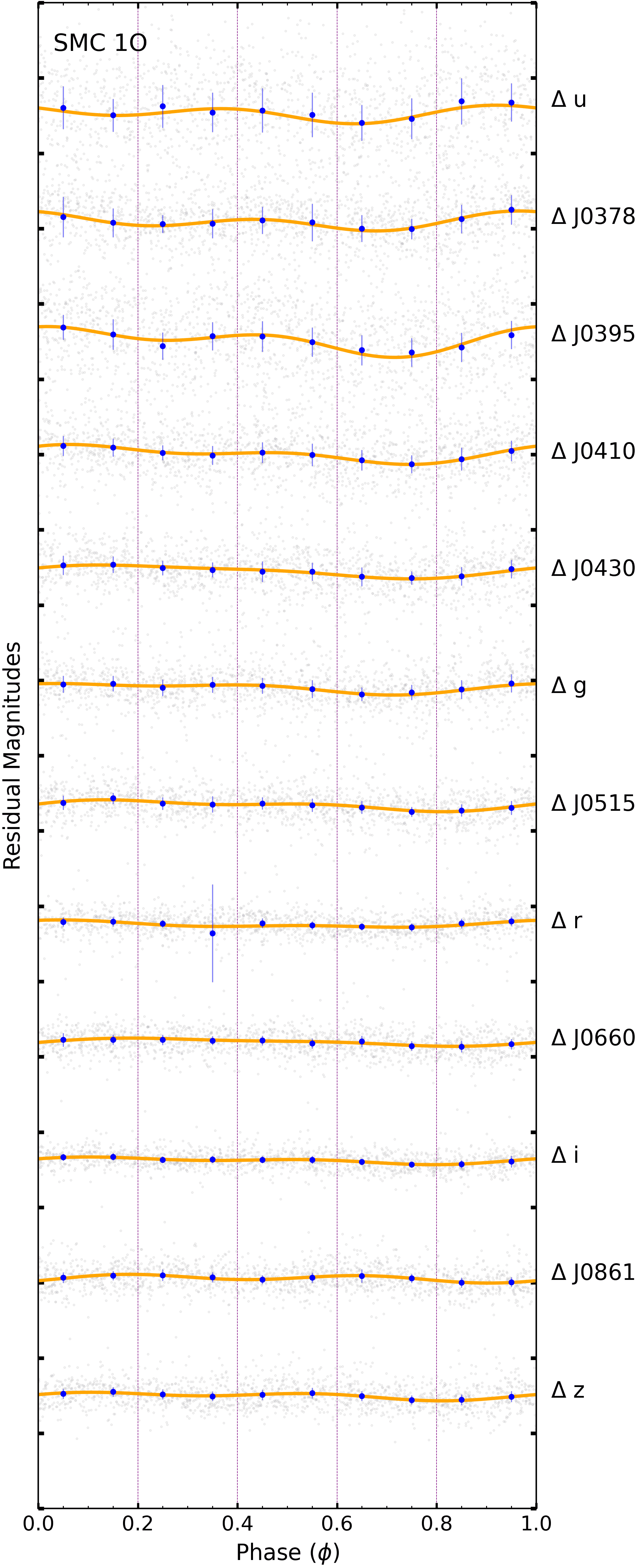}
    \caption{Variation of correction term $\Omega_{\lambda}(\phi)$ for LMC and SMC Cepheids across the 12 S-PLUS bands. The gray dots in the background are the individual residuals, blue points are averages with standard deviation for 10 equally spaced phase bins, and the orange curves represent the fitted $\Omega_{\lambda}(\phi)$.}
    \label{fig:omegaplot_lmc}
\end{figure*}
 
\begin{figure*}
    \centering
    \includegraphics[width=0.23\linewidth]{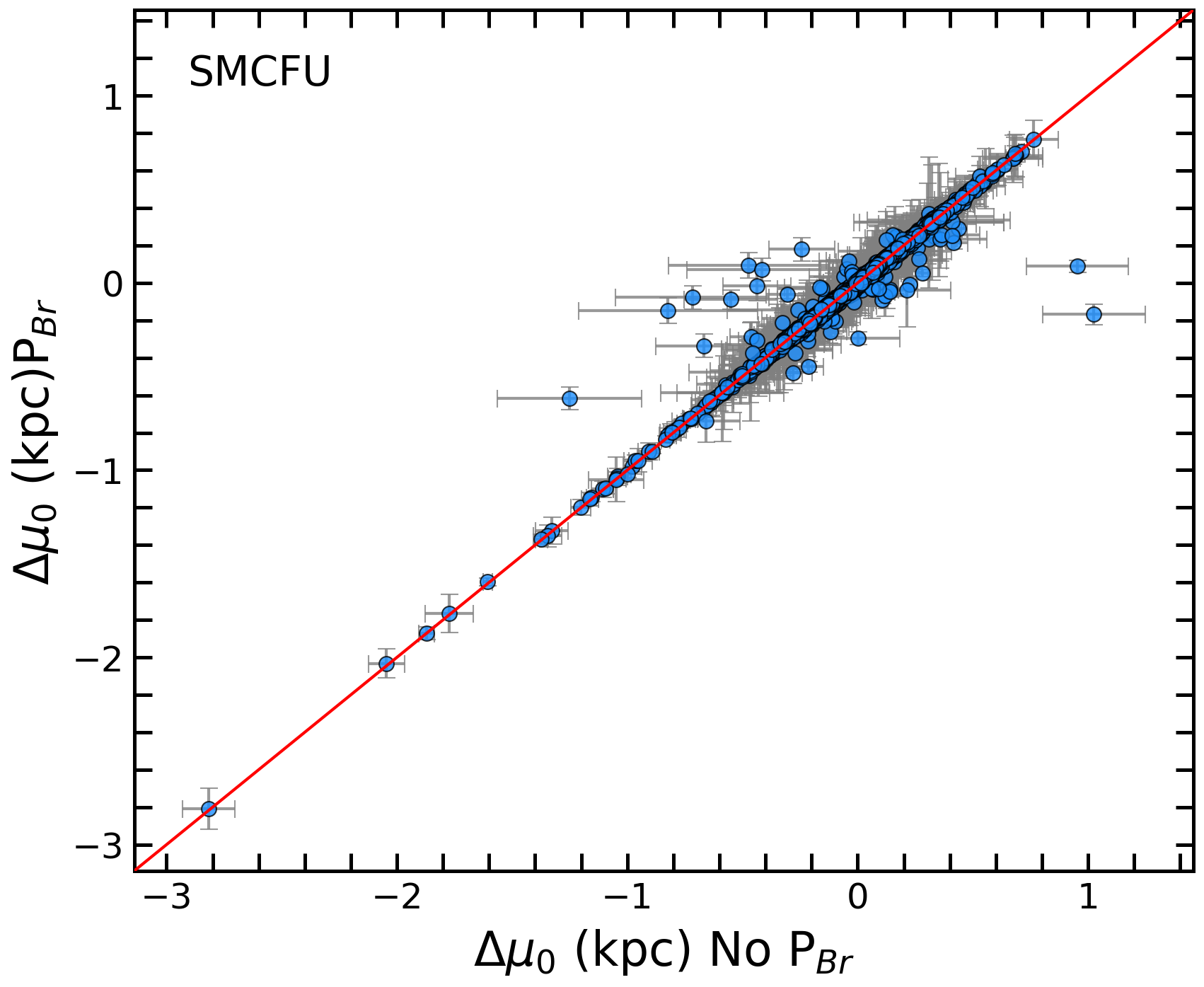}
    \includegraphics[width=0.23\linewidth]{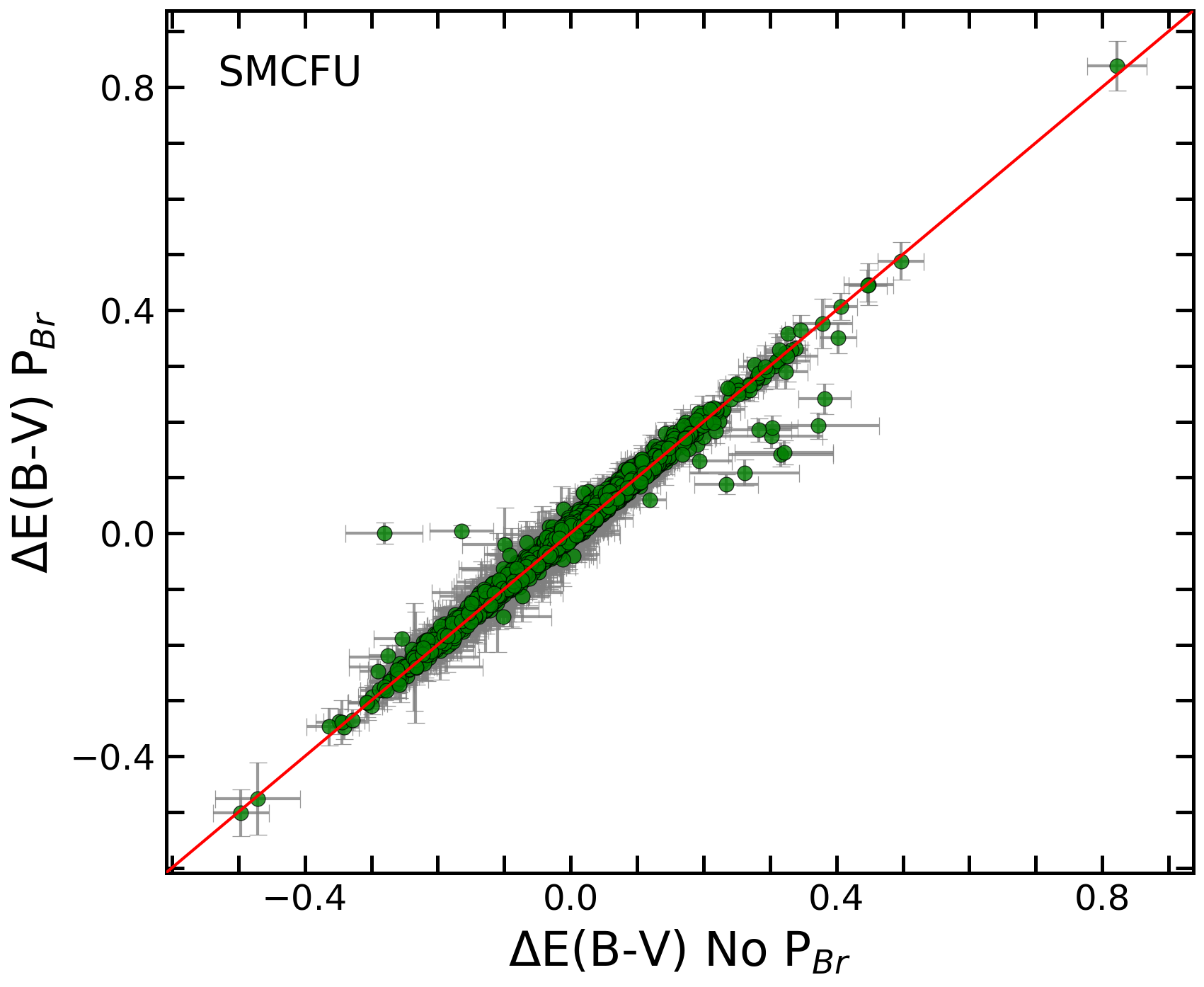}
    \includegraphics[width=0.23\linewidth]{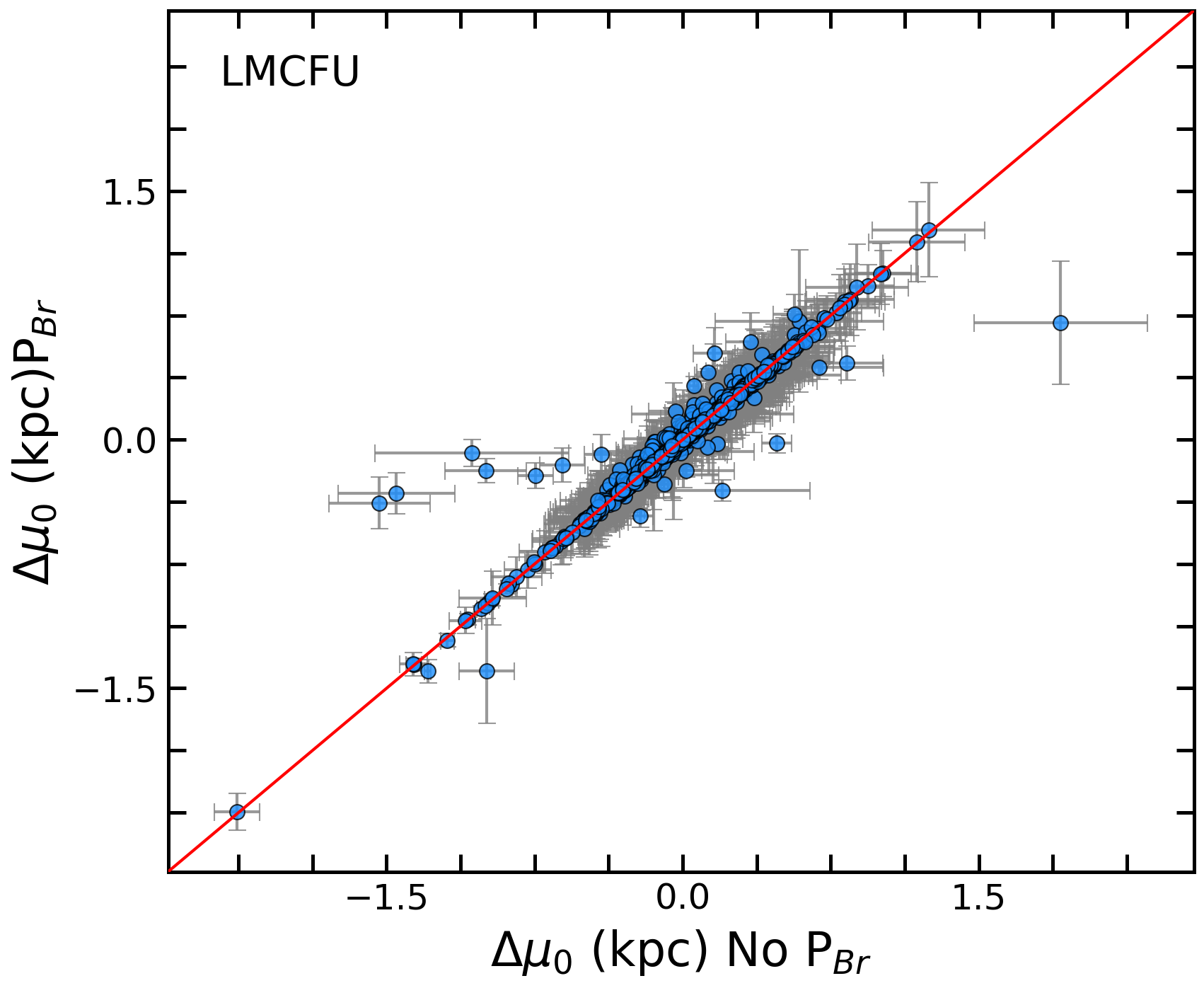}
    \includegraphics[width=0.23\linewidth]{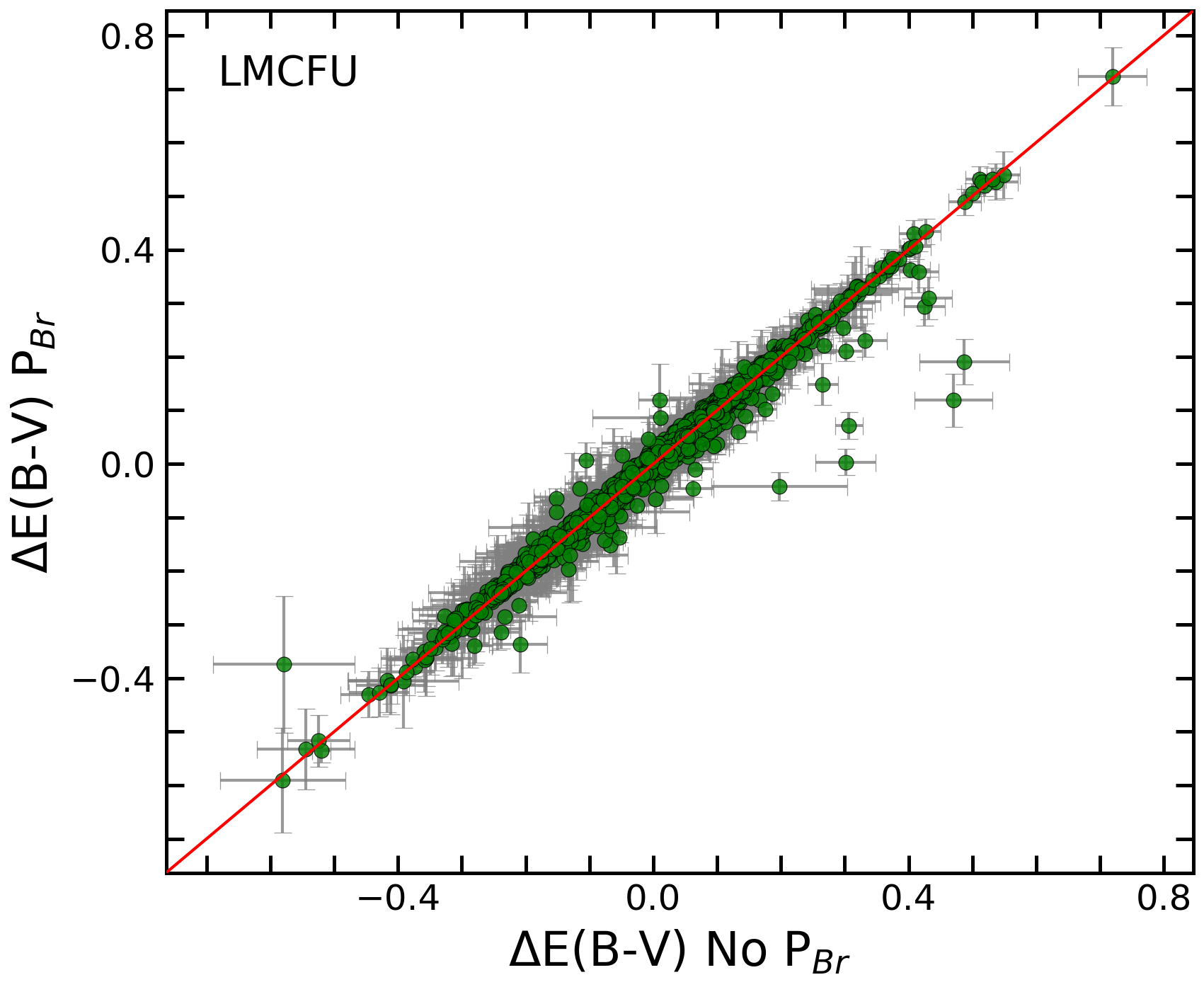}
    \caption{Comparison of $\Delta \mu_{0}$  and $\Delta$ E(B-V)  values calculated with and without considering the break point in the period for SMC FU and LMC FU Cepheids.}
    \label{fig:nobr_vs_br}
\end{figure*}

\onecolumn
\section{Data Tables}
\label{app:table}

\begin{table}[!htbp]
\centering
\caption{Corrected mean magnitudes of classical Cepheids in S-PLUS, distance modulus and reddening offsets relative to the mean values of the host galaxies derived from without considering the period breaks in P-L relations.}
\scriptsize

\resizebox{\columnwidth}{!}{
\begin{tabular}{l l c c c c c c c c c c c c c}
\hline
SPLUS-IDs & Sample & RAJ2000 & DEJ2000 & DMoff & EBVoff & e\_DMoff & e\_EBVoff &
cor\_u & cor\_J0378 & $\cdots$ &
err\_u & err\_J0378 & $\cdots$ & N \\
\hline

MC0114\_000004527 & SMCFU & 9.9998827 & -73.92684174 & 0.678 & 0.049 & 0.039 & 0.014 &
17.667 & 16.860 & $\cdots$ &
0.082 & 0.075 & $\cdots$ & 12 \\

MC0115\_000029862 & SMCFU & 11.84041882 & -73.10324097 & 0.657 & -0.068 & 0.045 & 0.017 &
17.924 & 17.488 & $\cdots$ &
0.103 & 0.099 & $\cdots$ & 12 \\

MC0094\_000035963 & SMCFU & 14.36598587 & -72.56373596 & 0.639 & 0.045 & 0.041 & 0.015 &
18.636 & 17.823 & $\cdots$ &
0.091 & 0.085 & $\cdots$ & 12 \\

MC0134\_000028838 & SMCFU & 9.95636177 & -74.51816559 & 0.630 & 0.020 & 0.042 & 0.015 &
18.176 & 17.409 & $\cdots$ &
0.090 & 0.083 & $\cdots$ & 12 \\

\ldots & \ldots & \ldots & \ldots & \ldots & \ldots & \ldots & \ldots &
\ldots & \ldots & \ldots &
\ldots & \ldots & \ldots & \ldots \\

\hline
\end{tabular}
}

\tablefoot{Columns 1--4 list the Cepheid identifier, Cepheid types in host galaxies, and equatorial coordinates (J2000). Columns \texttt{DMoff} and \texttt{EBVoff} list the distance modulus and reddening offsets relative to the host-galaxy mean values, with their corresponding uncertainties in \texttt{e\_DMoff} and \texttt{e\_EBVoff}. The remaining columns provide the corrected S-PLUS mean magnitudes and associated uncertainties after disentangling the effects of distance and reddening offsets. Column \texttt{N} lists the number of observed bands. The full table is available online at the CDS.}
\label{table:catalog}
\end{table}

\begin{table*}[!htbp]
\centering
\caption{Same as Table~\ref{tab:lmc_pl_FU} but for LMC 1O.}
\label{tab:lmc_pl_1O}
\resizebox{\linewidth}{!}{
\renewcommand{\arraystretch}{1.5}
\begin{tabular}{ccccccccccccccccc}
\multicolumn{17}{c}{\Large \textbf{LMC 1O}} \\
\toprule
\Large \textbf{Filter} & \multicolumn{4}{c}{\Large \textbf{without \(P_\text{br}\)}} & \multicolumn{4}{c}{\Large \textbf{with \( P \leq P_\text{br}{(=0.58)} \)}} & \multicolumn{4}{c}{\Large \textbf{with \( P_\text{br}^{(=0.58)} < P < P_\text{br}^{(=2.5)} \)}} & \multicolumn{4}{c}{\Large \textbf{with \( P > P_\text{br}{(=2.5)} \)}} \\
\cmidrule(lr){2-5} \cmidrule(lr){6-9} \cmidrule(lr){10-13} \cmidrule(lr){14-17}
& {\Large $\alpha$} & {\Large $\beta$} & {\Large rms} & {\Large N} 
& {\Large $\alpha$} & {\Large $\beta$} & {\Large rms} & {\Large N} 
& {\Large $\alpha$} & {\Large $\beta$} & {\Large rms} & {\Large N} 
& {\Large $\alpha$} & {\Large $\beta$} & {\Large rms} & {\Large N} \\
\midrule

\multicolumn{17}{c}{\textbf{\large Iteration 1}}\\
 u & $-2.814 \pm 0.109$ & $18.738 \pm 0.039$ & $0.365$ & 1513 & $-2.337 \pm 1.344$ & $18.745 \pm 0.588$ & $0.370$ & 57 & $-3.025 \pm 0.203$ & $18.807 \pm 0.053$ & $0.368$ & 1077 & $-2.345 \pm 0.581$ & $18.452 \pm 0.310$ & $0.339$ & 379 \\
\rowcolor[HTML]{EFEFEF}
 J0378 & $-2.940 \pm 0.105$ & $17.962 \pm 0.038$ & $0.391$ & 1527 & $-2.051 \pm 1.158$ & $18.237 \pm 0.514$ & $0.352$ & 68 & $-3.167 \pm 0.203$ & $18.027 \pm 0.053$ & $0.398$ & 1080 & $-2.042 \pm 0.584$ & $17.462 \pm 0.311$ & $0.360$ & 379 \\
 J0395 & $-2.752 \pm 0.106$ & $17.762 \pm 0.038$ & $0.401$ & 1520 & $-2.239 \pm 1.181$ & $17.824 \pm 0.534$ & $0.378$ & 66 & $-3.075 \pm 0.203$ & $17.848 \pm 0.053$ & $0.403$ & 1076 & $-1.689 \pm 0.582$ & $17.182 \pm 0.310$ & $0.379$ & 378 \\
\rowcolor[HTML]{EFEFEF}
 J0410 & $-2.964 \pm 0.102$ & $17.478 \pm 0.037$ & $0.380$ & 1549 & $-2.072 \pm 1.152$ & $17.805 \pm 0.515$ & $0.375$ & 77 & $-3.170 \pm 0.202$ & $17.536 \pm 0.053$ & $0.383$ & 1086 & $-1.911 \pm 0.575$ & $16.905 \pm 0.306$ & $0.357$ & 386 \\
 J0430 & $-3.010 \pm 0.103$ & $17.396 \pm 0.037$ & $0.365$ & 1509 & $-2.452 \pm 1.119$ & $17.546 \pm 0.511$ & $0.355$ & 78 & $-3.182 \pm 0.204$ & $17.449 \pm 0.053$ & $0.369$ & 1059 & $-2.153 \pm 0.592$ & $16.916 \pm 0.316$ & $0.344$ & 372 \\
\rowcolor[HTML]{EFEFEF}
 g & $-3.050 \pm 0.101$ & $17.243 \pm 0.036$ & $0.312$ & 1564 & $-2.384 \pm 1.048$ & $17.451 \pm 0.474$ & $0.295$ & 83 & $-3.168 \pm 0.202$ & $17.284 \pm 0.053$ & $0.320$ & 1094 & $-2.395 \pm 0.582$ & $16.867 \pm 0.310$ & $0.276$ & 387 \\
 J0515 & $-3.083 \pm 0.100$ & $17.105 \pm 0.036$ & $0.301$ & 1566 & $-2.650 \pm 1.073$ & $17.244 \pm 0.486$ & $0.298$ & 83 & $-3.191 \pm 0.202$ & $17.141 \pm 0.053$ & $0.308$ & 1095 & $-2.355 \pm 0.579$ & $16.697 \pm 0.309$ & $0.272$ & 388 \\
\rowcolor[HTML]{EFEFEF}
 r & $-3.163 \pm 0.101$ & $16.922 \pm 0.036$ & $0.247$ & 1554 & $-2.686 \pm 1.057$ & $17.070 \pm 0.478$ & $0.244$ & 82 & $-3.249 \pm 0.201$ & $16.953 \pm 0.052$ & $0.255$ & 1092 & $-2.601 \pm 0.589$ & $16.601 \pm 0.314$ & $0.214$ & 380 \\
 J0660 & $-3.193 \pm 0.100$ & $16.931 \pm 0.036$ & $0.236$ & 1566 & $-2.938 \pm 1.115$ & $16.986 \pm 0.507$ & $0.242$ & 83 & $-3.263 \pm 0.201$ & $16.959 \pm 0.052$ & $0.245$ & 1096 & $-2.831 \pm 0.575$ & $16.717 \pm 0.307$ & $0.196$ & 387 \\
\rowcolor[HTML]{EFEFEF}
 i & $-3.202 \pm 0.101$ & $16.820 \pm 0.036$ & $0.228$ & 1556 & $-2.975 \pm 1.071$ & $16.882 \pm 0.480$ & $0.228$ & 83 & $-3.290 \pm 0.202$ & $16.847 \pm 0.053$ & $0.206$ & 1091 & $-2.778 \pm 0.579$ & $16.583 \pm 0.309$ & $0.277$ & 382 \\
 J0861 & $-3.228 \pm 0.100$ & $16.808 \pm 0.036$ & $0.182$ & 1570 & $-3.070 \pm 1.107$ & $16.823 \pm 0.506$ & $0.201$ & 83 & $-3.310 \pm 0.201$ & $16.834 \pm 0.052$ & $0.185$ & 1099 & $-2.898 \pm 0.575$ & $16.619 \pm 0.307$ & $0.160$ & 388 \\
\rowcolor[HTML]{EFEFEF}
 z & $-3.227 \pm 0.101$ & $16.779 \pm 0.036$ & $0.187$ & 1562 & $-3.195 \pm 1.078$ & $16.739 \pm 0.488$ & $0.212$ & 81 & $-3.330 \pm 0.201$ & $16.810 \pm 0.052$ & $0.193$ & 1096 & $-2.949 \pm 0.578$ & $16.619 \pm 0.309$ & $0.157$ & 385 \\

\multicolumn{17}{c}{\textbf{\large Iteration 2}}\\
 u & $-2.842 \pm 0.008$ & $18.747 \pm 0.003$ & $0.178$ & 1513 & $-2.562 \pm 0.144$ & $18.724 \pm 0.060$ & $0.194$ & 57 & $-3.021 \pm 0.014$ & $18.802 \pm 0.004$ & $0.175$ & 1077 & $-2.226 \pm 0.032$ & $18.391 \pm 0.017$ & $0.169$ & 379 \\
\rowcolor[HTML]{EFEFEF}
 J0378 & $-2.933 \pm 0.007$ & $17.955 \pm 0.003$ & $0.167$ & 1527 & $-2.259 \pm 0.108$ & $18.192 \pm 0.046$ & $0.180$ & 68 & $-3.188 \pm 0.013$ & $18.023 \pm 0.004$ & $0.174$ & 1080 & $-1.898 \pm 0.032$ & $17.387 \pm 0.017$ & $0.140$ & 379 \\
 J0395 & $-2.745 \pm 0.008$ & $17.750 \pm 0.003$ & $0.167$ & 1520 & $-2.045 \pm 0.117$ & $17.954 \pm 0.051$ & $0.205$ & 66 & $-3.092 \pm 0.014$ & $17.839 \pm 0.004$ & $0.167$ & 1076 & $-1.574 \pm 0.032$ & $17.116 \pm 0.017$ & $0.158$ & 378 \\
\rowcolor[HTML]{EFEFEF}
 J0410 & $-2.942 \pm 0.007$ & $17.465 \pm 0.003$ & $0.181$ & 1549 & $-2.121 \pm 0.099$ & $17.789 \pm 0.043$ & $0.234$ & 77 & $-3.174 \pm 0.013$ & $17.528 \pm 0.003$ & $0.176$ & 1086 & $-1.867 \pm 0.031$ & $16.877 \pm 0.017$ & $0.178$ & 386 \\
 J0430 & $-3.004 \pm 0.007$ & $17.391 \pm 0.003$ & $0.173$ & 1509 & $-2.512 \pm 0.092$ & $17.542 \pm 0.040$ & $0.224$ & 78 & $-3.203 \pm 0.012$ & $17.447 \pm 0.003$ & $0.171$ & 1059 & $-2.108 \pm 0.032$ & $16.894 \pm 0.017$ & $0.173$ & 372 \\
\rowcolor[HTML]{EFEFEF}
 g & $-3.068 \pm 0.006$ & $17.249 \pm 0.002$ & $0.173$ & 1564 & $-2.508 \pm 0.071$ & $17.451 \pm 0.031$ & $0.182$ & 83 & $-3.203 \pm 0.011$ & $17.290 \pm 0.003$ & $0.180$ & 1094 & $-2.326 \pm 0.030$ & $16.834 \pm 0.016$ & $0.148$ & 387 \\
 J0515 & $-3.083 \pm 0.006$ & $17.104 \pm 0.002$ & $0.169$ & 1566 & $-2.722 \pm 0.080$ & $17.245 \pm 0.035$ & $0.183$ & 83 & $-3.212 \pm 0.012$ & $17.142 \pm 0.003$ & $0.173$ & 1095 & $-2.330 \pm 0.031$ & $16.685 \pm 0.017$ & $0.157$ & 388 \\
\rowcolor[HTML]{EFEFEF}
 r & $-3.175 \pm 0.006$ & $16.928 \pm 0.002$ & $0.149$ & 1554 & $-2.765 \pm 0.071$ & $17.084 \pm 0.031$ & $0.149$ & 82 & $-3.256 \pm 0.011$ & $16.956 \pm 0.003$ & $0.157$ & 1092 & $-2.582 \pm 0.031$ & $16.592 \pm 0.017$ & $0.121$ & 380 \\
 J0660 & $-3.199 \pm 0.006$ & $16.936 \pm 0.002$ & $0.150$ & 1566 & $-3.165 \pm 0.077$ & $16.938 \pm 0.034$ & $0.156$ & 83 & $-3.262 \pm 0.011$ & $16.959 \pm 0.003$ & $0.159$ & 1096 & $-2.764 \pm 0.030$ & $16.685 \pm 0.016$ & $0.116$ & 387 \\
\rowcolor[HTML]{EFEFEF}
 i & $-3.218 \pm 0.006$ & $16.827 \pm 0.002$ & $0.161$ & 1556 & $-2.943 \pm 0.075$ & $16.927 \pm 0.032$ & $0.146$ & 83 & $-3.281 \pm 0.011$ & $16.850 \pm 0.003$ & $0.126$ & 1091 & $-2.759 \pm 0.031$ & $16.566 \pm 0.017$ & $0.209$ & 382 \\
 J0861 & $-3.231 \pm 0.007$ & $16.816 \pm 0.002$ & $0.105$ & 1570 & $-3.136 \pm 0.092$ & $16.832 \pm 0.040$ & $0.111$ & 83 & $-3.295 \pm 0.012$ & $16.838 \pm 0.003$ & $0.110$ & 1099 & $-2.839 \pm 0.031$ & $16.592 \pm 0.017$ & $0.095$ & 388 \\
\rowcolor[HTML]{EFEFEF}
 z & $-3.230 \pm 0.007$ & $16.785 \pm 0.002$ & $0.119$ & 1562 & $-3.189 \pm 0.089$ & $16.774 \pm 0.038$ & $0.119$ & 81 & $-3.297 \pm 0.012$ & $16.809 \pm 0.003$ & $0.127$ & 1096 & $-2.876 \pm 0.031$ & $16.581 \pm 0.017$ & $0.095$ & 385 \\

\bottomrule
\end{tabular}
}
\end{table*}

\begin{table*}[!ht] 
\centering
\caption{Same as Table~\ref{tab:lmc_pl_FU} but for SMC FU.}
\label{tab:smc_pl_FU}
\resizebox{0.97\textwidth}{!}{
\begin{tabular}{ccccccccccccc}

\multicolumn{13}{c}{\textbf{SMC FU}} \\
\toprule
\textbf{Filter} & 
\multicolumn{4}{c}{\textbf{without \(P_\text{br}\)}} & 
\multicolumn{4}{c}{\textbf{with \(P \leq P_\text{br}(=10.0)\)}} & 
\multicolumn{4}{c}{\textbf{with \(P > P_\text{br} (=10.0)\)}} \\
\cmidrule(lr){2-5} \cmidrule(lr){6-9} \cmidrule(lr){10-13}
 & \(\alpha\) & \(\beta\) & rms & N & 
\(\alpha\) & \(\beta\) & rms & N & 
\(\alpha\) & \(\beta\) & rms & N \\
\midrule

\multicolumn{13}{c}{\textbf{Iteration 1}}\\
 u & $-2.318 \pm 0.067$ & $19.521 \pm 0.033$ & $0.556$ & 2484 & $-2.408 \pm 0.090$ & $19.548 \pm 0.038$ & $0.480$ & 2364 & $-1.953 \pm 0.396$ & $19.147 \pm 0.506$ & $0.739$ & 120 \\
\rowcolor[HTML]{EFEFEF}
 J0378 & $-2.366 \pm 0.066$ & $18.764 \pm 0.032$ & $0.617$ & 2586 & $-2.480 \pm 0.088$ & $18.798 \pm 0.037$ & $0.533$ & 2467 & $-1.991 \pm 0.399$ & $18.408 \pm 0.512$ & $0.821$ & 119 \\
 J0395 & $-2.206 \pm 0.066$ & $18.490 \pm 0.033$ & $0.671$ & 2547 & $-2.330 \pm 0.088$ & $18.527 \pm 0.037$ & $0.582$ & 2427 & $-2.062 \pm 0.396$ & $18.453 \pm 0.506$ & $0.844$ & 120 \\
\rowcolor[HTML]{EFEFEF}
 J0410 & $-2.488 \pm 0.066$ & $18.367 \pm 0.032$ & $0.605$ & 2625 & $-2.599 \pm 0.087$ & $18.399 \pm 0.036$ & $0.529$ & 2505 & $-2.354 \pm 0.396$ & $18.330 \pm 0.506$ & $0.708$ & 120 \\
 J0430 & $-2.585 \pm 0.065$ & $18.322 \pm 0.032$ & $0.578$ & 2648 & $-2.688 \pm 0.087$ & $18.352 \pm 0.036$ & $0.508$ & 2528 & $-2.490 \pm 0.396$ & $18.326 \pm 0.506$ & $0.648$ & 120 \\
\rowcolor[HTML]{EFEFEF}
 g & $-2.743 \pm 0.066$ & $18.213 \pm 0.032$ & $0.510$ & 2653 & $-2.811 \pm 0.087$ & $18.233 \pm 0.036$ & $0.452$ & 2535 & $-2.549 \pm 0.446$ & $18.064 \pm 0.558$ & $0.509$ & 118 \\
 J0515 & $-2.814 \pm 0.065$ & $18.089 \pm 0.032$ & $0.483$ & 2664 & $-2.870 \pm 0.087$ & $18.105 \pm 0.036$ & $0.429$ & 2544 & $-2.820 \pm 0.396$ & $18.170 \pm 0.506$ & $0.475$ & 120 \\
\rowcolor[HTML]{EFEFEF}
 r & $-2.955 \pm 0.068$ & $17.931 \pm 0.032$ & $0.411$ & 2640 & $-2.995 \pm 0.087$ & $17.942 \pm 0.036$ & $0.367$ & 2527 & $-2.868 \pm 0.588$ & $17.891 \pm 0.716$ & $0.363$ & 113 \\
 J0660 & $-3.005 \pm 0.068$ & $17.943 \pm 0.033$ & $0.388$ & 2632 & $-3.038 \pm 0.087$ & $17.952 \pm 0.036$ & $0.346$ & 2519 & $-2.992 \pm 0.598$ & $17.981 \pm 0.735$ & $0.343$ & 113 \\
\rowcolor[HTML]{EFEFEF}
 i & $-3.044 \pm 0.069$ & $17.836 \pm 0.033$ & $0.350$ & 2637 & $-3.081 \pm 0.087$ & $17.847 \pm 0.036$ & $0.313$ & 2529 & $-2.905 \pm 0.623$ & $17.727 \pm 0.756$ & $0.294$ & 108 \\
 J0861 & $-3.085 \pm 0.067$ & $17.828 \pm 0.032$ & $0.336$ & 2655 & $-3.119 \pm 0.087$ & $17.838 \pm 0.036$ & $0.300$ & 2539 & $-2.908 \pm 0.471$ & $17.654 \pm 0.587$ & $0.292$ & 116 \\
\rowcolor[HTML]{EFEFEF}
 z & $-3.096 \pm 0.068$ & $17.804 \pm 0.032$ & $0.333$ & 2645 & $-3.118 \pm 0.087$ & $17.810 \pm 0.036$ & $0.297$ & 2532 & $-3.118 \pm 0.573$ & $17.869 \pm 0.704$ & $0.279$ & 113 \\

\multicolumn{13}{c}{\textbf{Iteration 2}}\\
 u & $-2.403 \pm 0.004$ & $19.587 \pm 0.002$ & $0.225$ & 2484 & $-2.488 \pm 0.006$ & $19.611 \pm 0.003$ & $0.190$ & 2364 & $-2.144 \pm 0.021$ & $19.317 \pm 0.027$ & $0.343$ & 120 \\
\rowcolor[HTML]{EFEFEF}
 J0378 & $-2.409 \pm 0.004$ & $18.786 \pm 0.002$ & $0.244$ & 2586 & $-2.517 \pm 0.006$ & $18.818 \pm 0.003$ & $0.206$ & 2467 & $-2.147 \pm 0.021$ & $18.530 \pm 0.027$ & $0.365$ & 119 \\
 J0395 & $-2.241 \pm 0.004$ & $18.495 \pm 0.002$ & $0.222$ & 2547 & $-2.354 \pm 0.006$ & $18.528 \pm 0.003$ & $0.184$ & 2427 & $-2.241 \pm 0.021$ & $18.598 \pm 0.027$ & $0.376$ & 120 \\
\rowcolor[HTML]{EFEFEF}
 J0410 & $-2.505 \pm 0.004$ & $18.371 \pm 0.002$ & $0.232$ & 2625 & $-2.602 \pm 0.005$ & $18.399 \pm 0.002$ & $0.200$ & 2505 & $-2.479 \pm 0.020$ & $18.427 \pm 0.026$ & $0.306$ & 120 \\
 J0430 & $-2.602 \pm 0.004$ & $18.327 \pm 0.002$ & $0.241$ & 2648 & $-2.691 \pm 0.005$ & $18.353 \pm 0.002$ & $0.210$ & 2528 & $-2.595 \pm 0.020$ & $18.403 \pm 0.026$ & $0.278$ & 120 \\
\rowcolor[HTML]{EFEFEF}
 g & $-2.790 \pm 0.004$ & $18.243 \pm 0.002$ & $0.230$ & 2653 & $-2.859 \pm 0.005$ & $18.263 \pm 0.002$ & $0.204$ & 2535 & $-2.695 \pm 0.023$ & $18.203 \pm 0.029$ & $0.223$ & 118 \\
 J0515 & $-2.852 \pm 0.004$ & $18.112 \pm 0.002$ & $0.215$ & 2664 & $-2.909 \pm 0.005$ & $18.128 \pm 0.002$ & $0.190$ & 2544 & $-2.871 \pm 0.020$ & $18.195 \pm 0.026$ & $0.209$ & 120 \\
\rowcolor[HTML]{EFEFEF}
 r & $-3.006 \pm 0.004$ & $17.966 \pm 0.002$ & $0.184$ & 2640 & $-3.050 \pm 0.005$ & $17.978 \pm 0.002$ & $0.166$ & 2527 & $-2.980 \pm 0.030$ & $17.999 \pm 0.037$ & $0.116$ & 113 \\
 J0660 & $-3.059 \pm 0.004$ & $17.980 \pm 0.002$ & $0.171$ & 2632 & $-3.097 \pm 0.005$ & $17.990 \pm 0.002$ & $0.154$ & 2519 & $-3.059 \pm 0.031$ & $18.036 \pm 0.038$ & $0.123$ & 113 \\
\rowcolor[HTML]{EFEFEF}
 i & $-3.092 \pm 0.004$ & $17.873 \pm 0.002$ & $0.137$ & 2637 & $-3.133 \pm 0.005$ & $17.884 \pm 0.002$ & $0.124$ & 2529 & $-2.897 \pm 0.032$ & $17.698 \pm 0.039$ & $0.064$ & 108 \\
 J0861 & $-3.122 \pm 0.004$ & $17.857 \pm 0.002$ & $0.137$ & 2655 & $-3.160 \pm 0.005$ & $17.869 \pm 0.002$ & $0.121$ & 2539 & $-2.866 \pm 0.025$ & $17.591 \pm 0.031$ & $0.136$ & 116 \\
\rowcolor[HTML]{EFEFEF}
 z & $-3.134 \pm 0.004$ & $17.835 \pm 0.002$ & $0.136$ & 2645 & $-3.160 \pm 0.005$ & $17.842 \pm 0.002$ & $0.122$ & 2532 & $-3.035 \pm 0.030$ & $17.759 \pm 0.036$ & $0.118$ & 113 \\
\bottomrule
\end{tabular}
}
\end{table*}

\begin{table*}[!ht] 
\centering
\caption{Same as Table~\ref{tab:lmc_pl_FU} but for SMC 1O.}
\label{tab:smc_pl_1O}
\resizebox{0.97\textwidth}{!}{
\begin{tabular}{ccccccccccccc}

\multicolumn{13}{c}{\textbf{SMC 1O}} \\
\toprule
\textbf{Filter} & 
\multicolumn{4}{c}{\textbf{without \(P_\text{br}\)}} & 
\multicolumn{4}{c}{\textbf{with \(P \leq P_\text{br}(=10.0)\)}} & 
\multicolumn{4}{c}{\textbf{with \(P > P_\text{br} (=10.0)\)}} \\
\cmidrule(lr){2-5} \cmidrule(lr){6-9} \cmidrule(lr){10-13}
 & \(\alpha\) & \(\beta\) & rms & N & 
\(\alpha\) & \(\beta\) & rms & N & 
\(\alpha\) & \(\beta\) & rms & N \\
\midrule

\multicolumn{13}{c}{\textbf{Iteration 1}}\\
 u & $-2.690 \pm 0.124$ & $18.936 \pm 0.030$ & $0.496$ & 1614 & $-2.784 \pm 0.164$ & $18.940 \pm 0.031$ & $0.443$ & 1415 & $-0.992 \pm 1.215$ & $18.150 \pm 0.591$ & $0.425$ & 199 \\
\rowcolor[HTML]{EFEFEF}
 J0378 & $-2.791 \pm 0.121$ & $18.126 \pm 0.029$ & $0.514$ & 1665 & $-2.934 \pm 0.159$ & $18.130 \pm 0.030$ & $0.454$ & 1464 & $-1.117 \pm 1.213$ & $17.373 \pm 0.591$ & $0.477$ & 201 \\
 J0395 & $-2.629 \pm 0.122$ & $17.838 \pm 0.029$ & $0.540$ & 1666 & $-2.780 \pm 0.160$ & $17.843 \pm 0.030$ & $0.480$ & 1465 & $-0.831 \pm 1.207$ & $17.032 \pm 0.588$ & $0.475$ & 201 \\
\rowcolor[HTML]{EFEFEF}
 J0410 & $-2.791 \pm 0.118$ & $17.686 \pm 0.029$ & $0.504$ & 1690 & $-2.943 \pm 0.154$ & $17.690 \pm 0.029$ & $0.446$ & 1488 & $-1.126 \pm 1.210$ & $16.950 \pm 0.589$ & $0.456$ & 202 \\
 J0430 & $-2.859 \pm 0.117$ & $17.639 \pm 0.029$ & $0.483$ & 1697 & $-2.994 \pm 0.153$ & $17.642 \pm 0.029$ & $0.428$ & 1495 & $-1.256 \pm 1.210$ & $16.927 \pm 0.589$ & $0.431$ & 202 \\
\rowcolor[HTML]{EFEFEF}
 g & $-2.973 \pm 0.116$ & $17.558 \pm 0.028$ & $0.441$ & 1707 & $-3.093 \pm 0.149$ & $17.560 \pm 0.028$ & $0.394$ & 1507 & $-1.735 \pm 1.232$ & $17.014 \pm 0.599$ & $0.379$ & 200 \\
 J0515 & $-3.001 \pm 0.116$ & $17.436 \pm 0.028$ & $0.418$ & 1709 & $-3.103 \pm 0.150$ & $17.438 \pm 0.028$ & $0.374$ & 1507 & $-1.790 \pm 1.202$ & $16.897 \pm 0.585$ & $0.355$ & 202 \\
\rowcolor[HTML]{EFEFEF}
 r & $-3.100 \pm 0.116$ & $17.314 \pm 0.028$ & $0.380$ & 1710 & $-3.175 \pm 0.149$ & $17.315 \pm 0.028$ & $0.343$ & 1509 & $-2.144 \pm 1.226$ & $16.888 \pm 0.596$ & $0.301$ & 201 \\
 J0660 & $-3.148 \pm 0.115$ & $17.337 \pm 0.028$ & $0.355$ & 1711 & $-3.213 \pm 0.148$ & $17.338 \pm 0.028$ & $0.321$ & 1509 & $-2.175 \pm 1.205$ & $16.893 \pm 0.588$ & $0.275$ & 202 \\
\rowcolor[HTML]{EFEFEF}
 i & $-3.183 \pm 0.115$ & $17.249 \pm 0.028$ & $0.334$ & 1715 & $-3.233 \pm 0.148$ & $17.250 \pm 0.028$ & $0.303$ & 1514 & $-2.339 \pm 1.228$ & $16.864 \pm 0.597$ & $0.248$ & 201 \\
 J0861 & $-3.212 \pm 0.116$ & $17.249 \pm 0.028$ & $0.324$ & 1714 & $-3.259 \pm 0.150$ & $17.250 \pm 0.028$ & $0.295$ & 1511 & $-2.518 \pm 1.202$ & $16.935 \pm 0.585$ & $0.235$ & 203 \\
\rowcolor[HTML]{EFEFEF}
 z & $-3.221 \pm 0.116$ & $17.225 \pm 0.028$ & $0.323$ & 1715 & $-3.265 \pm 0.149$ & $17.226 \pm 0.028$ & $0.295$ & 1514 & $-2.509 \pm 1.228$ & $16.901 \pm 0.597$ & $0.223$ & 201 \\

\multicolumn{13}{c}{\textbf{Iteration 2}}\\
 u & $-2.742 \pm 0.009$ & $18.963 \pm 0.002$ & $0.180$ & 1614 & $-2.877 \pm 0.012$ & $18.967 \pm 0.002$ & $0.160$ & 1415 & $-0.973 \pm 0.071$ & $18.165 \pm 0.035$ & $0.152$ & 199 \\
\rowcolor[HTML]{EFEFEF}
 J0378 & $-2.820 \pm 0.008$ & $18.137 \pm 0.002$ & $0.196$ & 1665 & $-2.983 \pm 0.011$ & $18.142 \pm 0.002$ & $0.161$ & 1464 & $-1.127 \pm 0.068$ & $17.384 \pm 0.033$ & $0.237$ & 201 \\
 J0395 & $-2.633 \pm 0.008$ & $17.838 \pm 0.002$ & $0.187$ & 1666 & $-2.800 \pm 0.011$ & $17.843 \pm 0.002$ & $0.157$ & 1465 & $-0.816 \pm 0.070$ & $17.026 \pm 0.034$ & $0.205$ & 201 \\
\rowcolor[HTML]{EFEFEF}
 J0410 & $-2.788 \pm 0.007$ & $17.688 \pm 0.002$ & $0.192$ & 1690 & $-2.951 \pm 0.010$ & $17.692 \pm 0.002$ & $0.158$ & 1488 & $-1.120 \pm 0.067$ & $16.951 \pm 0.033$ & $0.227$ & 202 \\
 J0430 & $-2.856 \pm 0.007$ & $17.642 \pm 0.002$ & $0.193$ & 1697 & $-3.003 \pm 0.010$ & $17.646 \pm 0.002$ & $0.161$ & 1495 & $-1.274 \pm 0.066$ & $16.941 \pm 0.032$ & $0.219$ & 202 \\
\rowcolor[HTML]{EFEFEF}
 g & $-2.968 \pm 0.006$ & $17.565 \pm 0.002$ & $0.196$ & 1707 & $-3.095 \pm 0.008$ & $17.567 \pm 0.002$ & $0.170$ & 1507 & $-1.737 \pm 0.064$ & $17.029 \pm 0.031$ & $0.191$ & 200 \\
 J0515 & $-2.997 \pm 0.007$ & $17.441 \pm 0.002$ & $0.181$ & 1709 & $-3.105 \pm 0.009$ & $17.443 \pm 0.002$ & $0.157$ & 1507 & $-1.802 \pm 0.064$ & $16.913 \pm 0.031$ & $0.179$ & 202 \\
\rowcolor[HTML]{EFEFEF}
 r & $-3.112 \pm 0.006$ & $17.325 \pm 0.002$ & $0.197$ & 1710 & $-3.192 \pm 0.008$ & $17.326 \pm 0.002$ & $0.179$ & 1509 & $-2.161 \pm 0.064$ & $16.905 \pm 0.031$ & $0.150$ & 201 \\
 J0660 & $-3.156 \pm 0.006$ & $17.349 \pm 0.002$ & $0.166$ & 1711 & $-3.221 \pm 0.008$ & $17.350 \pm 0.002$ & $0.151$ & 1509 & $-2.173 \pm 0.063$ & $16.904 \pm 0.031$ & $0.123$ & 202 \\
\rowcolor[HTML]{EFEFEF}
 i & $-3.184 \pm 0.007$ & $17.259 \pm 0.002$ & $0.161$ & 1715 & $-3.229 \pm 0.009$ & $17.259 \pm 0.002$ & $0.148$ & 1514 & $-2.379 \pm 0.064$ & $16.893 \pm 0.031$ & $0.107$ & 201 \\
 J0861 & $-3.207 \pm 0.007$ & $17.256 \pm 0.002$ & $0.161$ & 1714 & $-3.242 \pm 0.009$ & $17.256 \pm 0.002$ & $0.147$ & 1511 & $-2.544 \pm 0.065$ & $16.955 \pm 0.031$ & $0.118$ & 203 \\
\rowcolor[HTML]{EFEFEF}
 z & $-3.209 \pm 0.007$ & $17.230 \pm 0.002$ & $0.163$ & 1715 & $-3.238 \pm 0.009$ & $17.230 \pm 0.002$ & $0.151$ & 1514 & $-2.553 \pm 0.065$ & $16.930 \pm 0.032$ & $0.094$ & 201 \\

\bottomrule
\end{tabular}
}
\end{table*}

\end{appendix}
\end{document}